\documentclass[twocolumn]{aastex631}

\usepackage{xspace}
\usepackage{csvsimple}

\shorttitle{Dwarf outskirts}
\shortauthors{Pan et al.}

\graphicspath{{./}{figures/}}

\begin{document}

\title{Stellar Metallicities from DECam $u$-band Photometry: A Study of Milky Way Ultra-Faint Dwarf Galaxies}

\author{Yue Pan}
\affiliation{Department of Astronomy \& Astrophysics, University of Chicago, 5640 S Ellis Avenue, Chicago, IL 60637, USA}
\affiliation{Department of Astrophysical Sciences, Princeton University, Princeton, NJ 08544, USA}

\author{Anirudh Chiti}
\affiliation{Department of Astronomy \& Astrophysics, University of Chicago, 5640 S Ellis Avenue, Chicago, IL 60637, USA}
\affiliation{Kavli Institute for Cosmological Physics, University of Chicago, Chicago, IL 60637, USA}

\author{Alex Drlica-Wagner}
\affiliation{Department of Astronomy \& Astrophysics, University of Chicago, 5640 S Ellis Avenue, Chicago, IL 60637, USA}
\affiliation{Kavli Institute for Cosmological Physics, University of Chicago, Chicago, IL 60637, USA}
\affiliation{Fermi National Accelerator Laboratory, P.O. Box 500, Batavia, IL 60510, USA}

\author{Alexander P. Ji}
\affiliation{Department of Astronomy \& Astrophysics, University of Chicago, 5640 S Ellis Avenue, Chicago, IL 60637, USA}
\affiliation{Kavli Institute for Cosmological Physics, University of Chicago, Chicago, IL 60637, USA}

\author{Ting S. Li}
\affiliation{David A. Dunlap Department of Astronomy \& Astrophysics, University of Toronto, 50 St. George Street, Toronto, ON, M5S3H4 Canada}
\affiliation{Dunlap Institute for Astronomy \& Astrophysics, University of Toronto, 50 St George Street, Toronto, ON M5S 3H4, Canada}

\author{Guilherme Limberg}
\affiliation{Department of Astronomy \& Astrophysics, University of Chicago, 5640 S Ellis Avenue, Chicago, IL 60637, USA}
\affiliation{Kavli Institute for Cosmological Physics, University of Chicago, Chicago, IL 60637, USA}
\affiliation{Universidade de São Paulo, IAG, Departamento de Astronomia, SP 05508-090, São Paulo, Brazil}

\author{Douglas L. Tucker}
\affiliation{Fermi National Accelerator Laboratory, P.O. Box 500, Batavia, IL 60510, USA}

\author{Sahar Allam}
\affiliation{Fermi National Accelerator Laboratory, P.O. Box 500, Batavia, IL 60510, USA}

\begin{abstract}
We conducted an in-depth analysis of candidate member stars located in the peripheries of three ultra-faint dwarf (UFD) galaxy satellites of the Milky Way: Boötes I (Boo1), Boötes II (Boo2), and Segue I (Seg1). Studying these peripheral stars has previously been difficult due to contamination from the Milky Way foreground. We used $u$-band photometry from the Dark Energy Camera (DECam) to derive metallicities to efficiently select UFD candidate member stars. This approach was validated on Boo1, where we identified both previously known and new candidate member stars beyond five half-light radii. We then applied a similar procedure to Boo2 and Seg1. Our findings hinted at evidence for tidal features in Boo1 and Seg1, with Boo1 having an elongation consistent with its proper motion and Seg1 showing some distant candidate stars, a few of which are along its elongation and proper motion. We find two Boo2 stars at large distances consistent with being candidate member stars. Using a foreground contamination rate derived from the \emph{Besançon} Galaxy model, we ascribed purity estimates to each candidate member star. We recommend further spectroscopic studies on the newly identified high-purity members. Our technique offers promise for future endeavors to detect candidate member stars at large radii in other systems, leveraging metallicity-sensitive filters with the Legacy Survey of Space and Time and the new, narrow-band Ca HK filter on DECam.

\end{abstract}

\section{Introduction}  \label{sec:intro}

Over the last two decades, large digital sky surveys such as the Sloan Digital Sky Survey (SDSS), the Dark Energy Survey (DES), and PanSTARRS have allowed astronomers to identify many more low surface brightness stellar systems surrounding the Milky Way \citep[e.g.][]{Simon.2019}{}. Among these faint stellar systems, ultra-faint dwarf galaxies (UFDs), which were first detected in SDSS data two decades ago \citep{Willman.etal.2005a, Willman.etal.2005b}, have generated particular interest due to their extremely faint luminosities ($L < 10^5 L_\odot$), chemically primitive compositions ($[\rm{Fe}/\rm{H}]\lesssim -2.0$), large dark matter contents \citep[$\gtrsim 100\ M_\odot/L_\odot$; ][]{Simon.Geha.2007}, and old ages \citep[$\sim$ 13 Gyr; e.g.][]{Brown.etal.2014, Simon.2019}.

As per the hierarchical structure formation theory of galaxies, smaller galaxies are formed before larger ones \citep{White.Frenk.1991, Cole.etal.2000}. Therefore, UFDs are presumably the building blocks of more massive galaxies, and particularly interesting probes of the early chemical evolution of the universe since they are thought to have undergone only a few cycles of chemical enrichment \citep{Frebel.etal.2014}. The number and distribution of UFDs can constrain dark-matter models \citep[e.g. ][]{Kim.etal.2018, Nadler.etal.2021, Mau.etal.2022}. Moreover, the resolved kinematics of their member stars link to a number of questions related to dark matter and can be used to put constraints on the dark-matter annihilation cross-sections and the dark-matter distribution on small scales \citep{Calore.etal.2018, Abdallah.etal.2020, Malhan.etal.2022}. 

More than 50 UFDs have been discovered around the Milky Way (MW). While their relative proximity ($< 200 $\,kpc) makes them suitable for in-depth investigations, it is still extremely hard to study their resolved stellar populations due to their faint nature.  Traditionally, member stars of MW UFDs are identified by low- and medium-resolution spectroscopy, through the clustered radial velocities and low metallicities of UFD members. Subsequently, one can derive the dynamical mass from the velocity dispersion of member stars \citep{Walker.etal.2009b, Wolf.etal.2010} and use the estimated mass-to-light ratio to distinguish UFDs from globular clusters \cite[GCs; ][]{Willman.Strader.2012}. One can also distinguish UFDs from GCs through metallicity measurements since UFDs show significant metallicity spreads \citep[e.g. ][]{Frebel.etal.2014}, whereas GCs show minimal spreads \citep[$\lesssim 0.05$ dex; e.g.][]{Carretta.etal.2009}.

However, spectroscopy of UFDs is primarily devoted to their central regions ($\lesssim 2$ half-light radii), since member stars in the outskirts are sparsely distributed \citep{Munoz.etal.2018} and dominated by MW foreground stars. Simulations suggest that UFDs are formed within extended dark-matter halos and may be shaped by early galaxy mergers and supernova feedback \citep{Deason.etal.2014, Munshi.etal.2019, Wheeler.etal.2019}. Signatures of these effects would primarily reside in the outer regions of UFDs. Thus, a more efficient method to identify member stars in the outer regions of dwarf galaxies is needed to understand early dwarf galaxy evolution and probe the nature of their small-scale dark-matter halos.

It is possible to greatly improve the efficiency of spectroscopy by determining metallicities and identifying candidate member stars through photometry. Early research by \citet{Anthony-Twarog.etal.1991} demonstrated the efficacy of using photometry to detect metal-poor stars. More recently, 
\citet{Pace.Li.2019} showed that the use of metallicity-sensitive broadband colors from DES can enhance the efficiency of identifying member stars of UFDs, and recent studies have been successful in generalizing this technique with intermediate and narrow-band filters to identify UFD stars \citep[][]{Chiti.etal.2020, Chiti.etal.2021, Longeard.etal.2021, Longeard.etal.2022, Longeard.etal.2023, Fu.etal.2023}.
Current iterations of photometric metallicity work typically determine the metallicity by establishing a relation between a star's metallicity and its brightness in a narrowband imaging filter that encompasses a prominent metal absorption feature \citep[e.g., Ca\,\textsc{ii} K line; ][]{Keller.etal.2007, Starkenburg.etal.2017}. This technique offers the advantage of determining metallicities for all stars within the camera's field-of-view (FOV), unlike spectroscopy, which has to be more targeted and is constrained by factors like slit arrangements and the number of fibers.

In this paper, we combine targeted $u$-band observations from the Dark Energy Camera (DECam) with additional broad-band photometry from DECam Local Volume Exploration Survey (DELVE) Data Release 2 \citep[DR2,][]{Drlica-Wagner.etal.2022} and astrometry from the  \emph{Gaia} DR3 catalog \citep{gaiamission.2016, gaiadr3}  to identify candidate member stars of three MW UFDs: Boötes I (Boo1), Boötes II (Boo2), and Segue I (Seg1). The
DECam $u$-band covers the prominent Ca\,\textsc{ii} K metal absorption line, and thus can be used to derive photometric metallicities. 
Since Boo1 is the most widely studied MW UFD in our sample, we start by identifying candidate member stars of Boo1 and comparing our photometric catalog to existing spectroscopic catalogs.
We then statistically evaluate the purity and completeness of our sample and generalize our method to the other two systems to identify new candidate member stars out to large radii. 
We demonstrate the power of DECam $u$-band photometry to identify low-metallicity stars, we identify low-metallicity stars in the outskirts of each UFD, and we publicly release our selected low-metallicity UFD candidate member stars to facilitate spectroscopic follow-up work. 

This paper is organized as follows. Section~\ref{sec:obs_reduce} outlines our DECam $u$-band observations, data reduction, and zero-point calibration. Section~\ref{sec:mem_sel} presents our candidate member star selection process, which combines an isochrone selection, a proper motion selection, and a photometric metallicity selection. Section~\ref{sec:mem_cat_validate} presents our investigation of stars in the outskirts of Boo1, Boo2, and Seg1, along with a discussion of the completeness and purity of the sample. Finally, Section~\ref{sec:discussions} discusses the relation of the candidate member star spatial distribution to the evolution nature of UFDs, and Section~\ref{sec:conclusions} summarizes our main findings.

\section{Observations and data reduction}\label{sec:obs_reduce}

In this section, we discuss our observations and the process of producing source catalogs used in our subsequent analysis. 
We present the DECam $u$-band observations in Section~\ref{sec:observation}, the data reduction and generation of source catalogs from individual $u$-band exposures in Section~\ref{sec:data_reduc}, and the procedure for calibrating $u$-band zero-points and generating a combined source catalog in Section~\ref{sec:zpt_cali}.
Our final, combined $u$-band catalog is then cross-matched with DELVE DR2 \citep{Drlica-Wagner.etal.2022} for broadband $g, i$ photometry that is used in subsequent analyses.

\begin{deluxetable*}{c c c c c c c c c}[t]
\tablecolumns{9}
\tablewidth{\textwidth}
\tablecaption{\label{tab:observation} Observations of Boo1, Boo2, and Seg1}
\tablehead{   
  \colhead{Name} &
  \colhead{RA} & 
  \colhead{DEC} &
  \colhead{UT Observation Dates} &
  \colhead{Instrument} &
  \colhead{Filter} &
  \colhead{Seeing} &
  \colhead{Exposure time}\\
  \colhead{} &
  \colhead{h:m:s} & 
  \colhead{d:m:s} &
  \colhead{} &
  \colhead{} &
  \colhead{} &
  \colhead{} &
  \colhead{s}  
}
\startdata
Boo1 & 14:00:06.0  & +14:30:00  & 12 Mar 2021 & DECam & $u$ & 1\farcs0  & $20\times300$\,s \\
Boo2 & 13:58:00.0 &  +12:51:00  & 10 Jun 2021 & DECam  & $u$ & 1\farcs0  & $20\times300$\,s \\
Seg1 & 10:07:04.0  & +16:04:55 & 10 Jun 2021 & DECam  & $u$ & 1\farcs0  & $20\times300$\,s 
\enddata
\end{deluxetable*}

\begin{deluxetable*}{c c c c c c c c} 
\tablecolumns{8}
\tablewidth{\textwidth}
\tablecaption{\label{tab:basic_info} Properties of Boo1, Boo2, and Seg1}
\tablehead{   
  \colhead{Name} &
  \colhead{$r_h (')$} &
  \colhead{$\mu_{\alpha}\cos\delta$ (mas yr$^{-1}$)} &
  \colhead{$\mu_{\delta}$ (mas yr$^{-1}$)} &
  \colhead{$(m-M)_o$} &
  \colhead{$\theta (^\circ)$} & 
  \colhead{$\epsilon$} & 
  \colhead{References$^1$}
}
\startdata
Boo1  & $9.97 \pm 0.27$  & $-0.39 \pm 0.01$  & $-1.06 \pm 0.01$  & 19.11  & $6.0\pm 3.0$ & $0.30 \pm 0.03$ & a, b, b, c, a, a \\
Boo2 & $3.17 \pm 0.42$  & $-2.33^{+0.09}_{-0.08}$  & $-0.41 \pm 0.06$  & 18.10  & $-68.0 \pm 27.0$ & $0.25 \pm 0.11$ & a, b, b, d, a, a \\
Seg1  & $3.62\pm0.42$  & $-2.21 \pm 0.06$  & $-3.34 \pm 0.05$  & 16.80  & $77.0\pm 15.0$ & $0.33 \pm 0.10$ & a, b, b, e, a, a
\enddata
\tablenotetext{1}{{\bf{References}}: 
a. \citet{Munoz.etal.2018}; b. \citet{McConnachie.Venn.2020}; c. \citet{Dall'Ora.etal.2016}; d. \citet{Walsh.etal.2008}; e. \citet{Belokurov.etal.2007}}
\end{deluxetable*}

\subsection{Observations} \label{sec:observation}
We obtained $u$-band photometry of Boo1, Boo2, and Seg1 using DECam 
 \citep{Flaugher:2015} on the 4\,m Blanco Telescope located at Cerro Tololo Inter-American Observatory (CTIO) in Chile (PI: Ji, PropID: 2021A-0272). 
The DECam hosts 62 $2048\times4096$ pixel science CCDs that have a pixel scale of $\sim$0\farcs26/pixel.
This grants DECam a large field-of-view ($\sim$3 sq.\ degrees, $\sim$2.2\,degrees across), making it ideally suited for wide-field photometric studies. 
We observed each of the aforementioned dwarf galaxies using sequences of 300\,s exposures dithered by 30\farcs\  to cover chip gaps.
Total exposure times were calculated to target 0.03\,mag precision at $u=23$.
Conditions were clear and the seeing was stable at $\sim$1\farcs0 throughout observations.
Table~\ref{tab:observation} displays the full details of our observations. 

\subsection{DECam $u$-band data reduction \& photometry} \label{sec:data_reduc}

The DECam data were processed with the DES Data Management Final Cut pipeline \citep{Morganson:2018}.
This pipeline includes includes bias subtraction, non-linearity corrections, bad pixel masking, gain correction, brighter-fatter correction, and flat fielding.
Bias corrections and flat-fielding made use of ``super-cal'' assembled by combining flats and biases taken over several nights developed for $u$-band data processing for the DES Deep Fields \citep{Hartley.etal.2022}.
In addition, the DES Final Cut pipeline includes identification and masking of cosmic rays, bleed trails from saturated stars, and trails from artificial satellites was performed. 
Sky background estimation is performed through principal component analysis decomposition over the entire focal plane following \citet{Bernstein:2017}, and then adjusted CCD-by-CCD during source extraction and measurement.

Astronomical source detection and measurement were performed on a per CCD basis using the \texttt{PSFEx} and
\texttt{SourceExtractor} routines \citep{ Bertin:1996,Bertin:2011}. 
As part of this step, astrometric calibration was
performed with \texttt{SCAMP} \citep{Bertin:2006} by matching objects to the Gaia DR2 catalog \citep{GaiaDR2}. 
The \texttt{SourceExtractor} source detection threshold was set to detect sources with S/N $> 5$. Photometric fluxes and magnitudes refer to the \texttt{SourceExtractor} Point Spread Function (PSF) model fit.

\subsection{DECam $u$-band zero-point calibration} \label{sec:zpt_cali}
We calibrate the photometric zero-points for the DECam $u$-band source catalogs generated in Section~\ref{sec:data_reduc} by comparing instrumental DECam $u$-band magnitudes with predicted DECam $u$-band magnitudes derived from $u$-, $g$-, $r$- band photometry from the SDSS DR16 catalog \citep{Ahumada.etal.2020}. 
Specifically, DECam $u$-band magnitudes can be related to SDSS $u$-, $g$-, $r$-band magnitudes using the following equation (Allam, Tucker, et al., in prep.) : 
\begin{equation}
u_{\rm{DECam}} = u_{\rm{SDSS}} - 0.479 + 0.466\cdot(g-r)_{\rm{SDSS}} - 0.35\cdot(g-r)_{\rm{SDSS}}^2
\end{equation}
for objects with $0.2 \leq (g-r)_{\rm{sdss}} \leq 1.2$. This process is similar to the SDSS $g$-, $r$-, $i$-, $z$-band magnitude conversion to DES magnitudes in Appendix B.1 of \citet{Abbott.etal.2021}, but with a second order term. 
The zero-point can then be estimated to be the median difference between the DECam $u$-band magnitudes in our catalogs and the DECam $u$-band predicted from SDSS photometry.
 
\begin{figure}
    \centering
    \includegraphics[width=\columnwidth]{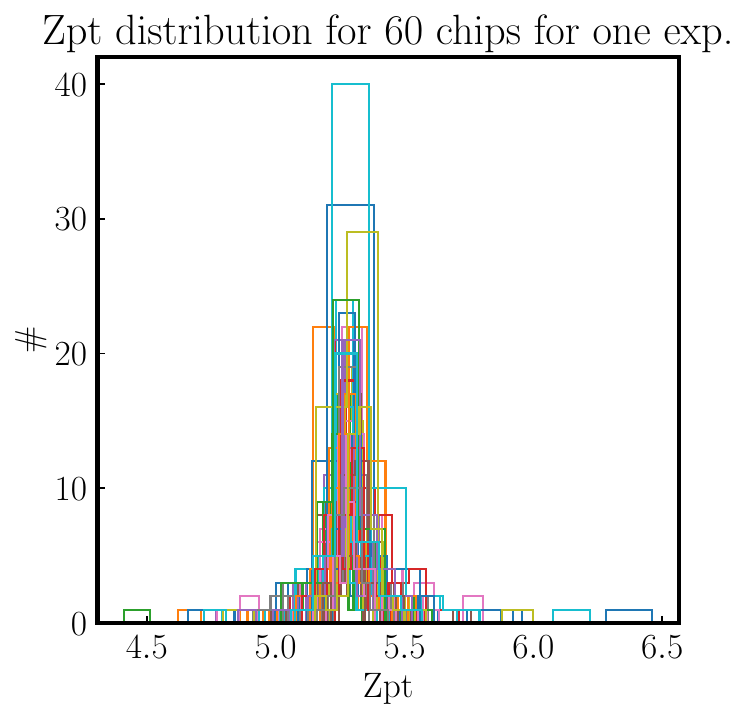}
    \caption{Zero-point distribution inferred from individual stars in the Boo1 FOV across all 60 chips in a single exposure. 
    The zero-point is defined as the theoretical DECam $u$-band magnitude from SDSS DR16 catalog minus the observed DECam $u$-band magnitude. 
    Each histogram represents the zero-point distribution for all stars on one chip at a single exposure. 
    The distribution is fairly narrow for most chips, but for some chips there are individual stars with large deviations which are sigma-clipped before computing the zeropoint.}\label{fig:zpt_dist}
\end{figure}

To obtain the SDSS $u$-, $g$-, $r$-band magnitudes for sources in our catalogs, we 
crossmatch our catalogs from Section~\ref{sec:data_reduc} with the SDSS DR16 catalog \citep{Ahumada.etal.2020} using R.A. and Decl. with a radius of 1\farcs.
To ensure a clean estimate of the zero-point, we (1) retain objects that have a $u$-band magnitude error of less than 0.1\,mag in both our catalog and the SDSS DR16 catalog, (2) retain objects with no photometric flags in either catalog (\texttt{FLAGS} = 0, \texttt{FLAGS\_WEIGHT} = 0, and \texttt{IMAFLAGS\_ISO} = 0), and (3) exclude galaxies with $|\texttt{SPREAD\_MODEL}| < 0.003 + \texttt{SPREADERR\_MODEL}$, \citep{Koposov.etal.2015, Slater.etal.2020}.
After this selection, we calculate a first-pass $u$-band zero-point offset for each individual chip in each exposure by taking the median offset between the $u$-band magnitudes in Section~\ref{sec:data_reduc} and those predicted by SDSS photometry. 
We then clip all sources that are $>3$\,$\sigma$ outliers relative to this zero-point and re-calculate as before to obtain zero-points for each chip and exposure. The median (${\rm{ZP}_i}$) and standard deviation (${\rm{ZP}}_{\rm{err,i}}$) of zero-points after sigma clipping are the zero-point value and error for each exposure.
The typical number of stars on each chip used for the zero-point estimate is $\sim 30-40$.
Figure~\ref{fig:zpt_dist} shows the distribution of inferred zero-points from individual stars over all chips for one 300s exposure. 
There are some variations across different chips for a given exposure time due to different responses, and across exposures due to e.g., weather condition.

For each source in our catalog, we take the weighted average of its zero-point corrected magnitude across 20 exposures to obtain a final $u$-band magnitude. Note that this is just a weighted average of the measured magnitudes for each star instead photometering a coadded image. 
To start with, for each exposure, we concatenate the catalog of all sources in each chip.
We then crossmatch between the 20 exposures to generate a final catalog. 
We note that not every star may be detected in each exposure, due to varying weather conditions at the time of each exposure, pixel artifacts, or statistical fluctuations for sources near the detection threshold.

We calculate the calibrated DECam $u$-band magnitude based on the zero-point correction in each exposure ($\tt{MAG\_PSF}$+${\rm{ZP}_i}$) for each selected star in all 20 exposures. Typical values of $\rm{ZP}_i$ varies between $\sim$5.1 and $\sim$5.3 depending on the weather condition. The random uncertainty associated with this calibrated magnitude is computed as the quadratic sum of the uncertainties from the DECam $u$-band measurement ($\tt{MAG\_PSF\_ERR}$) and the zero-point calculation (${\rm{ZP}}_{\rm{err,i}}$) for each exposure. In the bright limit $g < 18.5$, the stars' $u$-band magnitude error will be dominated by the zero-point calibration precision.
We find the standard deviation of $u$-band magnitudes in this regime to be $\sim 0.035$ mags across the 20 exposures. We take this to be our zero-point calibration uncertainty, and we add this value in quadrature to our random $u$-band magnitude uncertainty to derive a total photometric uncertainty, which will be propagated when computing photometric metallicity uncertainties. 

We compute the mean of the calibrated magnitudes for all selected objects across all exposures using the inverse-variance weighting method \citep{Strutz.2010}:
\begin{equation}
\mu = \frac{\sum_{i=1}^{20} x_i/\sigma_i^2}{\sum_{i=1}^{20} 1/\sigma_i^2}
\end{equation}
where $x_i$ and $\sigma_i$ are the calibrated DECam $u$-band magnitude ($\tt{MAG\_PSF}$+${\rm{ZP}_i}$) and the quadratic sum of $\tt{MAG\_PSF\_ERR}$ and ${\rm{ZP}}_{\rm{err,i}}$, respectively, for each exposure. The variance among all weighted average is then computed as in \citet{Strutz.2010}:
\begin{equation}
\sigma^2(\mu) = \frac{1}{\sum_{i=1}^{20} 1/\sigma_i^2}
\end{equation}

\begin{figure*}
    \centering
    \includegraphics[width=\textwidth]{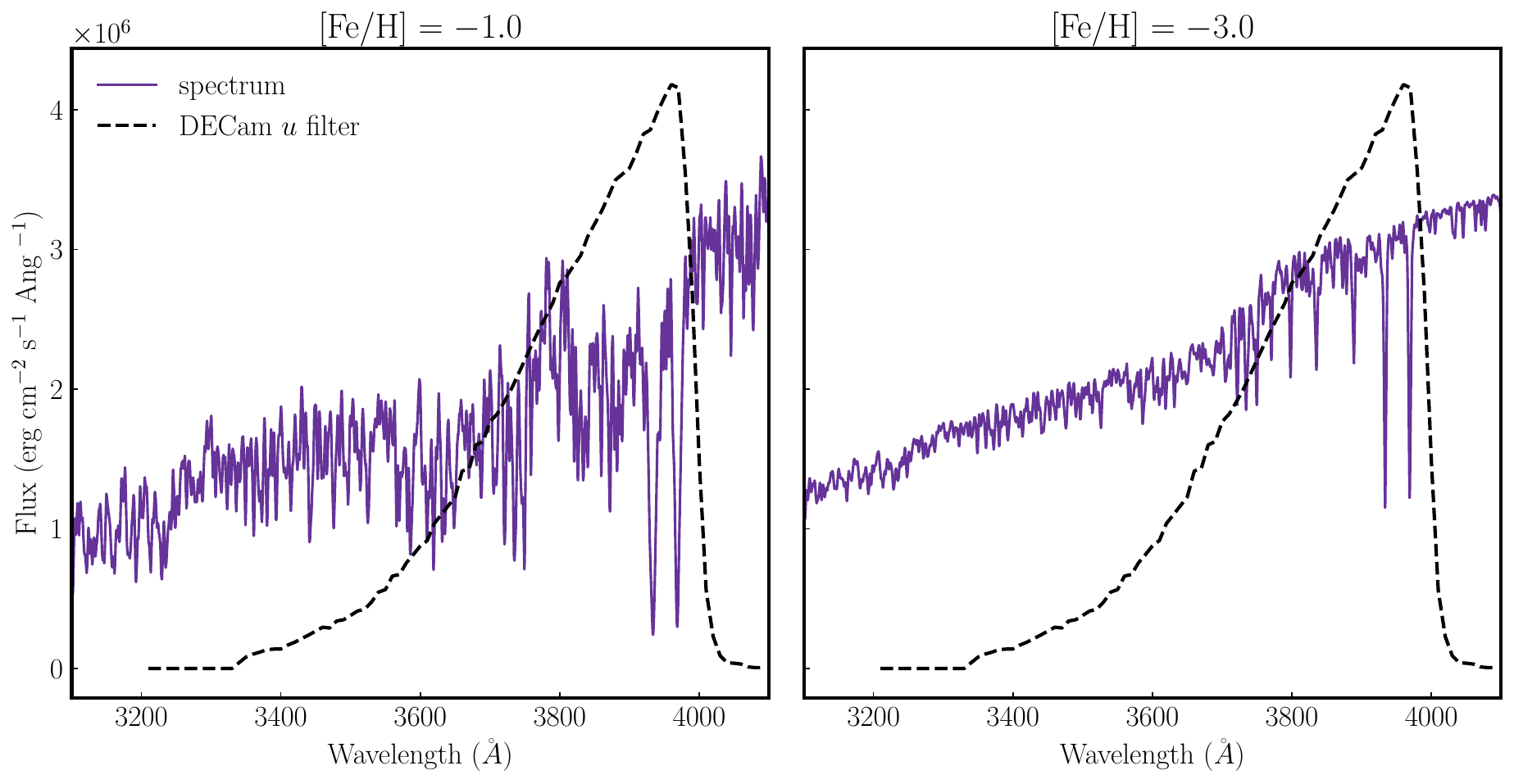}
    \caption{An illustration of why the DECam $u$-band is sensitive to metallicity. \emph{Left:} a synthetic spectrum generated by \texttt{Turbospectrum} \citep{Alvarez.Plez.1998, Plez.2012} for a sample cool red giant star with metallicity [Fe/H] = $-1$ overlaid with the bandpass of the DECam $u$ filter. \emph{Right:} the synthetic spectrum for a sample star with the same surface gravity and surface temperature but with metallicity [Fe/H] = $-3$. The Ca\,\textsc{ii} K metal absorption lines are more prominent for more metal-rich stars (left), and the DECam $u$-band filter completely covers this feature. Therefore, the more metal-rich a star is, the dimmer it appears in the DECam $u$-band. This correlation provides a way to associate the DECam $u$-band magnitude of a star with its metallicity, as shown in Figure~\ref{fig:grid}.}
    \label{fig:illustration_spectra}
    \centering
\end{figure*}

\begin{figure*}
    \centering
    \includegraphics[width=\textwidth]{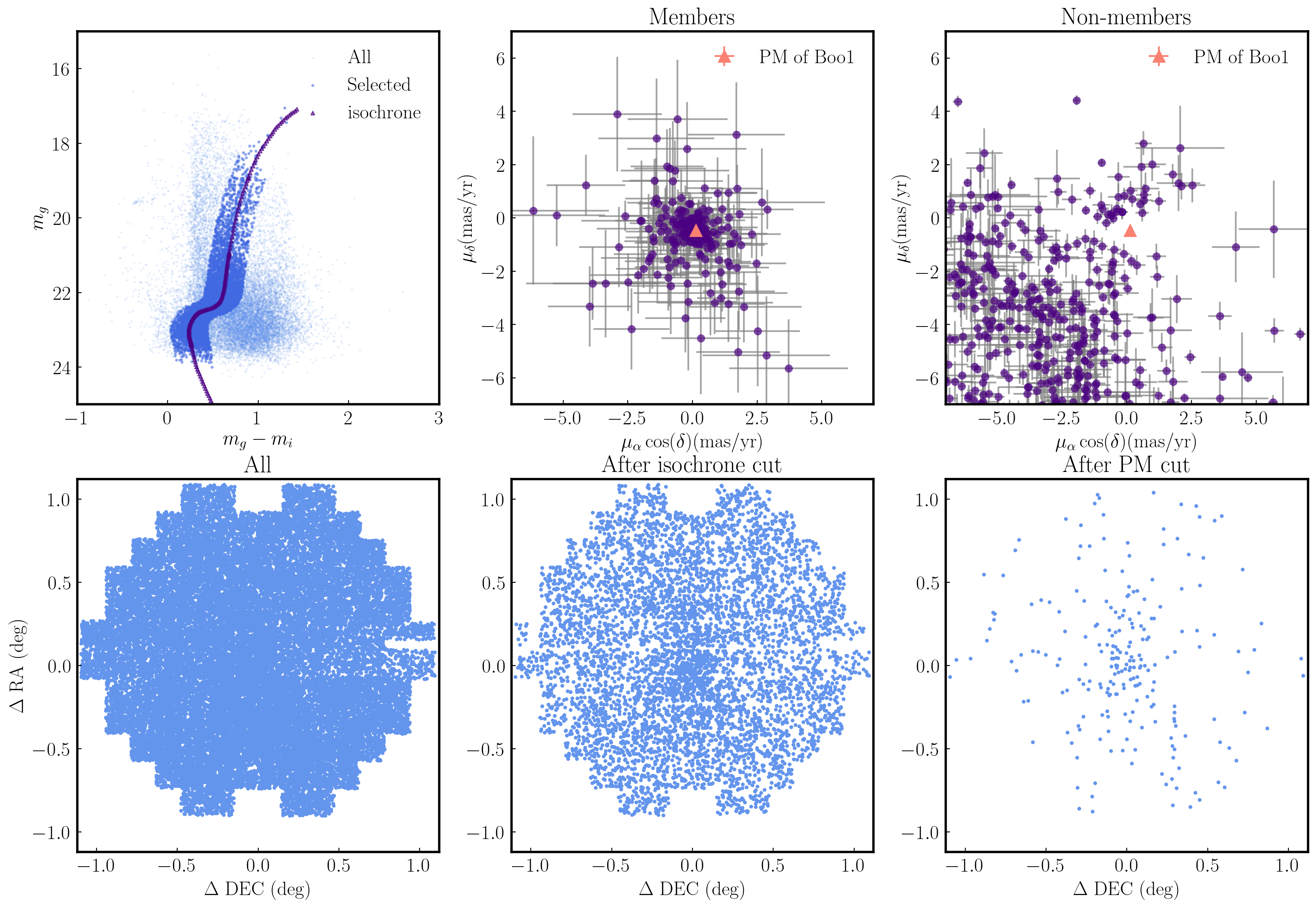}
    \caption{\emph{Top}: Isochrone and proper motion selection of candidate member stars. Top left: The CMD of all stars in our catalog for Boo1 (light blue) and selected stars (dark blue) using the isochrone cuts illustrated in Section~\ref{sec:mem_sel}. The isochrone of metallicity [Fe/H] = $-2.49$ and age = 13 Gyr is shown as purple. Top middle: Proper motion of Boo1 candidate member stars selected using proper motion and isochrone cuts. Top right: Stars that pass the isochrone cut but not the proper motion cut. The proper motion of Boo1 is shown as a triangle. \emph{Bottom}: spatial distribution of all stars in our original catalog (left), stars that pass the isochrone cut (middle), and stars that pass the isochrone and proper motion cut (right) centered at the position of Boo1.}
    \label{fig:iso_pm}
    \centering
\end{figure*}

\section{Selecting candidate member stars using photometric metallicity}
\label{sec:mem_sel}

After obtaining the calibrated DECam $u$-band magnitudes for all objects in our catalog, we apply several selection criteria to identify a reasonably pure sample of potential UFD member stars.
First, we apply {\texttt{parallax\_over\_error}}$<3$ to weed out sources with resolved parallax values (i.e., nearby foreground stars); second, we perform an isochrone cut along the color-magnitude diagram (CMD) of each UFD using DELVE DR2 photometry \citep{Drlica-Wagner.etal.2022} (Section~\ref{sec:iso_cut}); third, we apply a proper motion cut using the \textit{Gaia} DR3 catalog \citep{gaiadr3} to identify stars with kinematics similar to each UFD (Section~\ref{sec:pm_cut}); and finally, we identify low-metallicity stars using the DECam $u$-band magnitude and corresponding $g,i$ magnitudes since UFD stars are more metal-poor compared to Milky Way foreground stars (Section~\ref{sec:feh_cut}). Figure~\ref{fig:illustration_spectra} illustrates a synthetic spectrum overlaid with the DECam $u$-band, showing that the DECam $u$-band total throughput covers the prominent Ca\,\textsc{ii} K absorption feature at 3933\,\AA. 
Therefore, the higher a star's metallicity, the fainter it will appear in the DECam $u$-band at a given effective temperature and surface gravity.

To relate the metallicity of a star to its DECam $u$-band magnitude, we utilize a grid of synthetic spectra similar to \citet{Chiti.etal.2020}, who used such a grid to establish the relationship between a star's metallicity and SkyMapper $v$-band magnitude with $g,i$ photometry. 
They noted that for a given surface gravity, a star's metallicity can be interpolated from its $v-g-0.9\times(g-i)$ color. In this study, we use a comparable grid of synthetic spectra to establish the relationship between a star's metallicity and $u-g-0.9\times(g-i)$ color. A detailed discussion of this is presented in Section~\ref{sec:feh_cut}.
In this section, we outline the steps of our selection technique on Boo1 as an illustrative example. 
The same steps were performed for Boo2 and Seg1.

\subsection{Isochrone cut} \label{sec:iso_cut}

We begin by applying an isochrone cut to select stars that exhibit a consistent color-magnitude diagram (CMD) with that of an old, metal-poor population at the distance of Boo1 (see Table~\ref{tab:basic_info}). 
We first use {\tt{STILTS}} to crossmatch the final catalog obtained in Section~\ref{sec:zpt_cali} with the DELVE DR2 catalog \citep{Drlica-Wagner.etal.2022} to obtain precise DECam $g$- and $i$- band magnitudes ($\tt{MAG\_PSF\_g},\ \tt{MAG\_PSF\_i}$). 
To determine the reddening correction for each star, we use dust maps from \citet{Schlegel.Finkbeiner.Marc.1998} to obtain $E(B-V)$ value. 
The reddening coefficients for $u$-, $g$- and $i$- bands are taken as $R_u=3.995$\footnote{taken from https://www.legacysurvey.org/dr8/catalogs/}, $R_g=3.186$ and $R_i=1.569$, respectively, following \citet{Abbott.etal.2018} to convert $E(B-V)$ to the reddening corrections $A_g$ and $A_i$. 
We apply this correction to the DELVE magnitudes to obtain the dereddened $g$- and $i$- band magnitudes.
We perform the same procedure to get dereddened $u$- band magnitudes.

After obtaining dereddened $g$- and $i$-band magnitudes for all selected stars, we fit the $g, g-i$ CMD with an isochrone of [Fe/H] = $-2.49$ and age = 13 Gyr from the Dartmouth Stellar Evolution Database \citep{Dotter.etal.2008} at a distance modulus of $m - M = 19.1$ \citep{McConnachie.etal.2012}, which has been shown to represent the stellar population of Boo1 well. 
We then select stars that have a $g-i$ color within 0.2\,mag of this isochrone. 
The left panel of Figure~\ref{fig:iso_pm} illustrates the selection procedure based on the isochrone.

\begin{figure*}
    \centering
    \includegraphics[width=\textwidth]{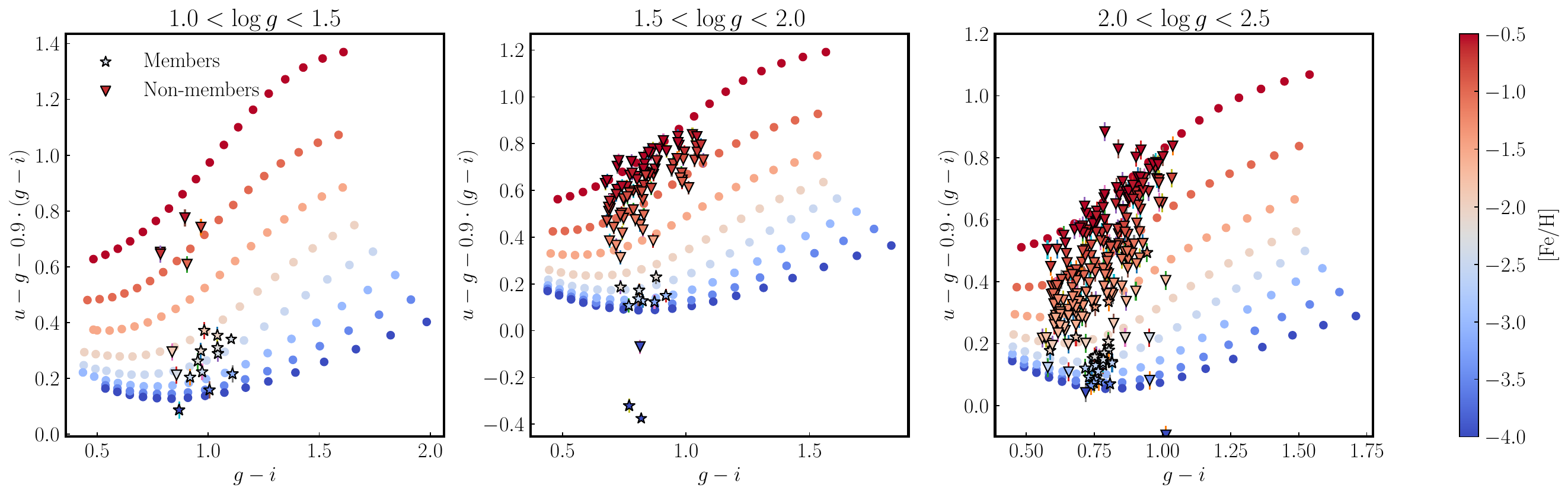}
    \caption{$u-g-0.9\times(g-i)$ as a function of $g-i$ color for three different surface gravity bins, color coded by metallicity [Fe/H]. We manually shift the contours down by 0.07 to match our photometric metallicity with spectroscopic results (Section~\ref{sec:derive_phot_feh}). Stars (triangles) are selected candidate member stars (non-member stars) that pass (do not pass) our isochrone, proper motion, and metallicity cuts. In general, the grid shows that for stars with surface  gravity $\log g=1, 2$, a higher $g-i$ color is associated with a higher $u-g-0.9\times(g-i)$ value, and stars with a higher metallicity tends to have higher $u-g-0.9\times(g-i)$ values. The $u-g-0.9\times(g-i)$ vs $g-i$ parameter space cleanly separates metal-poor stars from metal-rich ones. }
    \label{fig:grid}
    \centering
\end{figure*}

\begin{figure}[h]
    \centering
    \includegraphics[width=\columnwidth]{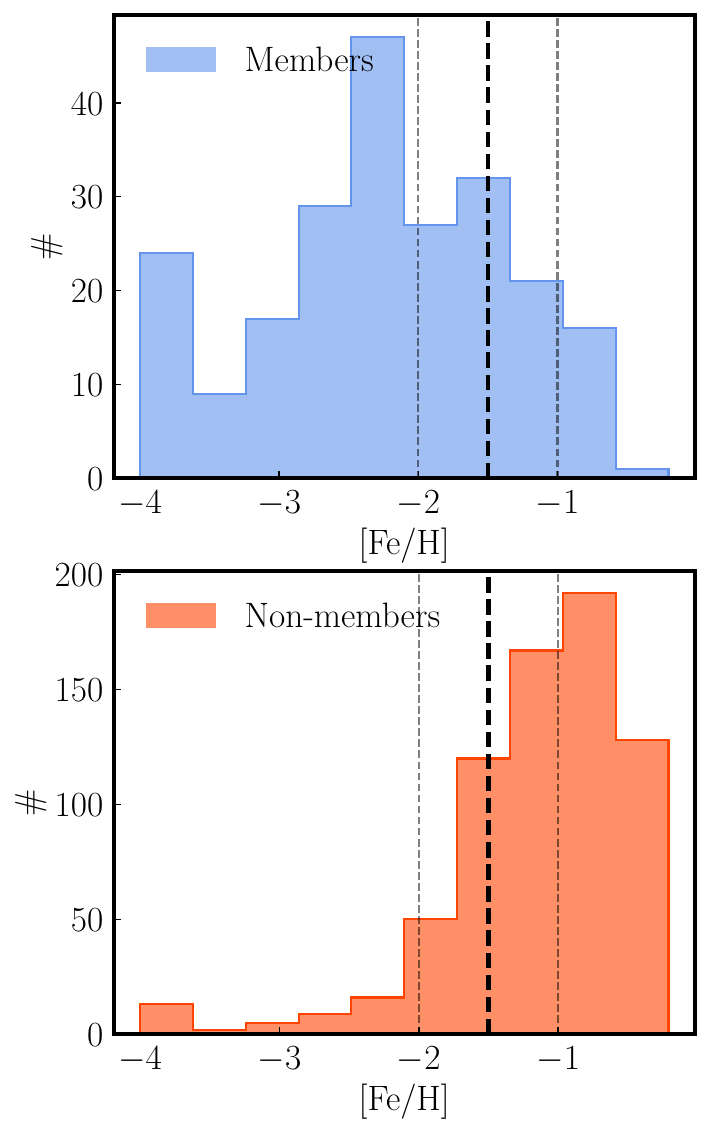}
    \caption{The distribution of photometric metallicity for candidate member (blue) and non-member stars of Boo1 (orange). It is evident that the two groups are clearly distinguished by their metallicity, with candidate member stars having low-metallicity ([Fe/H] $\leq 1.5$) and non-members having high metallicity ([Fe/H] $\geq 1.5$). However, it should be noted that some non-member stars appear in the low-metallicity range. In Section~\ref{sec:feh_cut}, we discuss that this might be due to those non-member stars being off-grid in the parameter space shown in Figure~\ref{fig:grid}, and the photometric metallicity calculation is inaccurate for off-grid stars.
    Such off-grid stars can be seen in the build-up at [Fe/H] = $-4.0$ in both histograms.}
    \label{fig:feh_dist}
    \centering
\end{figure}

\begin{figure*}
    \centering
    \includegraphics[width=17cm]{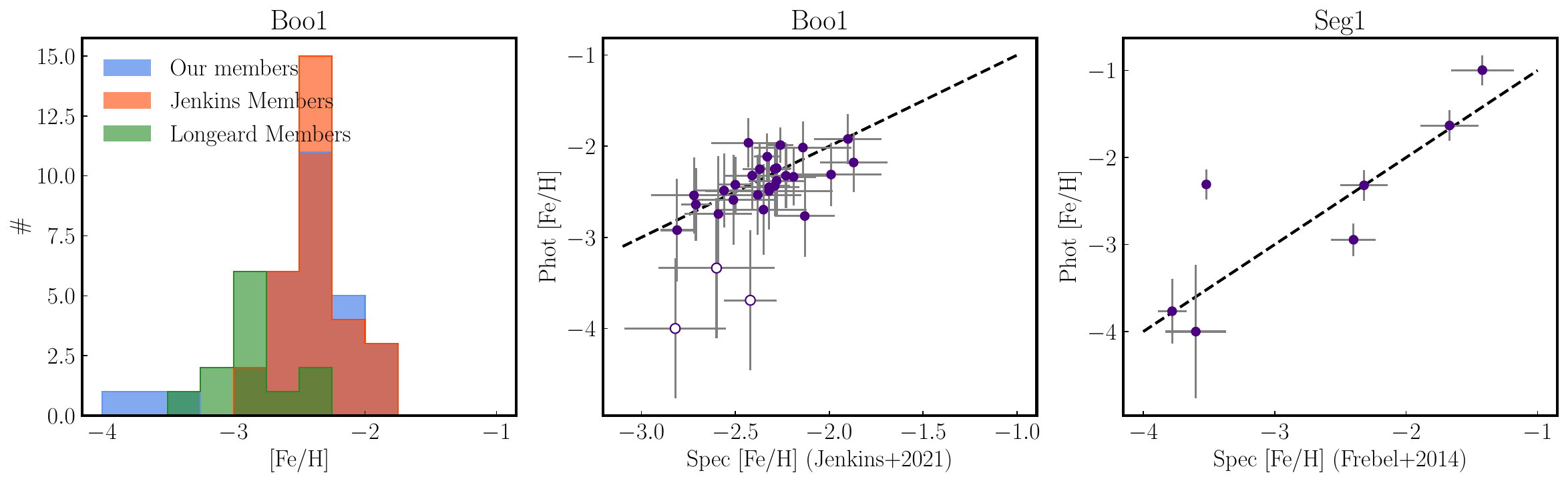}
    \caption{\emph{Left:} The distribution of photometric metallicity for our selected candidate member stars (blue), member stars from \citet{Jenkins.etal.2021} (red), and \citet{Longeard.etal.2022} (green). Note that we exclude horizontal branch stars in our analysis, and stars from our catalog and from the Jenkins et al. catalog are of the same magnitude range ($m_g < 21.5$). \emph{Middle:} The photometric metallicities obtained for Boo1 stars in this work are compared with the metallicities obtained by \citet{Jenkins.etal.2021} for the same sample of stars. 
    Our error includes both the systematic uncertainty of our calculation method (0.17 dex) and the uncertainty propagated from the photometric uncertainty. The points follow a clean one-to-one relationship with a $1\sigma$ scatter $< 0.2$ dex, except for three stars in the bottom left (open circles). See discussion in Section~\ref{sec:validate_phot_feh} on these stars. \emph{Right:} Comparison between our photometric metallicities  and spectroscopic metallicities from \citet{Frebel.etal.2014} of Seg1 stars. 
    The points also follow a clean one-to-one relationship with minimal scatter, further validating the accuracy of our photometric metallicity calculation apart from one outlier. 
    Note that the outlier (at [Fe/H]$_{\mathrm{phot}}\approx -2$) is Segue 1-7 \citep{Norris.etal.2010}, an extremely carbon-enhanced star with [C/Fe] = 2.3 at [Fe/H] = $-3.57$ \citep{Frebel.etal.2014}.}
    \label{fig:feh_comparison}
    \centering
\end{figure*}

\subsection{Proper motion cut} \label{sec:pm_cut}

To further identify candidate member stars, we select stars that have proper motions consistent with the systemic proper motion of Boo1.
We crossmatch the catalog of stars that passed our isochrone cut in Section~\ref{sec:iso_cut} with the Gaia DR3 catalog \citep{gaiadr3} to obtain proper motion data and corresponding uncertainties for our selected stars. We apply a reflex correction on both the proper motions of stars and the systemic proper motion of the UFD from \citet{McConnachie.Venn.2020}. The systematic proper motion after our reflex correction is visually consistent with that in Figure 16 of \citet{Pace.Erkal.Li.2022} for Boo1.
We then evaluate the consistency of these proper motions with the proper motion of Boo1 obtained from \citet{McConnachie.Venn.2020} using the Mahalanobis distance:
\begin{equation}
d_{\rm Mahalanobis} = \sqrt{(\vec{x}-\vec{y})^T\cdot\Sigma^{-1}\cdot(\vec{x}+\vec{y})}
\end{equation}
where $\Sigma$ is the covariance matrix between pmRA and pmDEC obtained from the correlation matrix and the proper motion errors in \textit{Gaia}, and $\vec{x} = (\rm{pm_{RA, Boo1}}$,$\rm{pm_{DEC, Boo1}}), \vec{y} = (\rm{pm_{RA}},\rm{pm_{DEC}})$ for each source. We select stars that have $d_{\rm{Mahalanobis}} < 3$. 
The right two panels of Figure~\ref{fig:iso_pm} illustrate this selection procedure.

\subsection{Photometric metallicities}
\label{sec:feh_cut}
We outline a photometric method to obtain a star's metallicity based on its DECam $u$-band magnitude and $g-i$ color, which is an efficient way to estimate metallicities for spatially complete samples of stars without requiring traditional spectroscopy. 
Such a method has been used in studies of dwarf galaxies in the past with e.g., the SkyMapper $v$ filter \citep{Chiti.etal.2020, Chiti.etal.2021} and the Pristine Ca HK filter \citep{Longeard.etal.2021,Longeard.etal.2022,Longeard.etal.2023}.
Here we demonstrate that a similar analysis can be performed with DECam $u$-band photometry due to the Ca\,\textsc{ii} K
 absorption feature encompassing a fraction of the bandpass of the filter (see Figure~\ref{fig:illustration_spectra}).
Consequently, the strength of the feature (as a function of the metallicity of the star, among other stellar parameters) affects the brightness of the star in the DECam $u$-band.
We describe our analysis with Boo1 here since it is the most extensively studied system regarding stellar metallicities among the three UFDs we focused on in this study and has evidence of extra-tidal features \citep{Belokurov.etal.2006, Martin.etal.2007, Norris.etal.2008, Roderick.etal.2016, Jenkins.etal.2021, filion.Wyse2021, Longeard.etal.2022}. 
\subsubsection{Deriving photometric metallicities}
\label{sec:derive_phot_feh}

To illustrate how we compute metallicity from photometry, we present a grid in Figure~\ref{fig:grid} that shows $u-g-0.9\times(g-i)$ and $g-i$ colors separated by four different surface gravities, where we determine the surface gravity of each star by interpolating along the same isochrone as in Section~\ref{sec:iso_cut}. This grid is a modified version of the grid computed by \citet{Chiti.etal.2020}, with the DECam $u$-band filter used to generate synthetic photometry as opposed to the SkyMapper $v$ filter in that study. The general trend of the grid shows that metallicity decreases as the $u-g$ color and $g-i$ color become bluer at a fixed surface gravity. The data points of each star are located in this parameter space for a given surface gravity based on their dereddened DECam $u$-, $g$-, $i$-band magnitudes. We interpolate the metallicity of each star using this grid. Note that we adjust the synthetic $u$-band magnitudes down by 0.07\,mag to better match our photometric metallicities with spectroscopic metallicities in the literature.
This shift is roughly consistent with the shift applied in \citet{Chiti.etal.2020} to their synthetic SkyMapper $v$ magnitudes, which is an empirically driven correction that accounts for the difference between photometric and spectroscopic metallicities due to potential imperfections in the grid of synthetic spectra. 

We note also that the photometric metallicities derived for non-member stars in our catalog are inaccurate because these stars are not located at the distance of the system and have different surface gravity values than member stars on the red giant branch. 
We do not apply any metallicity corrections to get accurate values for these non-member stars in our catalogs as we need an accurate measure of their distance to derive a surface gravity. 
However, we note that we do extend the stellar parameter range of the MSTO grid of synthetic spectra in \citet{Chiti.etal.2020} to include main-sequence stars (3.5 $<$ log\,$g \leq 5$) and compute synthetic photometry for those. 
This main-sequence grid is later used to adjust for the photometric metallicity difference when a the star is assumed to be on the red giant branch as opposed to the main sequence in Section~\ref{sec:boo1_purity} when we estimate the contamination rate due to MW foreground main-sequence stars using the \emph{Besançon} model.

The metallicity distribution of candidate member and non-member stars is shown in Figure~\ref{fig:feh_dist}, where candidate member stars are defined as those that pass our proper motion and isochrone cuts in Section~\ref{sec:iso_cut} and Section~\ref{sec:pm_cut}, while non-member stars are defined as those that pass the isochrone cut but not the proper motion cut. We observe that candidate member stars have significantly lower metallicities than non-member MW foreground stars, indicating the accuracy of our photometric metallicities. We further apply a metallicity cut of [Fe/H] $ \leq-1.5$ to select metal-poor candidate member stars. We choose $-1.5$ to ensure both a more complete and pure sample, and we discuss the details in Section~\ref{sec:boo1_purity}.

However, there are some non-member stars in the low-metallicity regime. 
By plotting their photometric locations in $u-g-0.9\times(g-i)$ vs $g-i$ colors, we found that they are typically located out of range of the contours in Figure~\ref{fig:feh_dist} and thereby off-grid relative to our synthetic photometry, indicating that their photometric metallicity is likely inaccurate.

\subsubsection{Validating photometric metallicities}
\label{sec:validate_phot_feh}
To assess the accuracy of our photometric metallicity calculation, we compare the photometric metallicities of our candidate member stars with the spectroscopic metallicities from \citet{Jenkins.etal.2021} and 
\citet{Frebel.etal.2014} for Boo1 and Seg1, respectively. The error of our photometric metallicities is the quadratic sum of the systematic uncertainty of our method (0.17 dex, taken from analysis with the same grid from \citealt{Chiti.etal.2020}) and the error propagated from the photometric uncertainty. The right two panels of Figure~\ref{fig:feh_comparison} show that the mean offset of the two metallicities is small for both Boo1 and Seg1 stars, with 1$\sigma \lesssim 0.2$ dex. 
There are three outliers from the one-to-one line in the Boo1 sample (open circles), which we find out to be outliers in the $u-g-0.9(g-i)$ vs $g-i$ grid and thus have biased low photometric metallicities and higher photometric metallicity errors as can be seen in the middle panel of Figure~\ref{fig:feh_comparison}. The synthetic grid in Figure~\ref{fig:grid} shows that contours are more tightly clustered at lower metallicity, so a $u$-band uncertainty would translate to a larger photometric metallicity uncertainty at lower metallicity. 
Thus, low metallicity outliers with correspondingly larger uncertainties are more likely to appear than metal-rich outliers at the lowest metallicities, potentially explaining why the three outliers all have photometric metallicities lower than spectroscopic metallicites. Nevertheless, for our purposes, the three outliers do not affect our sample because they pass our metallicity cut and are members.
Our goal of this study is to select potential members using photometric metallicities, as opposed to characterizing metallicity distribution of the UFD.
Note that the outlier in the Seg1 comparison (right panel of Figure~\ref{fig:feh_comparison}) at [Fe/H]$_{\mathrm{phot}}\approx -2$ is Segue 1-7 \citep{Norris.etal.2010}, an extremely carbon-enhanced star with $\rm [C/Fe]= 2.3$ at $\rm [Fe/H] = -3.57$ \citep{Frebel.etal.2014}.
These stars are known to have photometric metallicities biased high due to the CN feature at $\sim 3800$ \AA \citep{Starkenburg.etal.2017}. 
Overall, the comparison demonstrates that our photometric metallicity calculation is robust.

\section{Candidate member star catalog and member star validation}\label{sec:mem_cat_validate}

In previous sections, we demonstrate that we can effectively select candidate member stars in the peripheries of UFDs using photometric metallicities from DECam $u$-band.
In this section, we show the spatial distribution of detected candidate member stars for Boo1, Boo2, and Seg1.
To gain a statistical understanding of our sample, we calculate the purity and completeness of our selected candidate member stars for all three systems using the \emph{Besançon model of stellar population synthesis of the Galaxy}\footnote{\href{https://model.obs-besancon.fr/modele_home.php}{https://model.obs-besancon.fr/modele\_home.php}} \citep{Robin.etal.2003, Czekaj.etal.2014} and literature samples of members. We also discuss the implications of the candidate member star spatial distribution considering their proper motions. Finally, for each identified star, we calculate an estimate of purity and use this as a measure to guide future spectroscopic follow-up.
\subsection{Boo1}\label{se:boo1}
\begin{figure*}

    \includegraphics[width=\textwidth]{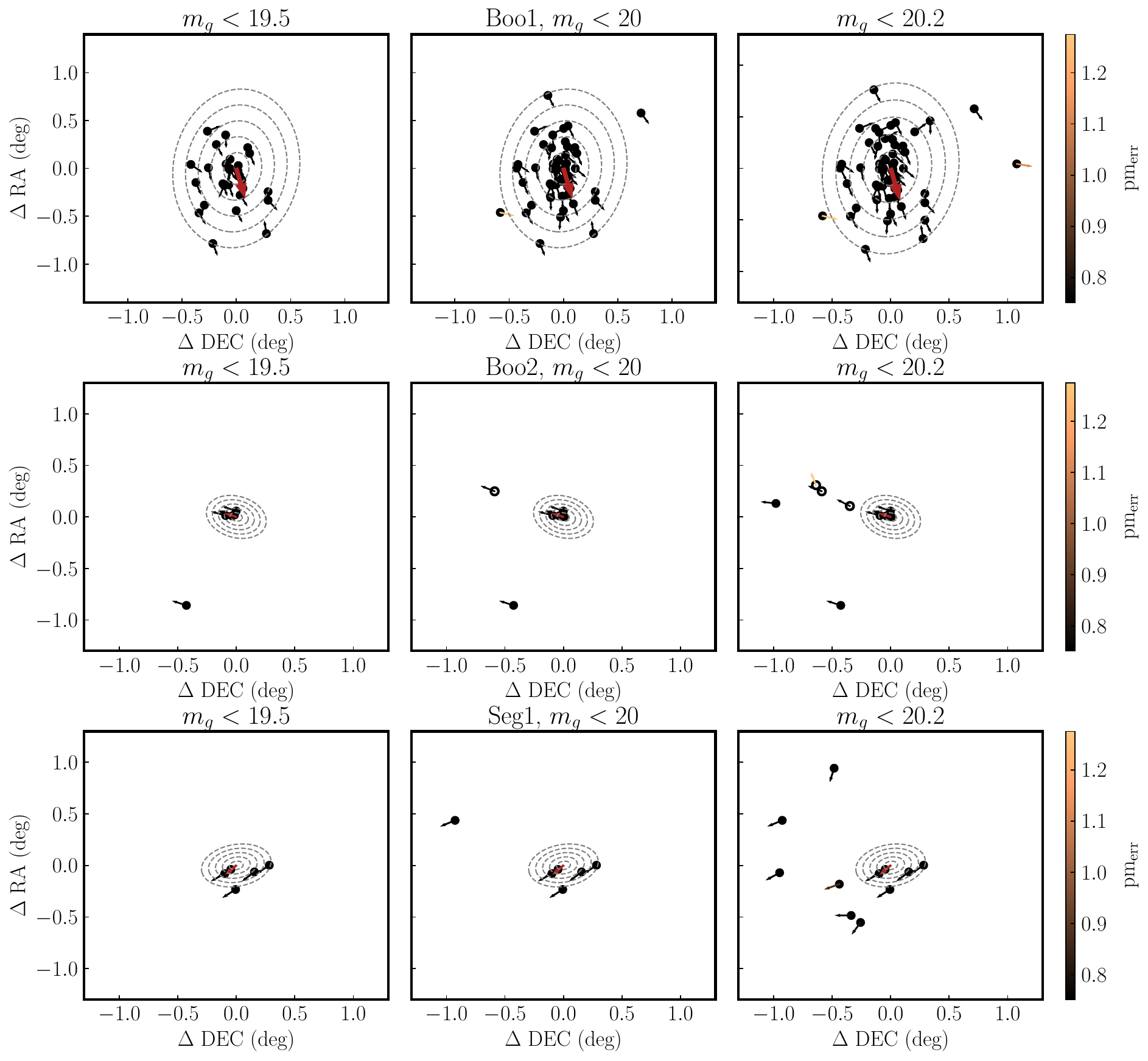}
    \caption{Spatial distribution of candidate member stars in our catalog for Boo1 (top), Boo2 (middle), and Seg1 (bottom). We divide the candidate member stars by three magnitudes: $M_g < 19.5$ (left), $M_g < 20$ (middle), and $M_g < 20.2$ (right). The dashed grey ellipses denote 1, 2, 3, 4, and 5 half-light radii of each system with ellipticity and position angle listed as in Table~\ref{tab:basic_info}. Each arrow represents the reflex-corrected velocity vector of a star calculated from its proper motion. For Boo1 (top), our candidate member star population extends beyond $\sim 5 r_h$, and they form a streak in the same direction as the proper motion of the system in the brightest sample (top left panel), indicating possible tidal disruption. For Boo2 (middle rows), most of our candidate member stars lie within 2 $r_h$. There exists an extension of stars north-east of the system, but three of them (open black circles) are outliers in CMD, indicating they may not be members. The other two solid black points outside 5 $r_h$ could be targets for future spectroscopic follow-up. For Seg1 (bottom), we apply an additional cut of [Fe/H] $< -2$ to weed out potential 300S member stars (Section~\ref{sec:spa_dist_boo2}). There is also a handful of candidates south-east of the system, in addition to potential contamination from the Sagittarius stream or other substructures (Section~\ref{sec:spa_dist_boo2}). } 
    \label{fig:spa_dist_all}

\end{figure*}

\begin{figure*}

    \includegraphics[width=18cm]{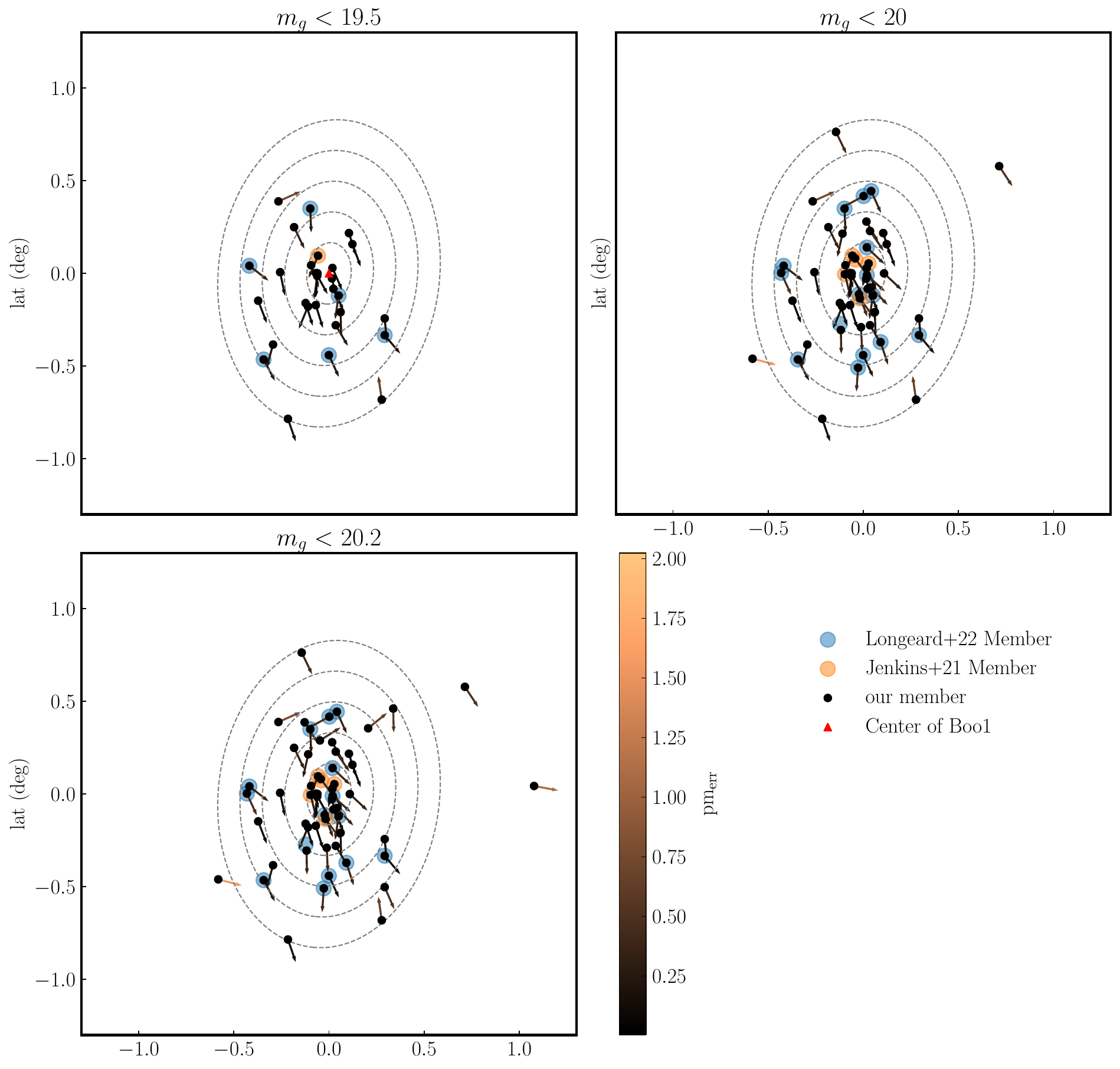}
    \caption{Spatial distribution of candidate member stars for Boo1 with a metallicity cut [Fe/H] $<-1.5$ compared with member stars in \citet{Longeard.etal.2022} (blue) and \citet{Jenkins.etal.2021} (orange). Other symbols are the same as in Figure~\ref{fig:spa_dist_all}. Most of the member stars in the Jenkins et al. catalog are located within 1 half-light radius, while the member stars in the Longeard et al. catalog extend out to approximately three to 4 half-light radii for stars brighter than $m_g = 20.2$. We were able to recover all of the member stars in the Jenkins et al. and Longeard et al. catalogs except for one that has a photometric metallicity of $-1.47$ and thus does not pass our $-1.5$ photometric metallicity cut. We discuss the completeness and purity of our selected candidate member star sample compared to the combined Jenkins et al. and Longeard et al. catalog in Section~\ref{sec:boo1_completeness}.}
    \label{fig:L_J_spa_dist_boo1}

\end{figure*}

\subsubsection{Spatial distribution of Boo1 candidate members}\label{sec:spa_dist_boo1}

In Figure~\ref{fig:spa_dist_all}, we present the spatial distribution of candidate member stars in Boo1 that satisfy our isochrone, proper motion, and metallicity cuts as described in Section~\ref{sec:mem_sel}. For stars with $m_g < 20$, we observe that the majority of them form an inclined streak that aligns with the proper motion of most stars. 
This echoes suggestions in the literature that Boo1 is currently experiencing tidal disruption by the Milky Way \citep{Longeard.etal.2022, Pace.Erkal.Li.2022}. 
Moreover, we identify several new candidate member stars located at distances of $\sim3-5$ half-light radii from the center.

In Figure~\ref{fig:L_J_spa_dist_boo1}, we compare the spatial distribution of our selected candidate member stars to that of member and non-member stars in the \citet{Jenkins.etal.2021} and \citet{Longeard.etal.2022} catalogs.
We apply our isochrone and proper motion selections to determine members/non-members in the two catalogs to enable a fair comparison. 
For stars with $m_g < 19.5$, we identify more candidate member stars at large radii than in previous catalogs. We also identify more faint stars $(m_g < 20.2)$ than in previous catalogs; however, some of these stars may be non-members with large proper motion uncertainty, allowing them to pass the proper motion cut, or MW foreground stars with low-metallicity. 
The contamination rate increases as we go to fainter magnitudes (e.g. $m_g > 20.5$), but for our sample of stars with $m_g < 20.2$, we still retain an obvious elongated feature in the lower-left panel of Figure~\ref{fig:L_J_spa_dist_boo1}.
\begin{figure*}

    \includegraphics[width=18cm]{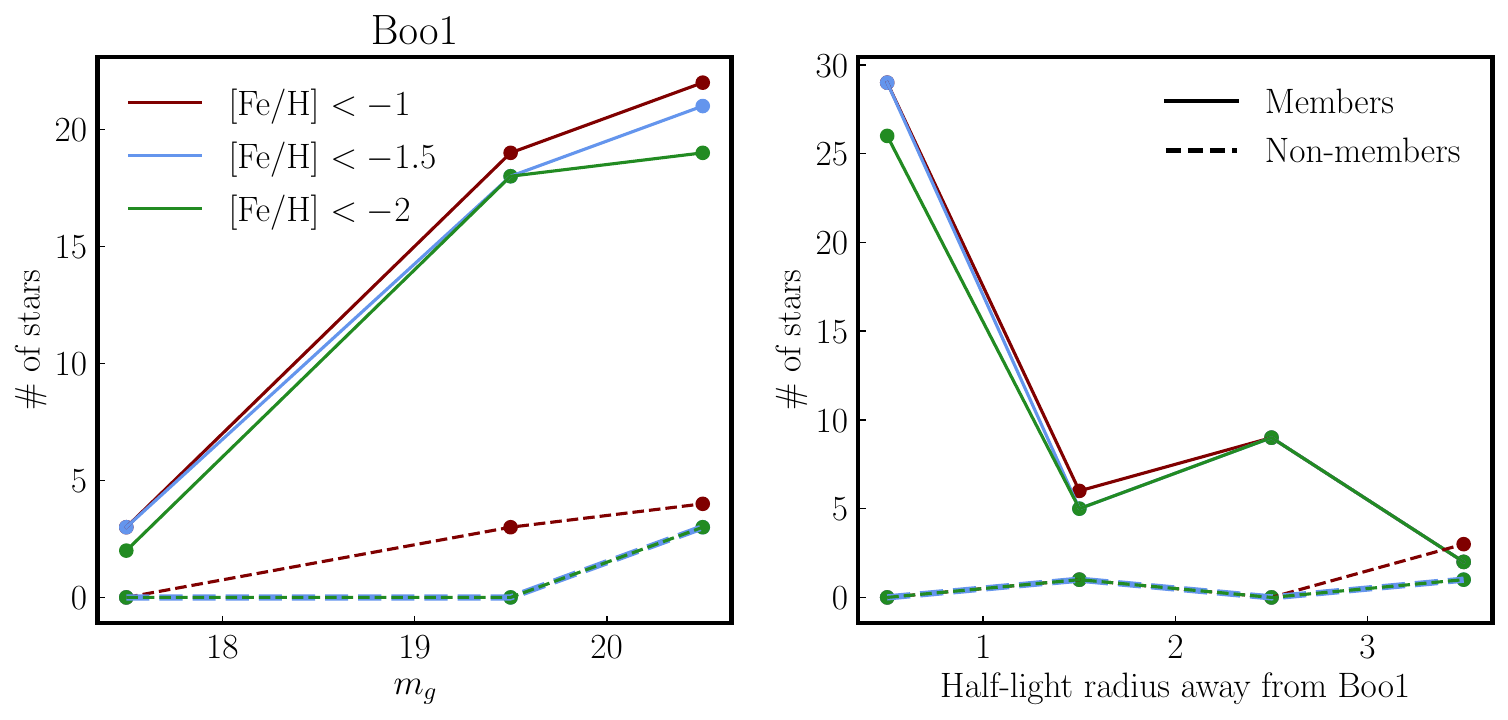}
    \caption{Number of member and non-member stars as a function of $m_g$ and half-light radius $r_h$ from the center for three different metallicity cuts. Stricter metallicity cuts result in a decrease of both candidate member stars and non-member stars. Looser metallicity cuts result in an increase in both candidate member stars and non-member stars. As discussed in Section~\ref{sec:boo1_purity}, we choose [Fe/H]$<-1.5$ as our final photometric metallicity cut given joint completeness and purity considerations. }
    \label{fig:purity_boo1}

\end{figure*}

\subsubsection{Completeness of Boo1 sample}
\label{sec:boo1_completeness}
We now calculate the completeness of our selected star sample, which is an essential metric for assessing the accuracy of our candidate member star selection. Boo1 is a particularly suitable target for this analysis, as it is the only MW UFD that has an extensive sample of literature member stars in this study. 
By quantifying the completeness and purity of our selection of candidate Boo1 member, we can obtain a preliminary estimate of the potential bias and uncertainty in our selection criteria, which could be applied to other systems in this study.

The completeness of our selected sample of candidate member stars for Boo1 is determined by calculating the ratio of the number of stars in our catalog that are also present in the combined \citet{Jenkins.etal.2021} and \citet{Longeard.etal.2022} catalog to the total number of stars in the combined catalog. To compute this, we first crossmatch the two catalogs, which yields a total of 48 member stars. Subsequently, we crossmatch these 48 catalog member stars with our selected stars in Section~\ref{sec:mem_sel} that satisfy the metallicity cuts ${\rm [Fe/H]}<-1$, ${\rm [Fe/H]}<-1.5$, and ${\rm [Fe/H]}<-2$. We obtain 47, 45, and 42 overlaps, respectively. Hence, the completeness rates are 97.9\%, 93.8\%, and 87.5\%, respectively. These results indicate that lower metallicity cuts may exclude more true members. Note that we only consider red-giant branch (RGB) stars here and exclude horizontal branch stars.

\subsubsection{Purity of Boo1 sample}
\label{sec:boo1_purity}

Purity is defined as the ratio of the number of our selected candidate member stars that are actually members in the combined catalog to the total number of our selected candidate member stars. We first combine all the stars (members and non-members) in the \citet{Jenkins.etal.2021} and \citet{Longeard.etal.2022} catalogs. Next, we crossmatch these stars with our selected stars that pass the isochrone, proper motion, and metallicity cuts ${\rm [Fe/H]}< -1$, ${\rm [Fe/H]}< -1.5$, and ${\rm [Fe/H]}< -2$ to obtain three different samples. We compute the purity rate for different metallicity cuts using these overlaps.

As we only have two catalogs available to compute the completeness and purity rates, and only the \citet{Longeard.etal.2022} catalog extends beyond 1 half-light radius of Boo1, we require an independent method to estimate contamination rate from MW foreground stars. We use the \emph{Besançon} model, which enables us to simulate the number of MW foreground stars within 5 half-light radii from Boo1. We adopt the SDSS+JHK photometric system in our simulations, and our field of view encompasses 150 kpc, with the frame centered on Boo1 and a solid angle of 100 deg$^2$. We use the large field instead of the small field (10 deg$^2$) since the small field contains only a few foreground stars that satisfy our selection criteria, leading to a biased estimate of the foreground contamination rate, particularly at larger radii. We verified that the proper motion and radial velocity distributions do not show spatial variations in the 100 deg$^2$ field that was selected.

We select all stars with $g$-band magnitudes between g=16 and g=23 and generate model stars without applying a proper motion error function in the query. We use 
the correlation between proper motion uncertainties and the dereddened g-band magnitudes in our sample of observed non-members and fit an exponential function $0.01+e^{1.2\cdot m_g-25}$, and use this to derive the proper motion error for simulated stars. The model generates a total of 311,416 stars, and we apply our isochrone, proper motion, and metallicity cut [Fe/H] $<-1.5$ to obtain 1712 stars that pass all our selection criteria. Since most of the MW foreground stars are main sequence stars, their metallicities will be inaccurate if we use the synthetic photometry grid obtained for giant stars in Section~\ref{sec:derive_phot_feh}. To get accurate metallicities for the MW foreground stars, we use the synthetic photometry grid described in Section~\ref{sec:derive_phot_feh} for main sequence stars (3.5 $< $log\,$g \leq 5.0$). 
From this grid, we derive the metallicity differential between assuming a star is on the main-sequence versus the red giant branch as a function of the star's uncorrected metallicity and SDSS $g-r$ color.
We apply this correction to the main-sequence stars in the Besancon catalog to mimic how their metallicities would appear in our analysis. We further apply a magnitude cut $m_g < 21.5$, as all our candidate member stars have $m_g < 21.5$. 

We assume a uniform spatial \emph{density} distribution of MW foreground stars and calculate the MW contamination rate as the total number of stars that pass all our selection criteria (1712) divided by the total FOV (100 deg$^2$). We divide the area into several radial bins: 0-1 $r_h$, 1-2 $r_h$, 2-3 $r_h$, 3-4 $r_h$, etc, and for each bin, the estimated MW contamination number is the area of the annulus times the contamination rate. For each annulus, we further divide it into magnitude bins $m_g = (16, 19.5, 20, 20.5, 21, 21.5)$. We record the number of candidate member stars and MW foreground stars that fall into each radial and magnitude bin.  Then the purity of each star in a specific radial and magnitude bin is computed as (\# of candidate members - \# MW foreground)/(\# of candidate members) in that bin. Stars that fall into the same radial and magnitude bin will have the same purity. We use purity as a guide for spectroscopic studies to decide which stars should be targets given limited telescope time.

As shown in Figure~\ref{fig:purity_boo1}, both the number of candidate member stars and the number of MW foreground stars increase from bright to faint magnitudes. From small to large radial distances, the number of candidate member stars decreases significantly and the number of foreground stars increases steadily. Going from metallicity cut [Fe/H] $< -1$ to [Fe/H] $< -1.5$ decreases the foreground number with minimal decrease of candidate member stars, so to maximize both completeness and purity, we have chosen the metallicity cut [Fe/H] $< -1.5$ and apply this metallicity cut on all subsequent analyses.

\begin{figure*}

    \includegraphics[width=\textwidth]{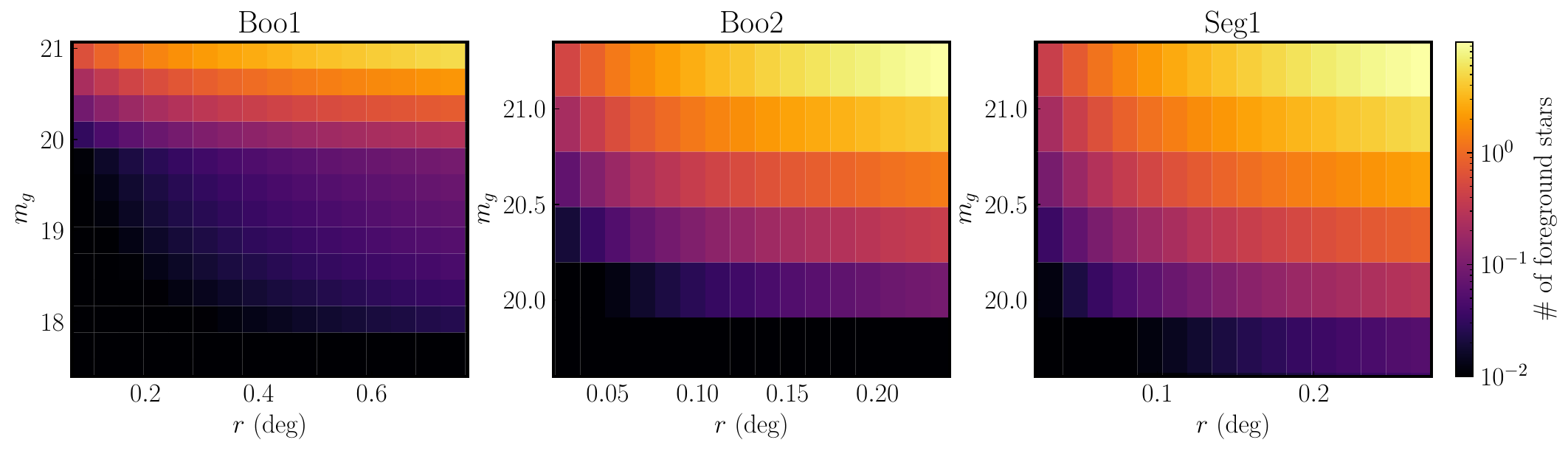}
    \caption{A 2D histogram displaying the estimated number of MW foreground stars for Boo1, Boo2, and Seg1 as a function of radius (in degrees) and $m_g$. The number of MW foreground stars increases as the magnitude becomes fainter and the radius grows larger.}
    \label{fig:num_mg_r_all}

\end{figure*}

We fit a function to the number of foreground stars for each system, as a function of both radius and $g$-band magnitude, and display the color-mesh interpolator in Figure~\ref{fig:num_mg_r_all}. 
The number of MW foreground stars increases as the magnitude becomes fainter and the radius grows larger. For Boo1, at a $g$-band magnitude of approximately 21 and a radius of around 0.2 degree, approximately 10 foreground stars pass all of our cuts, potentially contaminating our sample.

To ensure the robustness of our method, we perform one final check by incorporating the uncertainty in metallicity into our foreground contamination estimate. We use the same sample of large-field model stars as in Section~\ref{sec:boo1_purity}, where we calculated the theoretical purity rate. We fit an exponential function of the form $y=0.0001+e^{1.08x-25}$ to the metallicity error and $g$-band magnitude of our sample of non-member stars. We then use this fitted function to get the metallicity error ($y$) of model stars based on their $g$-band magnitudes ($x$). Assuming a Gaussian distribution of metallicity for each model star with a mean equal to the metallicity returned by the model and a standard deviation equal to the metallicity error we calculated using the fit, we generate 1000 ensembles for each star and run 1000 simulation experiments to estimate, on average, how many stars pass all selection cuts. The standard deviation of the number of stars passing all selection cuts for the 1000 runs is approximately 2.5, indicating that there will be a variation of only around 2.5 stars passing all selection cuts if we include the metallicity uncertainty. This demonstrates that our selection is robust against metallicity uncertainties.

We use purity as a measure of how likely our identified star is a true member to guide future spectroscopic studies, and report this our tables of candidate members (Tables~\ref{tab:boo1mem},~\ref{tab:boo2mem},~\ref{tab:seg1mem}). {\tt{MEM\_FLAG = 0}} means this star is an identified member star in literature. {\tt{MEM\_FLAG = 1}} means this star is an identified non-member star in literature, but we flag it as a candidate member star in our catalog.   {\tt{MEM\_FLAG = 2}} means this star is not identified in previous studies, and we identify it as a bright candidate member star ($m_g < 20.2$). These are interesting candidates for future spectroscopic studies.  {\tt{MEM\_FLAG = 3}} means this star is not identified in previous studies, and we identify it as a candidate with lower purity due to its faintness ($m_g > 20.2$). Note that in some radial and magnitude bins, the number of predicted foreground stars exceed the observed number of stars, so we assign a purity of 0 for these cases.
In these tables, we also exclude stars that are $\gtrsim 0.05$\,mag from the edge of our grid of synthetic photometry, have discrepant $g-r$ vs. $r-i$ colors, or have entries in the \textit{Gaia} DR3 variable source catalog \citep{gaia.variable.2023}.
We additionally indicate stars that may be outliers relative to the CMD under a tighter selection, but still pass our isochrone tolerance in Section~\ref{sec:iso_cut} through the {\tt{CMD\_outlier\_flag}} column.

\begin{figure*}

    \includegraphics[width=\textwidth]{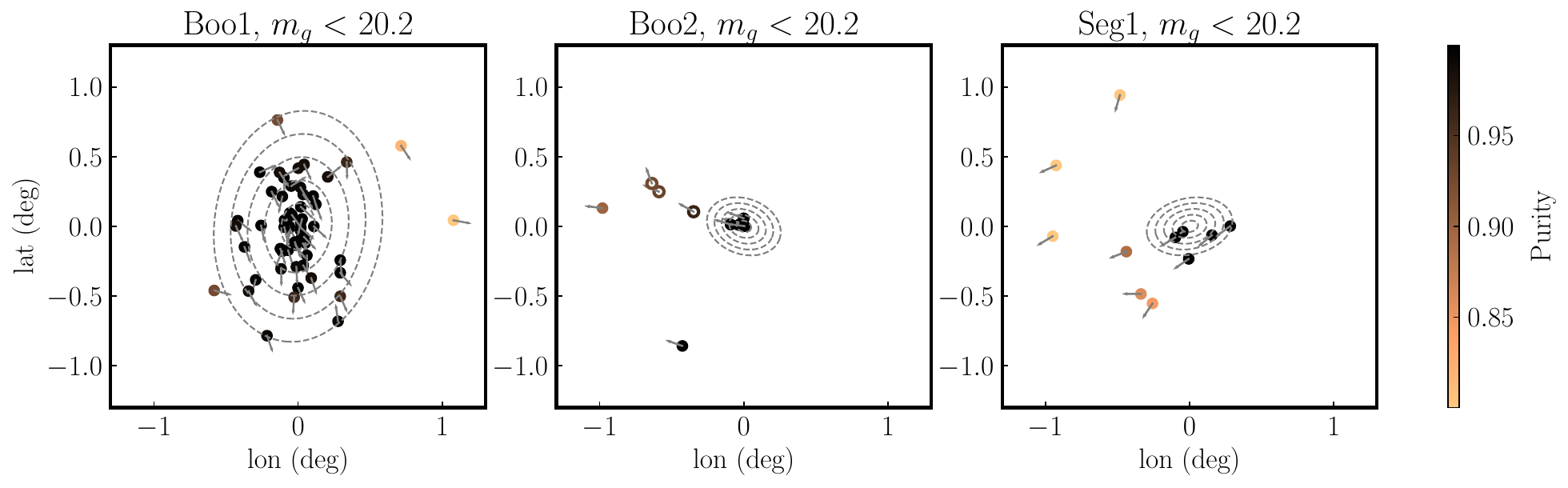}
    \caption{The spatial distribution of Boo1, Boo2, and Seg1 candidate member stars brighter than $m_g = 20.2$ color-coded according to their purity. Purity decreases as the radius increases and magnitude becomes fainter. We use purity as a measure to guide future spectroscopic follow-up of the newly found potential member stars.}
    \label{fig:mem_prob_all}

\end{figure*}

In Figure~\ref{fig:mem_prob_all}, we show a spatial distribution of candidate member stars with $m_g < 20.2$ color-coded by their purity. Most of the stars we identified have high purity $>95\%$, and all stars have purity $>80\%$. We propose these stars ({\tt{MEM\_FLAG = 2}}) for future spectroscopic studies. Table~\ref{tab:boo1mem} shows a representative sample of identified candidate member stars in our catalog.

\begin{figure*}

    \includegraphics[width=18cm]{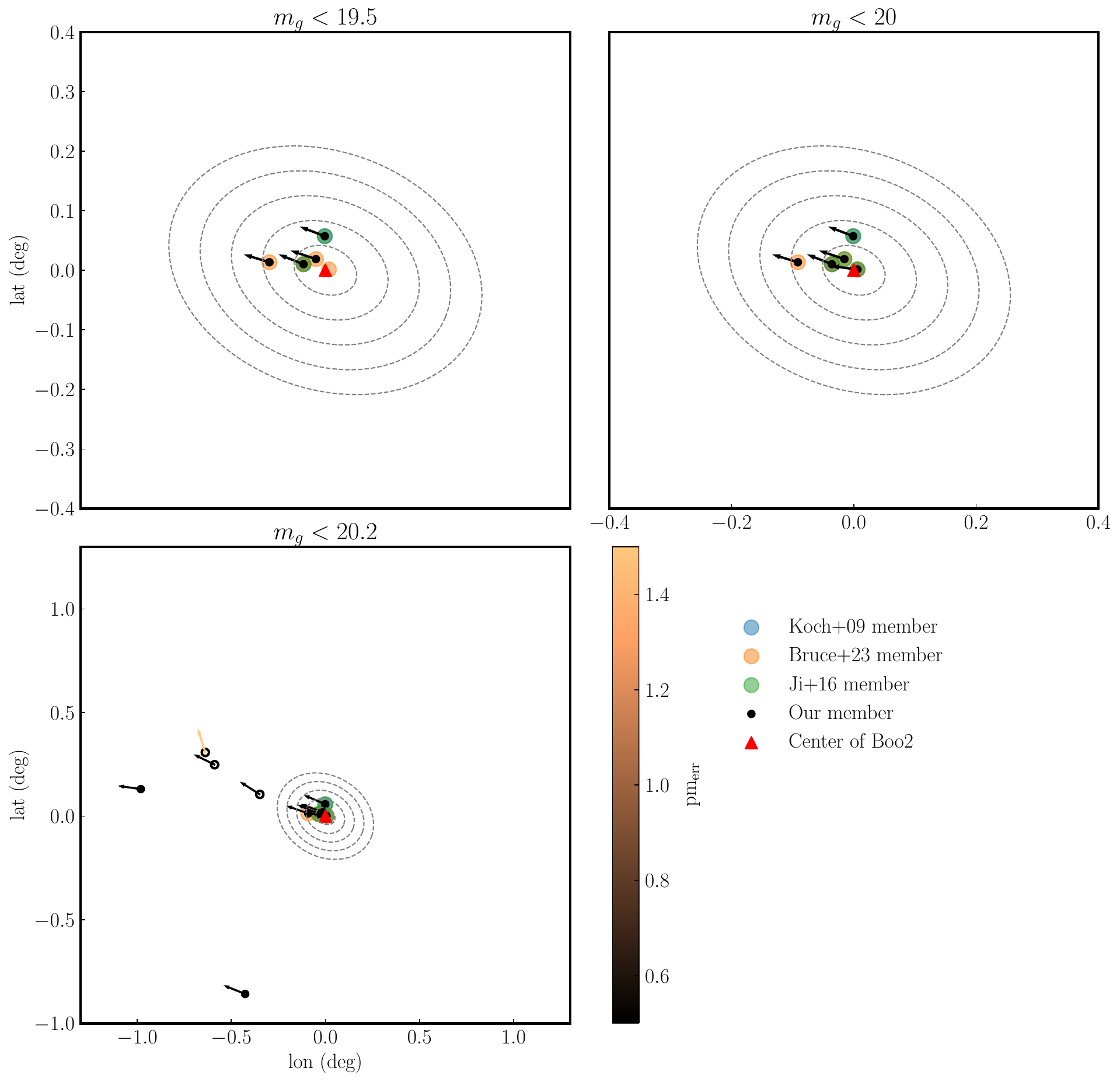}
    \caption{Spatial distribution of candidate member stars for Boo2 with a metallicity cut [Fe/H] $<-1.5$ compared with member stars in \citet{Koch.etal.2009} (blue), \citet{Bruce.etal.2023} (orange), and \citet{Ji.etal.2016} (green). Other symbols are the same as in Figure~\ref{fig:spa_dist_all}. We recover all the member stars in all three catalogs, reaching a 100\% completeness rate. For stars with $m_g < 20.2$ (lower left), we zoom out to show the entire DECam FOV ($\sim 4 $deg$^2$). There is a hint of metal-poor candidate member stars to the south and to the east  of the system, two of which (solid circles) are not outliers in CMD and thus could be interesting targets for future spectroscopic follow-up. }
    \label{fig:K_J_our_spa_dist_boo2}

\end{figure*}

\subsection{Boo2 and Seg1}
\subsubsection{Spatial distribution}\label{sec:spa_dist_boo2}

In Figure~\ref{fig:spa_dist_all}, we present the spatial distribution of candidate member stars in Boo2 that satisfy our isochrone, proper motion, and metallicity cuts. Most of our identified stars with $m_g < 20$ lie within 2 $r_h$ of the system. For the sample brighter than $m_g = 20.2$, we detect 4 stars with consistent proper motions in the DECam FOV north-east of the system at large radial distances. Three of them, indicated by hollow circles in Figure~\ref{fig:spa_dist_all} appear as outliers on the CMD despite passing our selection, so we flag them as unlikely members. The remaining candidate is a potential target for future spectroscopic follow-up studies. 

In Figure~\ref{fig:K_J_our_spa_dist_boo2}, we compare the spatial distribution of our selected candidate member stars of Boo2 to that of member and non-member stars in the \citet{Koch.etal.2009}, \citet{Ji.etal.2016}, and \citet{Bruce.etal.2023} catalogs. We apply our isochrone and proper motion cuts to all members in the three catalogs to enable a fair comparison. Note that Boo2 has a relatively small size, but we show the entire DECam FOV ($\sim 3 $deg$^2$) for the faint sample with $m_g < 20.2$ to include all detected candidate member stars in the DECam FOV. The four stars north-east of the system are not detected in any previous spectroscopic studies.

\begin{figure*}

    \includegraphics[width=18cm]{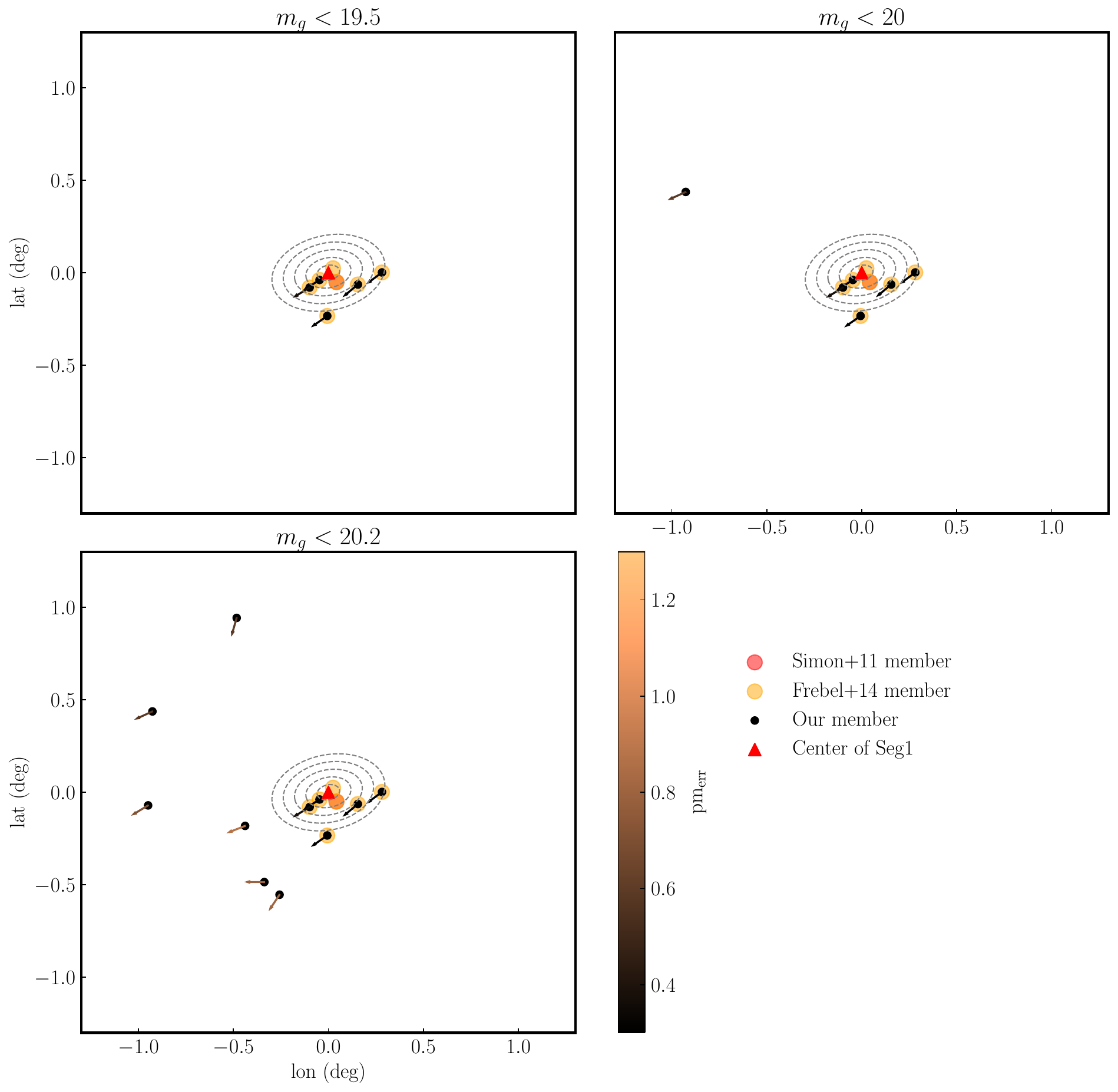}
    \caption{Spatial distribution of candidate member stars for Seg1 with a metallicity cut [Fe/H] $<-2$ compared with member stars in \citet{Simon.etal.2011} (red) and \citet{Frebel.etal.2014} (orange). We recover all but two member star in the two catalogs, SDSS J100714+160154 and SDSS J100710+160623, which have unusually high spectroscopic metallicities of $-1.42$ and $-1.67$ respectively \citep{Frebel.etal.2014}, and photometric metallicities $-1.00$ and $-1.63$, so they are cut off from our photometric metallicity cut of [Fe/H] = $-$2. We find a handful more candidate members to the south of the system compared to previous studies, in addition to a few east and north.
    As discussed in Sections~\ref{sec:spa_dist_boo2} and~\ref{sec:boo2_completeness}, the presence of the 300S and Sagittarius streams in the field-of-view may affect the purity of these candidates.}
    \label{fig:Simon_our_spa_dist_seg1}

\end{figure*}

For Seg1, Figure~\ref{fig:spa_dist_all} suggests that similar to Boo1, the spatial distribution of several new candidates in Seg1 suggest an elongation consistent with the direction of the proper motion.
This may indicate signs of Seg1 undergoing tidal disruption by the MW, but further kinematic studies of distant members are needed to fully ascertain the dynamical state of the system. By comparing with the \citet{Simon.etal.2011} and \citet{Frebel.etal.2014} samples in Figure~\ref{fig:Simon_our_spa_dist_seg1}, we find a handful more candidate member stars outside of 5 $r_h$ to the south and to the east of the system in addition to several directly east and north.
We note that the Sagittarius stream has increasing presence toward the north of the field \citep[e.g.,][]{Belokurov.etal.2006b,geha.etal.2009} and the 300S stream passes east-west \citep{Fu.etal.2018}, suggesting some distant stars in these directions may be members of those streams.  

\subsubsection{Completeness \& purity}\label{sec:boo2_completeness}

The completeness of our selected sample of candidate member stars for Boo2 and Seg1 is determined in the same way as in Section~\ref{sec:boo1_completeness}. For Boo2, every candidate member star determined in previous spectroscopic studies is identified in our photometric sample. For Seg1, to compare with \citet{Simon.etal.2011} and \citet{Frebel.etal.2014} catalog, we weed out potential 300S stream members in the field of view \citep{Fu.etal.2018} by applying an additional metallicity cut of [Fe/H] $<-2$ \citep{Li.etal.2022,Usman.etal.2024}. This cut also removes all Sagittarius stream members apart from its faint component \citep{Limberg.etal.2023}. Two member stars of Seg1, SDSS J100714+160154 and SDSS J100710+160623 \citep{Frebel.etal.2014}, in the two catalogs that have unusually high spectroscopic metallicities of $-1.42$ and $-1.67$ respectively, and photometric metallicities $-1.00$ and $-1.63$, do not show up in our catalog because they do not pass our [Fe/H] $<-2$ cut.

The purity for Boo2 and Seg1 candidate member stars is calculated in the same way as in Section~\ref{sec:boo1_purity}. For Boo2, Figure~\ref{fig:num_mg_r_all} shows that our sample is essentially clean of contamination across all radii for stars brighter than 20.0 mag, and for Seg1, our sample is essentially clean of contamination across all radii for stars brighter than 20.0 mag. 
However, the region of sky around Seg1 has many substructures like the Sagittarius stream and 300S stream, so the \emph{Besançon} model might underestimate the foreground contamination. Table~\ref{tab:boo2mem} sand Table~\ref{tab:seg1mem} show a representative sample of candidate member stars in our Boo2 and Seg1 catalogs, respectively, with same columns as in Table~\ref{tab:boo1mem}.

\section{Discussion}\label{sec:discussions}
The processes that drive the formation and evolution of UFDs, including the prominence of early dwarf-dwarf mergers, the influence of supernova feedback, and the strength of tidal disruption by a massive host, are still relatively poorly constrained. Observationally, these phenomena manifest in the outskirts of these galaxies (beyond $4 r_h$), which can be probed through detailed studies of their resolved stellar populations at large distances. Observational efforts have concentrated on the outskirts of the MW UFDs to identify distant member stars. This approach has unveiled extended stellar populations in many UFDs. For example, RR Lyrae (RRL) stars outside of the tidal radius of fourteen UFDs (Boötes I, Boötes III, Sagittarius II, Pegasus IV, Reticulum III, Eridanus II, Eridanus III, DELVE 2, Tucana II, Tucana III, Tucana IV, Hercules, Grus II, and Ursa Major I) have been identified \citep{Garling.etal.2018,Vivas.etal.2020, Tau.etal.2024}. \citet{Chiti.etal.2021} showed that there are member stars of Tucana II out to distances of $\sim 1$\,kpc, and \citet{filion.Wyse2021, Longeard.etal.2022, Longeard.etal.2023, Ou.etal.2024} have shown similar distant members and candidates in Hercules and Boötes~I. 
\citet{Jensen.etal.2024} find evidence for low-density outer profiles in nine dwarf galaxies, and \citet{Martinex-vazquez.etal.2021} reported a RRL star at $6r_h$ and another blue horizontal branch star at $>9r_h$ in Centaurus I. Here, we report multiple new bright candidate member stars $(m_g < 20.2)$ of Boo1 located at $>4r_h$, five candidate member stars of Boo2 located at $>5r_h$ (including one promising candidate member), and two candidate member stars of Seg1 at $>5r_h$. These are interesting targets for spectroscopic follow-up.

Current observations strongly support the conclusion that UFDs host stars at large distances; however, the physical mechanism that relocates stars from these low-mass systems to large distances is currently uncertain. 
One explanation is tidal disruption, which transports stars to large radii through interactions with an external, more massive galaxy. 
Indeed, a number of studies have found evidence or have suggested that some UFDs are undergoing tidal disruption by the MW \citep[e.g., ][]{Drlica-Wagner.etal.2015, Simon.etal.2017, Li.etal.2018, Mutlu-Pakdil.etal.2019, Ou.etal.2024}. 
In particular, the previously known orientation of distant members in Boo1 and its velocity gradient strongly suggest the system is undergoing tidal disruption \citep{filion.Wyse2021, Longeard.etal.2022, filion.Wyse2021, Pace.Erkal.Li.2022}. 
This is consistent with our finding of a larger number of candidates in Boo1 at large distances. 
The orbital parameters of Boo1, such as a comparison of half-light radius and tidal radius from modeling \citep[i.e., Fig 5 and 6 in][]{Pace.Erkal.Li.2022}, also indicates Boo1 as potentially being tidally disrupted, although it has a pericenter of $\sim 38$ kpc. For Seg1, it is less clear if MW tidal disruption is at play \citep[see e.g., Section 5.6 of][for a detailed discussion]{Jensen.etal.2024} even if it has a close pericenter of $\sim 20$ kpc \citep{Pace.Erkal.Li.2022}.
In particular, the intersection of many stellar streams with the Seg1 field makes a conclusive statement difficult. 
Our discovery of a few brighter ($g < 20.2$) candidate stars beyond $9 r_h$ is suggestive, but not conclusive, evidence of tidal features, and spectroscopic follow-up of these stars will be important to confirm any association and interpret the nature of this system. Boo2 has a pericenter of $\sim 36$\,kpc, and other orbital parameters do not show evidence of it being especially prone to tidally disruption \citep{Pace.Erkal.Li.2022}.

Another plausible mechanism to displace stars to large distances relies on processes during the formation of UFDs, such as early galaxy mergers and stellar feedback, which may kinematically heat the system and form an extended stellar halo \citep{Deason.etal.2014}. 
For example, the extended stellar profiles of some UFDs (e.g., Tucana II; \citealt{Chiti.etal.2021}) have been discussed as a potential result of the merging of two primordial galaxies at $z \gtrsim 2$ \citep{Deason.etal.2014, tarumi.etal.2021}, or early supernova feedback driven by a bursty star formation history \citep{Wheeler.etal.2019, Pan.Kravtsov.2023}. 
This is also related to claims that higher-mass dwarf spheroidals have cored DM profiles due to bursty star formation or supernova outflows \citep[e.g.,][]{Pontzen.etal.2012,Amorisco.etal.2014,Read.etal.2016, Orkney.etal.2021}. 

While we cannot claim a clear mechanism for the formation of the stellar outskirts of UFDs from this study alone, spectroscopic follow up of candidate members is one observational avenue to further investigate distinguish between potential formation processes (e.g., to investigate velocity gradients and chemical differences between the central/outer populations; \citealt{chiti.etal.2023, Waller.etal.2023}). 
We provide a list of candidate members in Tables~\ref{tab:boo1mem},~\ref{tab:boo2mem}, and~\ref{tab:seg1mem} to help guide such work, and also indicate stars that have already been observed by ongoing but unpublished work from the S5 Collaboration. 
In particular, we highlight that spectroscopic follow-up of brighter distant candidates of Boo2 or Seg1 may pin down the evolutionary nature of this system, given that their orbital parameters suggest that they may not be especially prone to disruption (e.g., \citealt{Pace.Erkal.Li.2022}).
If the distant stars do reflect tidal perturbations, then it may suggest that larger numbers of UFDs have potentially been tidally perturbed and lost a significant fraction of their DM halo mass due to interactions with the MW \citep[e.g.,][]{Penarrubia.etal.2008}.

\section{Conclusions}\label{sec:conclusions}

In this paper, we show that DECam $u$-band photometry, which covers the prominent Ca\,\textsc{ii} K metal absorption line, can be used to compute photometric metallicities and effectively select candidate member stars at large radial distances for three MW UFDs: Boo1, Boo2, and Seg1. 
We combine multi-band photometry information from DELVE \citep{Drlica-Wagner.etal.2022}, proper motion information from \emph{Gaia} \citep{gaiadr3}, and photometric metallicities from DECam to select candidate member stars. 
We present the spatial distribution of candidate member stars for all three systems. 
Two of the systems potentially show signs of tidal disruption, but further spectroscopy follow-up is necessary to confirm this. 
We assess the purity and completeness of our sample, and use purity as a measure to guide future spectroscopic follow-ups of the new candidate member stars we found. Our main findings can be summarized as follows:
\begin{enumerate}
    \item The DECam photometry can be used to efficiently distinguish between candidate member and non-member stars and to compute photometric metallicities (Figure~\ref{fig:grid},~\ref{fig:feh_dist}).  Our photometric metallicities generally agree spectroscopic metallicities from \citet{Jenkins.etal.2021} for Boo1 member stars and from \citet{Frebel.etal.2014} for Seg1 member stars. 
    \item The spatial distribution of Boo1 shows a prominent elongation along the direction of the proper motion of the sytem, indicating that Boo1 is possibly undergoing tidal disruption by the MW (Figures~\ref{fig:spa_dist_all} and \ref{fig:L_J_spa_dist_boo1}).
    This corroborates previous wide-field spectroscopic studies of the system \citep[e.g.,][]{Longeard.etal.2022}
    \item The spatial distribution of Boo2 shows that there are several candidate member stars in the north-east direction of the system consistent with the proper motion of the system, but three of them turn out to be outliers in CMD and may thus be unlikely to be members. (Figures~\ref{fig:spa_dist_all} and \ref{fig:K_J_our_spa_dist_boo2}).  The remaining candidates could be targets for future spectroscopic follow-up.
    \item Seg1 is contaminated by the stellar streams 300S and Sagittarius, so we add an additional metallicity cut of [Fe/H] $<-2$ to lessen substructure contamination. The spatial distribution of Seg1 also suggests a hint of tidal disruption feature (Figure~\ref{fig:spa_dist_all},~\ref{fig:Simon_our_spa_dist_seg1}), but spectroscopic follow-up is needed of identified candidate members.
    \item We present tables of identified member stars and candidate members to aid future spectroscopic studies, in addition to providing a literature compilation of members. We note that several candidates already have spectroscopy from the Southern Stellar Stream Spectroscopic Survey ($S^5$ collaboration, in prep), and this information is also indicated in the tables.
\end{enumerate}

\section{acknowledgments}

This work is supported by the Quad Undergraduate Research program at the University of Chicago, and the Department of Astronomy and Astrophysics at the University of Chicago.
We thank Sergey Koposov for providing his code for proper motion reflex corrections. The Proposal ID for the DECam observations is 2021A-0272.

A.C. is supported by a Brinson Prize Fellowship at the University of Chicago/KICP.
A.D.W acknowledges support from NSF grants AST-2006340, AST-2108168, and AST-2307126.
A.P.J. acknowledges support from NSF grant AST-2307599. G.L. acknowledges FAPESP (procs. 2021/10429-0 and 2022/07301-5).

This work has made use of data from the European Space Agency (ESA) mission {\it Gaia} (\url{https://www.cosmos.esa.int/gaia}), processed by the {\it Gaia} Data Processing and Analysis Consortium (DPAC, \url{https://www.cosmos.esa.int/web/gaia/dpac/consortium}). Funding for the DPAC has been provided by national institutions, in particular the institutions
participating in the {\it Gaia} Multilateral Agreement.

The DECam Local Volume Exploration Survey (DELVE; NOAO Proposal ID 2019A-0305, PI: Drlica-Wagner) is partially supported by Fermilab LDRD project L2019-011, the NASA Fermi Guest Investigator Program Cycle 9 No. 91201, and NSF AST-2307126.

This project used data obtained with the Dark Energy Camera (DECam), which was constructed by the Dark Energy Survey (DES) collaboration. Funding for the DES Projects has been provided by the U.S. Department of Energy, the U.S. National Science Foundation, the Ministry of Science and Education of Spain, the Science and Technology Facilities Council of the United Kingdom, the Higher Education Funding Council for England, the National Center for Supercomputing Applications at the University of Illinois at Urbana–Champaign, the Kavli Institute of Cosmological Physics at the University of Chicago, the Center for Cosmology and Astro-Particle Physics at the Ohio State University, the Mitchell Institute for Fundamental Physics and Astronomy at Texas A\&M University, Financiadora de Estudos e Projetos, Fundação Carlos Chagas Filho de Amparo à Pesquisa do Estado do Rio de Janeiro, Conselho Nacional de Desenvolvimento Científico e Tecnológico and the Ministério da Ciência, Tecnologia e Inovação, the Deutsche Forschungsgemeinschaft and the Collaborating Institutions in the Dark Energy Survey.

The Collaborating Institutions are Argonne National Laboratory, the University of California at Santa Cruz, the University of Cambridge, Centro de Investigaciones Enérgeticas, Medioambientales y Tecnológicas–Madrid, the University of Chicago, University College London, the DES-Brazil Consortium, the University of Edinburgh, the Eidgenössische Technische Hochschule (ETH) Zürich, Fermi National Accelerator Laboratory, the University of Illinois at Urbana-Champaign, the Institut de Ciències de l'Espai (IEEC/CSIC), the Institut de Física d'Altes Energies, Lawrence Berkeley National Laboratory, the Ludwig-Maximilians Universität München and the associated Excellence Cluster Universe, the University of Michigan, the National Optical Astronomy Observatory, the University of Nottingham, the Ohio State University, the OzDES Membership Consortium, the University of Pennsylvania, the University of Portsmouth, SLAC National Accelerator Laboratory, Stanford University, the University of Sussex, and Texas A\&M University.

Based in part on observations at Cerro Tololo Inter-American Observatory, National Optical Astronomy Observatory, which is operated by the Association of Universities for Research in Astronomy (AURA) under a cooperative agreement with the National Science Foundation.

Database access and other data services are hosted by the Astro Data Lab at the Community Science and Data Center (CSDC) of the National Science Foundation's National Optical Infrared Astronomy Research Laboratory, operated by the Association of Universities for Research in Astronomy (AURA) under a cooperative agreement with the National Science Foundation.

The software packages {\tt{NUMPY}} \citep{van.der.Walt.etal.2011}, {\tt{SCIPY}} \citep{Jones.etal.2001}, {\tt{Astropy}} \citep{astropy:2018}, {\tt{MATPLOTLIB}} \citep{Hunter.2007}, and GITHUB were invaluable in conducting the analyses presented in this paper. We also used the Starlink Tables Infrastructure Library Tool Set (\texttt{STILTS})\footnote{\href{http://www.starlink.ac.uk/stilts/}{\texttt{http://www.starlink.ac.uk/stilts/}}} \citep{Taylor.2006} extensively. Additionally, this research relied heavily on the Astrophysics Data Service (\href{https://ui.adsabs.harvard.edu/classic-form/}{\tt{ADS}}) and the preprint repository \href{https://arxiv.org/}{\tt{arXiv}}.
\\
\\
\\
\\
\\

\begin{rotate}
\begin{deluxetable*}{cccccccccccccc}
\tablecaption{Boo1 member star catalog}
\label{tab:boo1mem}
\setlength{\tabcolsep}{3pt}
\tablehead{
\colhead{\emph{Gaia} source ID} &
\colhead{RA} &
\colhead{DEC} &
\colhead{$u0$} &
\colhead{$g0$} &
\colhead{$i0$} &
\colhead{PMRA} &
\colhead{PMDEC} &
\colhead{[Fe/H]$_{\rm{phot}}$} &
\colhead{Purity} &
\colhead{MEM\_FLAG$^1$} &
\colhead{S5\_obs$^2$} &
\colhead{Ref.$^3$} & 
\colhead{CMD\_outlier\_flag}\\
\colhead{ } &
\colhead{(deg)} &
\colhead{(deg)} &
\colhead{ } &
\colhead{ } &
\colhead{ } &
\colhead{mas yr$^{-1}$} &
\colhead{mas yr$^{-1}$} &
\colhead{ } &
\colhead{ } &
\colhead{ } &
\colhead{ } &
\colhead{ } & 
\colhead{ }
}
\startdata
1231095269513901568 & 210.02998 & 15.2386 & $20.90\pm0.035$ & 20.17 & 19.50 & 0.302 & $-0.442$ & $-2.44\pm0.41$ & 0.88 & 0 & 1 & a & 0 \\
1230887118218565504 & 210.0679 & 14.9441 & $20.52\pm0.035$ & $19.75$ & $18.98$ & 0.149 & $-0.334$ & $-3.01\pm0.71$ & 0.97 & 0 & 1 & a & 0\\
1231090596588925056 & 209.7151 & 15.1449 & $21.41\pm0.035$ & $20.67$ & $19.99$ & 0.517 & $-0.766$ & $-2.31\pm0.32$ & 0.00 & 1 & 1 & a & 0 \\
1230874954871117824 & 210.1153 & 14.7380 & $21.26\pm0.035$ & $20.47$ & $19.78$ & 0.171 & $-0.505$ & $-2.03\pm0.26$ & 0.78 & 1 & 1 & a & 0\\
\enddata

\tablenotemark{(A full version of this table is in a machine readable format online)}
\tablenotetext{1}{The {\tt{MEM\_FLAG}} numbers are illustrated in Section~\ref{sec:boo1_purity}.}
\tablenotetext{2}{{\tt{S5\_obs}}: {\tt{S5\_obs}} $=1$ means this star was observed in S5, and {\tt{S5\_obs}} $=0$ means it was not observed in S5.}
\tablenotetext{3}{a.\citet{Longeard.etal.2022}; b.\citet{Jenkins.etal.2021}}
\end{deluxetable*}
\end{rotate}

\begin{rotate}
\begin{deluxetable*}{cccccccccccccc}
\tablecaption{Boo2 member stars catalog}
\label{tab:boo2mem}
\setlength{\tabcolsep}{3pt}
\tablehead{
\colhead{\emph{Gaia} source ID} &
\colhead{RA} &
\colhead{DEC} &
\colhead{$u0$} &
\colhead{$g0$} &
\colhead{$i0$} &
\colhead{PMRA} &
\colhead{PMDEC} &
\colhead{[Fe/H]$_{\rm{phot}}$} &
\colhead{Purity} &
\colhead{MEM\_FLAG} &
\colhead{S5\_obs} &
\colhead{Ref.$^1$} & 
\colhead{CMD\_outlier\_flag}\\
\colhead{ } &
\colhead{(deg)} &
\colhead{(deg)} &
\colhead{ } &
\colhead{ } &
\colhead{ } &
\colhead{mas yr$^{-1}$} &
\colhead{mas yr$^{-1}$} &
\colhead{ } &
\colhead{ } &
\colhead{ } &
\colhead{ } &
\colhead{ } & 
\colhead{ }
}
\startdata
3727837759579718272 & 209.4061 & 12.8635 & $18.84\pm0.035$ & $18.15$ & $17.48$ & $-1.67$ & 0.526 & $-3.61\pm0.75$ & 1. & 0 & 0 & b & 1\\
3727825076541007232 & 209.4323 & 12.7905 & $20.95\pm0.035$ & $20.21$ & $19.54$ & $-1.569$ & 0.370 & $-2.26\pm0.31$ & 0.99 & 0 & 0 & b & 0\\
3728628853900665856 & 208.4935 & 12.9786 & $20.66\pm0.035$ & $19.96$ & $19.32$ & $-2.076$ & 0.268 & $-2.25\pm0.33$ & 0.65 & 2 & 0 & ... & 0 \\
3727823118035927936 & 209.5725 & 12.8035 & $21.40\pm0.035$ & $20.71$ & $20.10$ & $-1.787$ & $-1.205$ & $-2.50\pm0.75$ & 0.69 & 3 & 1 &  ... & 0\\
1230683227532109056 & 209.2676 & 13.7516 & $21.37\pm0.035$ & $20.75$ & $20.19$ & $-0.596$ & 0.685 & $-2.5\pm0.75$ & 0.00 & 3 & 0 & ... & 0
\enddata
\tablenotemark{(A full version of this table is in a machine readable format online)}
\tablenotetext{1}{a.\citet{Koch.etal.2009}; b.\citet{Bruce.etal.2023}; c.\citet{Ji.etal.2016}}
\end{deluxetable*}
\end{rotate}
\begin{rotate}
\begin{deluxetable*}{cccccccccccccc}
\tablecaption{Seg1 member stars catalog}
\label{tab:seg1mem}
\setlength{\tabcolsep}{3pt}
\tablehead{
\colhead{\emph{Gaia} source ID} &
\colhead{RA} &
\colhead{DEC} &
\colhead{$u0$} &
\colhead{$g0$} &
\colhead{$i0$} &
\colhead{PMRA} &
\colhead{PMDEC} &
\colhead{[Fe/H]$_{\rm{phot}}$} &
\colhead{Purity} &
\colhead{MEM\_FLAG} &
\colhead{S5\_obs} &
\colhead{Ref.$^1$} & 
\colhead{CMD\_outlier\_flag}\\
\colhead{ } &
\colhead{(deg)} &
\colhead{(deg)} &
\colhead{ } &
\colhead{ } &
\colhead{ } &
\colhead{mas yr$^{-1}$} &
\colhead{mas yr$^{-1}$} &
\colhead{ } &
\colhead{ } &
\colhead{ } &
\colhead{ } &
\colhead{ } & 
\colhead{ }
}
\startdata
621943184658438784 & 151.7180 & 16.0433 & $19.35\pm0.035$ & $18.74$ & $18.10$ & $-2.017$ & $-1.662$ & $-4.\pm0.75$ & 0.99 & 0 & 1 & a,b & 0\\
621924492960734464 & 151.7603 & 15.8487 & $19.13\pm0.035$ & $18.36$ & $17.66$ & $-1.898$ & $-1.698$ & $-2.28\pm0.29$ & 0.99 & 0 & 1 & a,b & 0\\
621925901710013952 & 151.8407 & 15.9068 & $20.97\pm0.035$ & $20.50$ & $20.18$ & $-2.333$ & $-1.339$ & $-2.68\pm0.75$ & 0. & 1 & 0 & a & 0\\
622634296435669376 & 150.7791 & 16.0090 & $20.65\pm0.035$ & $20.04$ & $19.47$ & $-1.609$ & $1.352$ & $-2.50\pm0.75$ & 0.86 & 2 & 0 & ... & 0\\
621819764477687296 & 151.4164 & 15.5965 & $20.59\pm0.035$ & $20.01$ & $19.54$ & $-2.483$ & $-0.394$ & $-2.07\pm0.54$ & 0.91 & 2 & 0 & ... & 0\\
622820771030434816 & 151.2795 & 16.9198 & $21.33\pm0.035$ & $20.81$ & $20.42$ & $-1.155$ & $0.165$ & $-2.11\pm0.79$ & 0. & 3 & 0 & ... & 0\\
622790908122977920 & 151.1391 & 16.7076 & $21.32\pm0.035$ & $20.81$ & $20.43$ & $-4.231$ & $-0.581$ & $-2.17\pm0.88$ & 0. & 3 & 0 & ...  & 0
\enddata
\tablenotemark{(A full version of this table is in a machine readable format online)}
\tablenotetext{1}{a.\citet{Simon.etal.2011}; b.\citet{Frebel.etal.2014}; c.\citet{Norris.etal.2010}}
\end{deluxetable*}
\end{rotate}
\bibliography{main}{}

\begin{thebibliography}{}
\expandafter\ifx\csname natexlab\endcsname\relax\def\natexlab#1{#1}\fi
\providecommand{\url}[1]{\href{#1}{#1}}
\providecommand{\dodoi}[1]{doi:~\href{http://doi.org/#1}{\nolinkurl{#1}}}
\providecommand{\doeprint}[1]{\href{http://ascl.net/#1}{\nolinkurl{http://ascl.net/#1}}}
\providecommand{\doarXiv}[1]{\href{https://arxiv.org/abs/#1}{\nolinkurl{https://arxiv.org/abs/#1}}}

\bibitem[{{Abbott} {et~al.}(2018){Abbott}, {Abdalla}, {Allam}, {Amara},
  {Annis}, {Asorey}, {Avila}, {Ballester}, {Banerji}, {Barkhouse}, {Baruah},
  {Baumer}, {Bechtol}, {Becker}, {Benoit-L{\'e}vy}, {Bernstein}, {Bertin},
  {Blazek}, {Bocquet}, {Brooks}, {Brout}, {Buckley-Geer}, {Burke}, {Busti},
  {Campisano}, {Cardiel-Sas}, {Carnero Rosell}, {Carrasco Kind}, {Carretero},
  {Castander}, {Cawthon}, {Chang}, {Chen}, {Conselice}, {Costa}, {Crocce},
  {Cunha}, {D'Andrea}, {da Costa}, {Das}, {Daues}, {Davis}, {Davis}, {De
  Vicente}, {DePoy}, {DeRose}, {Desai}, {Diehl}, {Dietrich}, {Dodelson},
  {Doel}, {Drlica-Wagner}, {Eifler}, {Elliott}, {Evrard}, {Farahi}, {Fausti
  Neto}, {Fernandez}, {Finley}, {Flaugher}, {Foley}, {Fosalba}, {Friedel},
  {Frieman}, {Garc{\'\i}a-Bellido}, {Gaztanaga}, {Gerdes}, {Giannantonio},
  {Gill}, {Glazebrook}, {Goldstein}, {Gower}, {Gruen}, {Gruendl}, {Gschwend},
  {Gupta}, {Gutierrez}, {Hamilton}, {Hartley}, {Hinton}, {Hislop}, {Hollowood},
  {Honscheid}, {Hoyle}, {Huterer}, {Jain}, {James}, {Jeltema}, {Johnson},
  {Johnson}, {Kacprzak}, {Kent}, {Khullar}, {Klein}, {Kovacs}, {Koziol},
  {Krause}, {Kremin}, {Kron}, {Kuehn}, {Kuhlmann}, {Kuropatkin}, {Lahav},
  {Lasker}, {Li}, {Li}, {Liddle}, {Lima}, {Lin}, {L{\'o}pez-Reyes}, {MacCrann},
  {Maia}, {Maloney}, {Manera}, {March}, {Marriner}, {Marshall}, {Martini},
  {McClintock}, {McKay}, {McMahon}, {Melchior}, {Menanteau}, {Miller},
  {Miquel}, {Mohr}, {Morganson}, {Mould}, {Neilsen}, {Nichol}, {Nogueira},
  {Nord}, {Nugent}, {Nunes}, {Ogando}, {Old}, {Pace}, {Palmese},
  {Paz-Chinch{\'o}n}, {Peiris}, {Percival}, {Petravick}, {Plazas}, {Poh},
  {Pond}, {Porredon}, {Pujol}, {Refregier}, {Reil}, {Ricker}, {Rollins},
  {Romer}, {Roodman}, {Rooney}, {Ross}, {Rykoff}, {Sako}, {Sanchez}, {Sanchez},
  {Santiago}, {Saro}, {Scarpine}, {Scolnic}, {Serrano}, {Sevilla-Noarbe},
  {Sheldon}, {Shipp}, {Silveira}, {Smith}, {Smith}, {Smith}, {Soares-Santos},
  {Sobreira}, {Song}, {Stebbins}, {Suchyta}, {Sullivan}, {Swanson}, {Tarle},
  {Thaler}, {Thomas}, {Thomas}, {Troxel}, {Tucker}, {Vikram}, {Vivas},
  {Walker}, {Wechsler}, {Weller}, {Wester}, {Wolf}, {Wu}, {Yanny}, {Zenteno},
  {Zhang}, {Zuntz}, {DES Collaboration}, {Juneau}, {Fitzpatrick}, {Nikutta},
  {Nidever}, {Olsen}, {Scott}, \& {NOAO Data Lab}}]{Abbott.etal.2018}
{Abbott}, T.~M.~C., {Abdalla}, F.~B., {Allam}, S., {et~al.} 2018, \apjs, 239,
  18, \dodoi{10.3847/1538-4365/aae9f0}

\bibitem[{{Abbott} {et~al.}(2021){Abbott}, {Adam{\'o}w}, {Aguena}, {Allam},
  {Amon}, {Annis}, {Avila}, {Bacon}, {Banerji}, {Bechtol}, {Becker},
  {Bernstein}, {Bertin}, {Bhargava}, {Bridle}, {Brooks}, {Burke}, {Carnero
  Rosell}, {Carrasco Kind}, {Carretero}, {Castander}, {Cawthon}, {Chang},
  {Choi}, {Conselice}, {Costanzi}, {Crocce}, {da Costa}, {Davis}, {De Vicente},
  {DeRose}, {Desai}, {Diehl}, {Dietrich}, {Drlica-Wagner}, {Eckert},
  {Elvin-Poole}, {Everett}, {Evrard}, {Ferrero}, {Fert{\'e}}, {Flaugher},
  {Fosalba}, {Friedel}, {Frieman}, {Garc{\'\i}a-Bellido}, {Gaztanaga},
  {Gelman}, {Gerdes}, {Giannantonio}, {Gill}, {Gruen}, {Gruendl}, {Gschwend},
  {Gutierrez}, {Hartley}, {Hinton}, {Hollowood}, {Honscheid}, {Huterer},
  {James}, {Jeltema}, {Johnson}, {Kent}, {Kron}, {Kuehn}, {Kuropatkin},
  {Lahav}, {Li}, {Lidman}, {Lin}, {MacCrann}, {Maia}, {Manning}, {Maloney},
  {March}, {Marshall}, {Martini}, {Melchior}, {Menanteau}, {Miquel}, {Morgan},
  {Myles}, {Neilsen}, {Ogando}, {Palmese}, {Paz-Chinch{\'o}n}, {Petravick},
  {Pieres}, {Plazas}, {Pond}, {Rodriguez-Monroy}, {Romer}, {Roodman}, {Rykoff},
  {Sako}, {Sanchez}, {Santiago}, {Scarpine}, {Serrano}, {Sevilla-Noarbe},
  {Smith}, {Smith}, {Soares-Santos}, {Suchyta}, {Swanson}, {Tarle}, {Thomas},
  {To}, {Tremblay}, {Troxel}, {Tucker}, {Turner}, {Varga}, {Walker},
  {Wechsler}, {Weller}, {Wester}, {Wilkinson}, {Yanny}, {Zhang}, {Nikutta},
  {Fitzpatrick}, {Jacques}, {Scott}, {Olsen}, {Huang}, {Herrera}, {Juneau},
  {Nidever}, {Weaver}, {Adean}, {Correia}, {de Freitas}, {Freitas},
  {Singulani}, {Vila-Verde}, \& {Linea Science Server}}]{Abbott.etal.2021}
{Abbott}, T.~M.~C., {Adam{\'o}w}, M., {Aguena}, M., {et~al.} 2021, \apjs, 255,
  20, \dodoi{10.3847/1538-4365/ac00b3}

\bibitem[{{Abdallah} {et~al.}(2020){Abdallah}, {Adam}, {Aharonian}, {Ait
  Benkhali}, {Ang{\"u}ner}, {Arakawa}, {Arcaro}, {Armand}, {Armstrong},
  {Ashkar}, {Backes}, {Baghmanyan}, {Barbosa Martins}, {Barnacka}, {Barnard},
  {Becherini}, {Berge}, {Bernl{\"o}hr}, {B{\"o}ttcher}, {Boisson}, {Bolmont},
  {Bonnefoy}, {Breuhaus}, {Bregeon}, {Brun}, {Brun}, {Bryan}, {B{\"u}chele},
  {Bulik}, {Bylund}, {Caroff}, {Carosi}, {Casanova}, {Chand}, {Chandra},
  {Chen}, {Cotter}, {Cury{\l}o}, {Davids}, {Davies}, {Deil}, {Devin}, {deWilt},
  {Dirson}, {Djannati-Ata{\"\i}}, {Dmytriiev}, {Donath}, {Doroshenko}, {Dyks},
  {Egberts}, {Eichhorn}, {Emery}, {Ernenwein}, {Eschbach}, {Feijen}, {Fegan},
  {Fiasson}, {Fontaine}, {Funk}, {F{\"u}{\ss}ling}, {Gabici}, {Gallant},
  {Giavitto}, {Giunti}, {Glawion}, {Glicenstein}, {Gottschall}, {Grondin},
  {Hahn}, {Haupt}, {Hermann}, {Hinton}, {Hofmann}, {Hoischen}, {Holch},
  {Holler}, {H{\"o}rbe}, {Horns}, {Huber}, {Iwasaki}, {Jamrozy}, {Jankowsky},
  {Jankowsky}, {Jardin-Blicq}, {Joshi}, {Jung-Richardt}, {Kastendieck},
  {Katarzy{\'n}ski}, {Katsuragawa}, {Katz}, {Khangulyan}, {Kh{\'e}lifi},
  {Klepser}, {Klu{\'z}niak}, {Komin}, {Konno}, {Kosack}, {Kostunin}, {Kreter},
  {Lamanna}, {Lemi{\`e}re}, {Lemoine-Goumard}, {Lenain}, {Leser}, {Levy},
  {Lohse}, {Lypova}, {Mackey}, {Majumdar}, {Malyshev}, {Malyshev}, {Marandon},
  {Marchegiani}, {Marcowith}, {Mares}, {Mart{\i}-Devesa}, {Marx}, {Maurin},
  {Meintjes}, {Moderski}, {Mohamed}, {Mohrmann}, {Moore}, {Morris}, {Moulin},
  {Muller}, {Murach}, {Nakashima}, {Nakashima}, {de Naurois}, {Ndiyavala},
  {Niederwanger}, {Niemiec}, {Oakes}, {O'Brien}, {Odaka}, {Ohm}, {de Ona
  Wilhelmi}, {Ostrowski}, {Panter}, {Parsons}, {Peyaud}, {Piel}, {Pita},
  {Poireau}, {Priyana Noel}, {Prokhorov}, {Prokoph}, {P{\"u}hlhofer}, {Punch},
  {Quirrenbach}, {Raab}, {Rauth}, {Reimer}, {Reimer}, {Remy}, {Renaud},
  {Rieger}, {Rinchiuso}, {Romoli}, {Rowell}, {Rudak}, {Ruiz-Velasco},
  {Sahakian}, {Sailer}, {Saito}, {Sanchez}, {Santangelo}, {Sasaki}, {Scalici},
  {Sch{\"u}ssler}, {Schutter}, {Schwanke}, {Schwemmer}, {Seglar-Arroyo},
  {Senniappan}, {Seyffert}, {Shafi}, {Shiningayamwe}, {Simoni}, {Sinha}, {Sol},
  {Specovius}, {Spencer}, {Spir-Jacob}, {Stawarz}, {Steenkamp}, {Stegmann},
  {Steppa}, {Takahashi}, {Tavernier}, {Taylor}, {Terrier}, {Tiziani},
  {Tluczykont}, {Tomankova}, {Trichard}, {Tsirou}, {Tsuji}, {Tuffs},
  {Uchiyama}, {van der Walt}, {van Eldik}, {van Rensburg}, {van Soelen},
  {Vasileiadis}, {Veh}, {Venter}, {Viana}, {Vincent}, {Vink}, {V{\"o}lk},
  {Vuillaume}, {Wadiasingh}, {Wagner}, {Watson}, {Werner}, {White},
  {Wierzcholska}, {Yang}, {Yoneda}, {Zacharias}, {Zanin}, {Zargaryan},
  {Zdziarski}, {Zech}, {Zhu}, {Zorn}, {{\.Z}ywucka}, \& {H.~E.~S.~S.
  Collaboration}}]{Abdallah.etal.2020}
{Abdallah}, H., {Adam}, R., {Aharonian}, F., {et~al.} 2020, \prd, 102, 062001,
  \dodoi{10.1103/PhysRevD.102.062001}

\bibitem[{{Ahumada} {et~al.}(2020){Ahumada}, {Prieto}, {Almeida}, {Anders},
  {Anderson}, {Andrews}, {Anguiano}, {Arcodia}, {Armengaud}, {Aubert}, {Avila},
  {Avila-Reese}, {Badenes}, {Balland}, {Barger}, {Barrera-Ballesteros}, {Basu},
  {Bautista}, {Beaton}, {Beers}, {Benavides}, {Bender}, {Bernardi}, {Bershady},
  {Beutler}, {Bidin}, {Bird}, {Bizyaev}, {Blanc}, {Blanton}, {Boquien},
  {Borissova}, {Bovy}, {Brandt}, {Brinkmann}, {Brownstein}, {Bundy}, {Bureau},
  {Burgasser}, {Burtin}, {Cano-D{\'\i}az}, {Capasso}, {Cappellari}, {Carrera},
  {Chabanier}, {Chaplin}, {Chapman}, {Cherinka}, {Chiappini}, {Doohyun Choi},
  {Chojnowski}, {Chung}, {Clerc}, {Coffey}, {Comerford}, {Comparat}, {da
  Costa}, {Cousinou}, {Covey}, {Crane}, {Cunha}, {Ilha}, {Dai}, {Damsted},
  {Darling}, {Davidson}, {Davies}, {Dawson}, {De}, {de la Macorra}, {De Lee},
  {Queiroz}, {Deconto Machado}, {de la Torre}, {Dell'Agli}, {du Mas des
  Bourboux}, {Diamond-Stanic}, {Dillon}, {Donor}, {Drory}, {Duckworth},
  {Dwelly}, {Ebelke}, {Eftekharzadeh}, {Davis Eigenbrot}, {Elsworth},
  {Eracleous}, {Erfanianfar}, {Escoffier}, {Fan}, {Farr},
  {Fern{\'a}ndez-Trincado}, {Feuillet}, {Finoguenov}, {Fofie},
  {Fraser-McKelvie}, {Frinchaboy}, {Fromenteau}, {Fu}, {Galbany}, {Garcia},
  {Garc{\'\i}a-Hern{\'a}ndez}, {Oehmichen}, {Ge}, {Maia}, {Geisler}, {Gelfand},
  {Goddy}, {Gonzalez-Perez}, {Grabowski}, {Green}, {Grier}, {Guo}, {Guy},
  {Harding}, {Hasselquist}, {Hawken}, {Hayes}, {Hearty}, {Hekker}, {Hogg},
  {Holtzman}, {Horta}, {Hou}, {Hsieh}, {Huber}, {Hunt}, {Chitham}, {Imig},
  {Jaber}, {Angel}, {Johnson}, {Jones}, {J{\"o}nsson}, {Jullo}, {Kim},
  {Kinemuchi}, {Kirkpatrick}, {Kite}, {Klaene}, {Kneib}, {Kollmeier}, {Kong},
  {Kounkel}, {Krishnarao}, {Lacerna}, {Lan}, {Lane}, {Law}, {Le Goff}, {Leung},
  {Lewis}, {Li}, {Lian}, {Lin}, {Long}, {Longa-Pe{\~n}a}, {Lundgren}, {Lyke},
  {Ted Mackereth}, {MacLeod}, {Majewski}, {Manchado}, {Maraston}, {Martini},
  {Masseron}, {Masters}, {Mathur}, {McDermid}, {Merloni}, {Merrifield},
  {M{\'e}sz{\'a}ros}, {Miglio}, {Minniti}, {Minsley}, {Miyaji}, {Mohammad},
  {Mosser}, {Mueller}, {Muna}, {Mu{\~n}oz-Guti{\'e}rrez}, {Myers}, {Nadathur},
  {Nair}, {Nandra}, {do Nascimento}, {Nevin}, {Newman}, {Nidever}, {Nitschelm},
  {Noterdaeme}, {O'Connell}, {Olmstead}, {Oravetz}, {Oravetz}, {Osorio},
  {Pace}, {Padilla}, {Palanque-Delabrouille}, {Palicio}, {Pan}, {Pan},
  {Parker}, {Paviot}, {Peirani}, {Ram{\'r}ez}, {Penny}, {Percival},
  {Perez-Fournon}, {P{\'e}rez-R{\`a}fols}, {Petitjean}, {Pieri},
  {Pinsonneault}, {Poovelil}, {Povick}, {Prakash}, {Price-Whelan}, {Raddick},
  {Raichoor}, {Ray}, {Rembold}, {Rezaie}, {Riffel}, {Riffel}, {Rix}, {Robin},
  {Roman-Lopes}, {Rom{\'a}n-Z{\'u}{\~n}iga}, {Rose}, {Ross}, {Rossi},
  {Rowlands}, {Rubin}, {Salvato}, {S{\'a}nchez}, {S{\'a}nchez-Menguiano},
  {S{\'a}nchez-Gallego}, {Sayres}, {Schaefer}, {Schiavon}, {Schimoia},
  {Schlafly}, {Schlegel}, {Schneider}, {Schultheis}, {Schwope}, {Seo},
  {Serenelli}, {Shafieloo}, {Shamsi}, {Shao}, {Shen}, {Shetrone}, {Shirley},
  {Aguirre}, {Simon}, {Skrutskie}, {Slosar}, {Smethurst}, {Sobeck}, {Sodi},
  {Souto}, {Stark}, {Stassun}, {Steinmetz}, {Stello}, {Stermer},
  {Storchi-Bergmann}, {Streblyanska}, {Stringfellow}, {Stutz}, {Su{\'a}rez},
  {Sun}, {Taghizadeh-Popp}, {Talbot}, {Tayar}, {Thakar}, {Theriault}, {Thomas},
  {Thomas}, {Tinker}, {Tojeiro}, {Toledo}, {Tremonti}, {Troup}, {Tuttle},
  {Unda-Sanzana}, {Valentini}, {Vargas-Gonz{\'a}lez}, {Vargas-Maga{\~n}a},
  {V{\'a}zquez-Mata}, {Vivek}, {Wake}, {Wang}, {Weaver}, {Weijmans}, {Wild},
  {Wilson}, {Wilson}, {Wolthuis}, {Wood-Vasey}, {Yan}, {Yang}, {Y{\`e}che},
  {Zamora}, {Zarrouk}, {Zasowski}, {Zhang}, {Zhao}, {Zhao}, {Zheng}, {Zheng},
  {Zhu}, \& {Zou}}]{Ahumada.etal.2020}
{Ahumada}, R., {Prieto}, C.~A., {Almeida}, A., {et~al.} 2020, \apjs, 249, 3,
  \dodoi{10.3847/1538-4365/ab929e10.48550/arXiv.1912.02905}

\bibitem[{{Alvarez} \& {Plez}(1998)}]{Alvarez.Plez.1998}
{Alvarez}, R., \& {Plez}, B. 1998, \aap, 330, 1109,
  \dodoi{10.48550/arXiv.astro-ph/9710157}

\bibitem[{{Amorisco} {et~al.}(2014){Amorisco}, {Zavala}, \& {de
  Boer}}]{Amorisco.etal.2014}
{Amorisco}, N.~C., {Zavala}, J., \& {de Boer}, T.~J.~L. 2014, \apjl, 782, L39,
  \dodoi{10.1088/2041-8205/782/2/L39}

\bibitem[{{Anthony-Twarog} {et~al.}(1991){Anthony-Twarog}, {Laird}, {Payne}, \&
  {Twarog}}]{Anthony-Twarog.etal.1991}
{Anthony-Twarog}, B.~J., {Laird}, J.~B., {Payne}, D., \& {Twarog}, B.~A. 1991,
  \aj, 101, 1902, \dodoi{10.1086/115815}

\bibitem[{{Astropy Collaboration} {et~al.}(2018){Astropy Collaboration},
  {Price-Whelan}, {Sip{\H{o}}cz}, {G{\"u}nther}, {Lim}, {Crawford}, {Conseil},
  {Shupe}, {Craig}, {Dencheva}, {Ginsburg}, {Vand erPlas}, {Bradley},
  {P{\'e}rez-Su{\'a}rez}, {de Val-Borro}, {Aldcroft}, {Cruz}, {Robitaille},
  {Tollerud}, {Ardelean}, {Babej}, {Bach}, {Bachetti}, {Bakanov}, {Bamford},
  {Barentsen}, {Barmby}, {Baumbach}, {Berry}, {Biscani}, {Boquien}, {Bostroem},
  {Bouma}, {Brammer}, {Bray}, {Breytenbach}, {Buddelmeijer}, {Burke},
  {Calderone}, {Cano Rodr{\'\i}guez}, {Cara}, {Cardoso}, {Cheedella}, {Copin},
  {Corrales}, {Crichton}, {D'Avella}, {Deil}, {Depagne}, {Dietrich}, {Donath},
  {Droettboom}, {Earl}, {Erben}, {Fabbro}, {Ferreira}, {Finethy}, {Fox},
  {Garrison}, {Gibbons}, {Goldstein}, {Gommers}, {Greco}, {Greenfield},
  {Groener}, {Grollier}, {Hagen}, {Hirst}, {Homeier}, {Horton}, {Hosseinzadeh},
  {Hu}, {Hunkeler}, {Ivezi{\'c}}, {Jain}, {Jenness}, {Kanarek}, {Kendrew},
  {Kern}, {Kerzendorf}, {Khvalko}, {King}, {Kirkby}, {Kulkarni}, {Kumar},
  {Lee}, {Lenz}, {Littlefair}, {Ma}, {Macleod}, {Mastropietro}, {McCully},
  {Montagnac}, {Morris}, {Mueller}, {Mumford}, {Muna}, {Murphy}, {Nelson},
  {Nguyen}, {Ninan}, {N{\"o}the}, {Ogaz}, {Oh}, {Parejko}, {Parley}, {Pascual},
  {Patil}, {Patil}, {Plunkett}, {Prochaska}, {Rastogi}, {Reddy Janga},
  {Sabater}, {Sakurikar}, {Seifert}, {Sherbert}, {Sherwood-Taylor}, {Shih},
  {Sick}, {Silbiger}, {Singanamalla}, {Singer}, {Sladen}, {Sooley},
  {Sornarajah}, {Streicher}, {Teuben}, {Thomas}, {Tremblay}, {Turner},
  {Terr{\'o}n}, {van Kerkwijk}, {de la Vega}, {Watkins}, {Weaver}, {Whitmore},
  {Woillez}, {Zabalza}, \& {Astropy Contributors}}]{astropy:2018}
{Astropy Collaboration}, {Price-Whelan}, A.~M., {Sip{\H{o}}cz}, B.~M., {et~al.}
  2018, \aj, 156, 123, \dodoi{10.3847/1538-3881/aabc4f}

\bibitem[{{Belokurov} {et~al.}(2006{\natexlab{a}}){Belokurov}, {Zucker},
  {Evans}, {Wilkinson}, {Irwin}, {Hodgkin}, {Bramich}, {Irwin}, {Gilmore},
  {Willman}, {Vidrih}, {Newberg}, {Wyse}, {Fellhauer}, {Hewett}, {Cole},
  {Bell}, {Beers}, {Rockosi}, {Yanny}, {Grebel}, {Schneider}, {Lupton},
  {Barentine}, {Brewington}, {Brinkmann}, {Harvanek}, {Kleinman}, {Krzesinski},
  {Long}, {Nitta}, {Smith}, \& {Snedden}}]{Belokurov.etal.2006}
{Belokurov}, V., {Zucker}, D.~B., {Evans}, N.~W., {et~al.} 2006{\natexlab{a}},
  \apjl, 647, L111, \dodoi{10.1086/507324}

\bibitem[{{Belokurov} {et~al.}(2006{\natexlab{b}}){Belokurov}, {Zucker},
  {Evans}, {Gilmore}, {Vidrih}, {Bramich}, {Newberg}, {Wyse}, {Irwin},
  {Fellhauer}, {Hewett}, {Walton}, {Wilkinson}, {Cole}, {Yanny}, {Rockosi},
  {Beers}, {Bell}, {Brinkmann}, {Ivezi{\'c}}, \&
  {Lupton}}]{Belokurov.etal.2006b}
---. 2006{\natexlab{b}}, \apjl, 642, L137, \dodoi{10.1086/504797}

\bibitem[{{Belokurov} {et~al.}(2007){Belokurov}, {Zucker}, {Evans}, {Kleyna},
  {Koposov}, {Hodgkin}, {Irwin}, {Gilmore}, {Wilkinson}, {Fellhauer},
  {Bramich}, {Hewett}, {Vidrih}, {De Jong}, {Smith}, {Rix}, {Bell}, {Wyse},
  {Newberg}, {Mayeur}, {Yanny}, {Rockosi}, {Gnedin}, {Schneider}, {Beers},
  {Barentine}, {Brewington}, {Brinkmann}, {Harvanek}, {Kleinman}, {Krzesinski},
  {Long}, {Nitta}, \& {Snedden}}]{Belokurov.etal.2007}
---. 2007, \apj, 654, 897, \dodoi{10.1086/509718}

\bibitem[{{Bernstein} {et~al.}(2017){Bernstein}, {Abbott}, {Desai}, {Gruen},
  {Gruendl}, {Johnson}, {Lin}, {Menanteau}, {Morganson}, {Neilsen}, {Paech},
  {Walker}, {Wester}, {Yanny}, \& {DES Collaboration}}]{Bernstein:2017}
{Bernstein}, G.~M., {Abbott}, T.~M.~C., {Desai}, S., {et~al.} 2017, \pasp, 129,
  114502, \dodoi{10.1088/1538-3873/aa858e}

\bibitem[{{Bertin}(2006)}]{Bertin:2006}
{Bertin}, E. 2006, in Astronomical Society of the Pacific Conference Series,
  Vol. 351, Astronomical Data Analysis Software and Systems XV, ed.
  C.~{Gabriel}, C.~{Arviset}, D.~{Ponz}, \& S.~{Enrique}, 112

\bibitem[{{Bertin}(2011)}]{Bertin:2011}
{Bertin}, E. 2011, in Astronomical Society of the Pacific Conference Series,
  Vol. 442, Astronomical Data Analysis Software and Systems XX, ed. I.~N.
  {Evans}, A.~{Accomazzi}, D.~J. {Mink}, \& A.~H. {Rots}, 435

\bibitem[{{Bertin} \& {Arnouts}(1996)}]{Bertin:1996}
{Bertin}, E., \& {Arnouts}, S. 1996, \aaps, 117, 393,
  \dodoi{10.1051/aas:1996164}

\bibitem[{{Brown} {et~al.}(2014){Brown}, {Tumlinson}, {Geha}, {Simon},
  {Vargas}, {VandenBerg}, {Kirby}, {Kalirai}, {Avila}, {Gennaro}, {Ferguson},
  {Mu{\~n}oz}, {Guhathakurta}, \& {Renzini}}]{Brown.etal.2014}
{Brown}, T.~M., {Tumlinson}, J., {Geha}, M., {et~al.} 2014, \apj, 796, 91,
  \dodoi{10.1088/0004-637X/796/2/91}

\bibitem[{{Bruce} {et~al.}(2023){Bruce}, {Li}, {Pace}, {Heiger}, {Song}, \&
  {Simon}}]{Bruce.etal.2023}
{Bruce}, J., {Li}, T.~S., {Pace}, A.~B., {et~al.} 2023, \apj, 950, 167,
  \dodoi{10.3847/1538-4357/acc943}

\bibitem[{{Calore} {et~al.}(2018){Calore}, {Serpico}, \&
  {Zaldivar}}]{Calore.etal.2018}
{Calore}, F., {Serpico}, P.~D., \& {Zaldivar}, B. 2018, \jcap, 2018, 029,
  \dodoi{10.1088/1475-7516/2018/10/029}

\bibitem[{{Carretta} {et~al.}(2009){Carretta}, {Bragaglia}, {Gratton},
  {D'Orazi}, \& {Lucatello}}]{Carretta.etal.2009}
{Carretta}, E., {Bragaglia}, A., {Gratton}, R., {D'Orazi}, V., \& {Lucatello},
  S. 2009, \aap, 508, 695, \dodoi{10.1051/0004-6361/200913003}

\bibitem[{{Chiti} {et~al.}(2020){Chiti}, {Frebel}, {Jerjen}, {Kim}, \&
  {Norris}}]{Chiti.etal.2020}
{Chiti}, A., {Frebel}, A., {Jerjen}, H., {Kim}, D., \& {Norris}, J.~E. 2020,
  \apj, 891, 8, \dodoi{10.3847/1538-4357/ab6d72}

\bibitem[{{Chiti} {et~al.}(2021){Chiti}, {Frebel}, {Simon}, {Erkal}, {Chang},
  {Necib}, {Ji}, {Jerjen}, {Kim}, \& {Norris}}]{Chiti.etal.2021}
{Chiti}, A., {Frebel}, A., {Simon}, J.~D., {et~al.} 2021, Nature Astronomy, 5,
  392, \dodoi{10.1038/s41550-020-01285-w}

\bibitem[{{Chiti} {et~al.}(2023){Chiti}, {Frebel}, {Ji}, {Mardini}, {Ou},
  {Simon}, {Jerjen}, {Kim}, \& {Norris}}]{chiti.etal.2023}
{Chiti}, A., {Frebel}, A., {Ji}, A.~P., {et~al.} 2023, \aj, 165, 55,
  \dodoi{10.3847/1538-3881/aca416}

\bibitem[{{Cole} {et~al.}(2000){Cole}, {Lacey}, {Baugh}, \&
  {Frenk}}]{Cole.etal.2000}
{Cole}, S., {Lacey}, C.~G., {Baugh}, C.~M., \& {Frenk}, C.~S. 2000, \mnras,
  319, 168, \dodoi{10.1046/j.1365-8711.2000.03879.x}

\bibitem[{{Czekaj} {et~al.}(2014){Czekaj}, {Robin}, {Figueras}, {Luri}, \&
  {Haywood}}]{Czekaj.etal.2014}
{Czekaj}, M.~A., {Robin}, A.~C., {Figueras}, F., {Luri}, X., \& {Haywood}, M.
  2014, \aap, 564, A102, \dodoi{10.1051/0004-6361/201322139}

\bibitem[{{Dall'Ora} {et~al.}(2006){Dall'Ora}, {Clementini}, {Kinemuchi},
  {Ripepi}, {Marconi}, {Di Fabrizio}, {Greco}, {Rodgers}, {Kuehn}, \&
  {Smith}}]{Dall'Ora.etal.2016}
{Dall'Ora}, M., {Clementini}, G., {Kinemuchi}, K., {et~al.} 2006, \apjl, 653,
  L109, \dodoi{10.1086/510665}

\bibitem[{{Deason} {et~al.}(2014){Deason}, {Wetzel}, \&
  {Garrison-Kimmel}}]{Deason.etal.2014}
{Deason}, A., {Wetzel}, A., \& {Garrison-Kimmel}, S. 2014, \apj, 794, 115,
  \dodoi{10.1088/0004-637X/794/2/115}

\bibitem[{{Dotter} {et~al.}(2008){Dotter}, {Chaboyer}, {Jevremovi{\'c}},
  {Kostov}, {Baron}, \& {Ferguson}}]{Dotter.etal.2008}
{Dotter}, A., {Chaboyer}, B., {Jevremovi{\'c}}, D., {et~al.} 2008, \apjs, 178,
  89, \dodoi{10.1086/589654}

\bibitem[{{Drlica-Wagner} {et~al.}(2015){Drlica-Wagner}, {Bechtol}, {Rykoff},
  {Luque}, {Queiroz}, {Mao}, {Wechsler}, {Simon}, {Santiago}, {Yanny},
  {Balbinot}, {Dodelson}, {Fausti Neto}, {James}, {Li}, {Maia}, {Marshall},
  {Pieres}, {Stringer}, {Walker}, {Abbott}, {Abdalla}, {Allam},
  {Benoit-L{\'e}vy}, {Bernstein}, {Bertin}, {Brooks}, {Buckley-Geer}, {Burke},
  {Carnero Rosell}, {Carrasco Kind}, {Carretero}, {Crocce}, {da Costa},
  {Desai}, {Diehl}, {Dietrich}, {Doel}, {Eifler}, {Evrard}, {Finley},
  {Flaugher}, {Fosalba}, {Frieman}, {Gaztanaga}, {Gerdes}, {Gruen}, {Gruendl},
  {Gutierrez}, {Honscheid}, {Kuehn}, {Kuropatkin}, {Lahav}, {Martini},
  {Miquel}, {Nord}, {Ogando}, {Plazas}, {Reil}, {Roodman}, {Sako}, {Sanchez},
  {Scarpine}, {Schubnell}, {Sevilla-Noarbe}, {Smith}, {Soares-Santos},
  {Sobreira}, {Suchyta}, {Swanson}, {Tarle}, {Tucker}, {Vikram}, {Wester},
  {Zhang}, {Zuntz}, \& {DES Collaboration}}]{Drlica-Wagner.etal.2015}
{Drlica-Wagner}, A., {Bechtol}, K., {Rykoff}, E.~S., {et~al.} 2015, \apj, 813,
  109, \dodoi{10.1088/0004-637X/813/2/109}

\bibitem[{{Drlica-Wagner} {et~al.}(2022){Drlica-Wagner}, {Ferguson},
  {Adam{\'o}w}, {Aguena}, {Allam}, {Andrade-Oliveira}, {Bacon}, {Bechtol},
  {Bell}, {Bertin}, {Bilaji}, {Bocquet}, {Bom}, {Brooks}, {Burke},
  {Carballo-Bello}, {Carlin}, {Carnero Rosell}, {Carrasco Kind}, {Carretero},
  {Castander}, {Cerny}, {Chang}, {Choi}, {Conselice}, {Costanzi},
  {Crnojevi{\'c}}, {da Costa}, {de Vicente}, {Desai}, {Esteves}, {Everett},
  {Ferrero}, {Fitzpatrick}, {Flaugher}, {Friedel}, {Frieman},
  {Garc{\'\i}a-Bellido}, {Gatti}, {Gaztanaga}, {Gerdes}, {Gruen}, {Gruendl},
  {Gschwend}, {Hartley}, {Hernandez-Lang}, {Hinton}, {Hollowood}, {Honscheid},
  {Hughes}, {Jacques}, {James}, {Johnson}, {Kuehn}, {Kuropatkin}, {Lahav},
  {Li}, {Lidman}, {Lin}, {March}, {Marshall}, {Mart{\'\i}nez-Delgado},
  {Mart{\'\i}nez-V{\'a}zquez}, {Massana}, {Mau}, {McNanna}, {Melchior},
  {Menanteau}, {Miller}, {Miquel}, {Mohr}, {Morgan}, {Mutlu-Pakdil},
  {Mu{\~n}oz}, {Neilsen}, {Nidever}, {Nikutta}, {Nilo Castellon}, {No{\"e}l},
  {Ogando}, {Olsen}, {Pace}, {Palmese}, {Paz-Chinch{\'o}n}, {Pereira},
  {Pieres}, {Plazas Malag{\'o}n}, {Prat}, {Riley}, {Rodriguez-Monroy}, {Romer},
  {Roodman}, {Sako}, {Sakowska}, {Sanchez}, {S{\'a}nchez}, {Sand},
  {Santana-Silva}, {Santiago}, {Schubnell}, {Serrano}, {Sevilla-Noarbe},
  {Simon}, {Smith}, {Soares-Santos}, {Stringfellow}, {Suchyta}, {Suson}, {Tan},
  {Tarle}, {Tavangar}, {Thomas}, {To}, {Tollerud}, {Troxel}, {Tucker}, {Varga},
  {Vivas}, {Walker}, {Weller}, {Wilkinson}, {Wu}, {Yanny}, {Zaborowski},
  {Zenteno}, {Delve Collaboration}, {Des Collaboration}, \& {Astro Data
  Lab}}]{Drlica-Wagner.etal.2022}
{Drlica-Wagner}, A., {Ferguson}, P.~S., {Adam{\'o}w}, M., {et~al.} 2022, \apjs,
  261, 38, \dodoi{10.3847/1538-4365/ac78eb}

\bibitem[{{Filion} \& {Wyse}(2021)}]{filion.Wyse2021}
{Filion}, C., \& {Wyse}, R. F.~G. 2021, \apj, 923, 218,
  \dodoi{10.3847/1538-4357/ac2df1}

\bibitem[{{Flaugher} {et~al.}(2015){Flaugher}, {Diehl}, {Honscheid}, {Abbott},
  {Alvarez}, {Angstadt}, {Annis}, {Antonik}, {Ballester}, {Beaufore},
  {Bernstein}, {Bernstein}, {Bigelow}, {Bonati}, {Boprie}, {Brooks},
  {Buckley-Geer}, {Campa}, {Cardiel-Sas}, {Castander}, {Castilla}, {Cease},
  {Cela-Ruiz}, {Chappa}, {Chi}, {Cooper}, {da Costa}, {Dede}, {Derylo},
  {DePoy}, {de Vicente}, {Doel}, {Drlica-Wagner}, {Eiting}, {Elliott}, {Emes},
  {Estrada}, {Fausti Neto}, {Finley}, {Flores}, {Frieman}, {Gerdes},
  {Gladders}, {Gregory}, {Gutierrez}, {Hao}, {Holland}, {Holm}, {Huffman},
  {Jackson}, {James}, {Jonas}, {Karcher}, {Karliner}, {Kent}, {Kessler},
  {Kozlovsky}, {Kron}, {Kubik}, {Kuehn}, {Kuhlmann}, {Kuk}, {Lahav}, {Lathrop},
  {Lee}, {Levi}, {Lewis}, {Li}, {Mandrichenko}, {Marshall}, {Martinez},
  {Merritt}, {Miquel}, {Mu{\~n}oz}, {Neilsen}, {Nichol}, {Nord}, {Ogando},
  {Olsen}, {Palaio}, {Patton}, {Peoples}, {Plazas}, {Rauch}, {Reil}, {Rheault},
  {Roe}, {Rogers}, {Roodman}, {Sanchez}, {Scarpine}, {Schindler}, {Schmidt},
  {Schmitt}, {Schubnell}, {Schultz}, {Schurter}, {Scott}, {Serrano}, {Shaw},
  {Smith}, {Soares-Santos}, {Stefanik}, {Stuermer}, {Suchyta}, {Sypniewski},
  {Tarle}, {Thaler}, {Tighe}, {Tran}, {Tucker}, {Walker}, {Wang}, {Watson},
  {Weaverdyck}, {Wester}, {Woods}, {Yanny}, \& {DES
  Collaboration}}]{Flaugher:2015}
{Flaugher}, B., {Diehl}, H.~T., {Honscheid}, K., {et~al.} 2015, \aj, 150, 150,
  \dodoi{10.1088/0004-6256/150/5/150}

\bibitem[{{Frebel} {et~al.}(2014){Frebel}, {Simon}, \&
  {Kirby}}]{Frebel.etal.2014}
{Frebel}, A., {Simon}, J.~D., \& {Kirby}, E.~N. 2014, \apj, 786, 74,
  \dodoi{10.1088/0004-637X/786/1/74}

\bibitem[{{Fu} {et~al.}(2018){Fu}, {Simon}, {Shetrone}, {Bovy}, {Beers},
  {Fern{\'a}ndez-Trincado}, {Placco}, {Zamora}, {Allende Prieto},
  {Garc{\'\i}a-Hern{\'a}ndez}, {Harding}, {Ivans}, {Lane}, {Nitschelm},
  {Roman-Lopes}, \& {Sobeck}}]{Fu.etal.2018}
{Fu}, S.~W., {Simon}, J.~D., {Shetrone}, M., {et~al.} 2018, \apj, 866, 42,
  \dodoi{10.3847/1538-4357/aad9f9}

\bibitem[{{Fu} {et~al.}(2023){Fu}, {Weisz}, {Starkenburg}, {Martin}, {Savino},
  {Boylan-Kolchin}, {Cote}, {Dolphin}, {Ji}, {Longeard}, {Mateo}, {Patel}, \&
  {Sandford}}]{Fu.etal.2023}
{Fu}, S.~W., {Weisz}, D.~R., {Starkenburg}, E., {et~al.} 2023, arXiv e-prints,
  arXiv:2306.06260, \dodoi{10.48550/arXiv.2306.06260}

\bibitem[{{Gaia Collaboration} {et~al.}(2016){Gaia Collaboration}, {Prusti},
  {de Bruijne}, {Brown}, {Vallenari}, {Babusiaux}, {Bailer-Jones}, {Bastian},
  {Biermann}, {Evans}, {Eyer}, {Jansen}, {Jordi}, {Klioner}, {Lammers},
  {Lindegren}, {Luri}, {Mignard}, {Milligan}, {Panem}, {Poinsignon},
  {Pourbaix}, {Randich}, {Sarri}, {Sartoretti}, {Siddiqui}, {Soubiran},
  {Valette}, {van Leeuwen}, {Walton}, {Aerts}, {Arenou}, {Cropper}, {Drimmel},
  {H{\o}g}, {Katz}, {Lattanzi}, {O'Mullane}, {Grebel}, {Holland}, {Huc},
  {Passot}, {Bramante}, {Cacciari}, {Casta{\~n}eda}, {Chaoul}, {Cheek}, {De
  Angeli}, {Fabricius}, {Guerra}, {Hern{\'a}ndez}, {Jean-Antoine-Piccolo},
  {Masana}, {Messineo}, {Mowlavi}, {Nienartowicz}, {Ord{\'o}{\~n}ez-Blanco},
  {Panuzzo}, {Portell}, {Richards}, {Riello}, {Seabroke}, {Tanga},
  {Th{\'e}venin}, {Torra}, {Els}, {Gracia-Abril}, {Comoretto},
  {Garcia-Reinaldos}, {Lock}, {Mercier}, {Altmann}, {Andrae}, {Astraatmadja},
  {Bellas-Velidis}, {Benson}, {Berthier}, {Blomme}, {Busso}, {Carry},
  {Cellino}, {Clementini}, {Cowell}, {Creevey}, {Cuypers}, {Davidson}, {De
  Ridder}, {de Torres}, {Delchambre}, {Dell'Oro}, {Ducourant}, {Fr{\'e}mat},
  {Garc{\'\i}a-Torres}, {Gosset}, {Halbwachs}, {Hambly}, {Harrison}, {Hauser},
  {Hestroffer}, {Hodgkin}, {Huckle}, {Hutton}, {Jasniewicz}, {Jordan},
  {Kontizas}, {Korn}, {Lanzafame}, {Manteiga}, {Moitinho}, {Muinonen},
  {Osinde}, {Pancino}, {Pauwels}, {Petit}, {Recio-Blanco}, {Robin}, {Sarro},
  {Siopis}, {Smith}, {Smith}, {Sozzetti}, {Thuillot}, {van Reeven}, {Viala},
  {Abbas}, {Abreu Aramburu}, {Accart}, {Aguado}, {Allan}, {Allasia},
  {Altavilla}, {{\'A}lvarez}, {Alves}, {Anderson}, {Andrei}, {Anglada Varela},
  {Antiche}, {Antoja}, {Ant{\'o}n}, {Arcay}, {Atzei}, {Ayache}, {Bach},
  {Baker}, {Balaguer-N{\'u}{\~n}ez}, {Barache}, {Barata}, {Barbier}, {Barblan},
  {Baroni}, {Barrado y Navascu{\'e}s}, {Barros}, {Barstow}, {Becciani},
  {Bellazzini}, {Bellei}, {Bello Garc{\'\i}a}, {Belokurov}, {Bendjoya},
  {Berihuete}, {Bianchi}, {Bienaym{\'e}}, {Billebaud}, {Blagorodnova},
  {Blanco-Cuaresma}, {Boch}, {Bombrun}, {Borrachero}, {Bouquillon}, {Bourda},
  {Bouy}, {Bragaglia}, {Breddels}, {Brouillet}, {Br{\"u}semeister},
  {Bucciarelli}, {Budnik}, {Burgess}, {Burgon}, {Burlacu}, {Busonero}, {Buzzi},
  {Caffau}, {Cambras}, {Campbell}, {Cancelliere}, {Cantat-Gaudin}, {Carlucci},
  {Carrasco}, {Castellani}, {Charlot}, {Charnas}, {Charvet}, {Chassat},
  {Chiavassa}, {Clotet}, {Cocozza}, {Collins}, {Collins}, {Costigan}, {Crifo},
  {Cross}, {Crosta}, {Crowley}, {Dafonte}, {Damerdji}, {Dapergolas}, {David},
  {David}, {De Cat}, {de Felice}, {de Laverny}, {De Luise}, {De March}, {de
  Martino}, {de Souza}, {Debosscher}, {del Pozo}, {Delbo}, {Delgado},
  {Delgado}, {di Marco}, {Di Matteo}, {Diakite}, {Distefano}, {Dolding}, {Dos
  Anjos}, {Drazinos}, {Dur{\'a}n}, {Dzigan}, {Ecale}, {Edvardsson}, {Enke},
  {Erdmann}, {Escolar}, {Espina}, {Evans}, {Eynard Bontemps}, {Fabre},
  {Fabrizio}, {Faigler}, {Falc{\~a}o}, {Farr{\`a}s Casas}, {Faye}, {Federici},
  {Fedorets}, {Fern{\'a}ndez-Hern{\'a}ndez}, {Fernique}, {Fienga}, {Figueras},
  {Filippi}, {Findeisen}, {Fonti}, {Fouesneau}, {Fraile}, {Fraser}, {Fuchs},
  {Furnell}, {Gai}, {Galleti}, {Galluccio}, {Garabato}, {Garc{\'\i}a-Sedano},
  {Gar{\'e}}, {Garofalo}, {Garralda}, {Gavras}, {Gerssen}, {Geyer}, {Gilmore},
  {Girona}, {Giuffrida}, {Gomes}, {Gonz{\'a}lez-Marcos},
  {Gonz{\'a}lez-N{\'u}{\~n}ez}, {Gonz{\'a}lez-Vidal}, {Granvik}, {Guerrier},
  {Guillout}, {Guiraud}, {G{\'u}rpide}, {Guti{\'e}rrez-S{\'a}nchez}, {Guy},
  {Haigron}, {Hatzidimitriou}, {Haywood}, {Heiter}, {Helmi}, {Hobbs},
  {Hofmann}, {Holl}, {Holland}, {Hunt}, {Hypki}, {Icardi}, {Irwin}, {Jevardat
  de Fombelle}, {Jofr{\'e}}, {Jonker}, {Jorissen}, {Julbe}, {Karampelas},
  {Kochoska}, {Kohley}, {Kolenberg}, {Kontizas}, {Koposov}, {Kordopatis},
  {Koubsky}, {Kowalczyk}, {Krone-Martins}, {Kudryashova}, {Kull}, {Bachchan},
  {Lacoste-Seris}, {Lanza}, {Lavigne}, {Le Poncin-Lafitte}, {Lebreton},
  {Lebzelter}, {Leccia}, {Leclerc}, {Lecoeur-Taibi}, {Lemaitre}, {Lenhardt},
  {Leroux}, {Liao}, {Licata}, {Lindstr{\o}m}, {Lister}, {Livanou}, {Lobel},
  {L{\"o}ffler}, {L{\'o}pez}, {Lopez-Lozano}, {Lorenz}, {Loureiro},
  {MacDonald}, {Magalh{\~a}es Fernandes}, {Managau}, {Mann}, {Mantelet},
  {Marchal}, {Marchant}, {Marconi}, {Marie}, {Marinoni}, {Marrese},
  {Marschalk{\'o}}, {Marshall}, {Mart{\'\i}n-Fleitas}, {Martino}, {Mary},
  {Matijevi{\v{c}}}, {Mazeh}, {McMillan}, {Messina}, {Mestre}, {Michalik},
  {Millar}, {Miranda}, {Molina}, {Molinaro}, {Molinaro}, {Moln{\'a}r},
  {Moniez}, {Montegriffo}, {Monteiro}, {Mor}, {Mora}, {Morbidelli}, {Morel},
  {Morgenthaler}, {Morley}, {Morris}, {Mulone}, {Muraveva}, {Musella},
  {Narbonne}, {Nelemans}, {Nicastro}, {Noval}, {Ord{\'e}novic},
  {Ordieres-Mer{\'e}}, {Osborne}, {Pagani}, {Pagano}, {Pailler}, {Palacin},
  {Palaversa}, {Parsons}, {Paulsen}, {Pecoraro}, {Pedrosa}, {Pentik{\"a}inen},
  {Pereira}, {Pichon}, {Piersimoni}, {Pineau}, {Plachy}, {Plum}, {Poujoulet},
  {Pr{\v{s}}a}, {Pulone}, {Ragaini}, {Rago}, {Rambaux}, {Ramos-Lerate},
  {Ranalli}, {Rauw}, {Read}, {Regibo}, {Renk}, {Reyl{\'e}}, {Ribeiro},
  {Rimoldini}, {Ripepi}, {Riva}, {Rixon}, {Roelens}, {Romero-G{\'o}mez},
  {Rowell}, {Royer}, {Rudolph}, {Ruiz-Dern}, {Sadowski}, {Sagrist{\`a}
  Sell{\'e}s}, {Sahlmann}, {Salgado}, {Salguero}, {Sarasso}, {Savietto},
  {Schnorhk}, {Schultheis}, {Sciacca}, {Segol}, {Segovia}, {Segransan},
  {Serpell}, {Shih}, {Smareglia}, {Smart}, {Smith}, {Solano}, {Solitro},
  {Sordo}, {Soria Nieto}, {Souchay}, {Spagna}, {Spoto}, {Stampa}, {Steele},
  {Steidelm{\"u}ller}, {Stephenson}, {Stoev}, {Suess}, {S{\"u}veges}, {Surdej},
  {Szabados}, {Szegedi-Elek}, {Tapiador}, {Taris}, {Tauran}, {Taylor},
  {Teixeira}, {Terrett}, {Tingley}, {Trager}, {Turon}, {Ulla}, {Utrilla},
  {Valentini}, {van Elteren}, {Van Hemelryck}, {van Leeuwen}, {Varadi},
  {Vecchiato}, {Veljanoski}, {Via}, {Vicente}, {Vogt}, {Voss}, {Votruba},
  {Voutsinas}, {Walmsley}, {Weiler}, {Weingrill}, {Werner}, {Wevers},
  {Whitehead}, {Wyrzykowski}, {Yoldas}, {{\v{Z}}erjal}, {Zucker}, {Zurbach},
  {Zwitter}, {Alecu}, {Allen}, {Allende Prieto}, {Amorim},
  {Anglada-Escud{\'e}}, {Arsenijevic}, {Azaz}, {Balm}, {Beck}, {Bernstein},
  {Bigot}, {Bijaoui}, {Blasco}, {Bonfigli}, {Bono}, {Boudreault}, {Bressan},
  {Brown}, {Brunet}, {Bunclark}, {Buonanno}, {Butkevich}, {Carret}, {Carrion},
  {Chemin}, {Ch{\'e}reau}, {Corcione}, {Darmigny}, {de Boer}, {de Teodoro}, {de
  Zeeuw}, {Delle Luche}, {Domingues}, {Dubath}, {Fodor}, {Fr{\'e}zouls},
  {Fries}, {Fustes}, {Fyfe}, {Gallardo}, {Gallegos}, {Gardiol}, {Gebran},
  {Gomboc}, {G{\'o}mez}, {Grux}, {Gueguen}, {Heyrovsky}, {Hoar}, {Iannicola},
  {Isasi Parache}, {Janotto}, {Joliet}, {Jonckheere}, {Keil}, {Kim},
  {Klagyivik}, {Klar}, {Knude}, {Kochukhov}, {Kolka}, {Kos}, {Kutka}, {Lainey},
  {LeBouquin}, {Liu}, {Loreggia}, {Makarov}, {Marseille}, {Martayan},
  {Martinez-Rubi}, {Massart}, {Meynadier}, {Mignot}, {Munari}, {Nguyen},
  {Nordlander}, {Ocvirk}, {O'Flaherty}, {Olias Sanz}, {Ortiz}, {Osorio},
  {Oszkiewicz}, {Ouzounis}, {Palmer}, {Park}, {Pasquato}, {Peltzer}, {Peralta},
  {P{\'e}turaud}, {Pieniluoma}, {Pigozzi}, {Poels}, {Prat}, {Prod'homme},
  {Raison}, {Rebordao}, {Risquez}, {Rocca-Volmerange}, {Rosen}, {Ruiz-Fuertes},
  {Russo}, {Sembay}, {Serraller Vizcaino}, {Short}, {Siebert}, {Silva},
  {Sinachopoulos}, {Slezak}, {Soffel}, {Sosnowska}, {Strai{\v{z}}ys}, {ter
  Linden}, {Terrell}, {Theil}, {Tiede}, {Troisi}, {Tsalmantza}, {Tur},
  {Vaccari}, {Vachier}, {Valles}, {Van Hamme}, {Veltz}, {Virtanen}, {Wallut},
  {Wichmann}, {Wilkinson}, {Ziaeepour}, \& {Zschocke}}]{gaiamission.2016}
{Gaia Collaboration}, {Prusti}, T., {de Bruijne}, J.~H.~J., {et~al.} 2016,
  \aap, 595, A1, \dodoi{10.1051/0004-6361/201629272}

\bibitem[{{Gaia Collaboration} {et~al.}(2018){Gaia Collaboration}, {Brown},
  {Vallenari}, {Prusti}, {de Bruijne}, {Babusiaux}, {Bailer-Jones}, {Biermann},
  {Evans}, {Eyer}, {Jansen}, {Jordi}, {Klioner}, {Lammers}, {Lindegren},
  {Luri}, {Mignard}, {Panem}, {Pourbaix}, {Randich}, {Sartoretti}, {Siddiqui},
  {Soubiran}, {van Leeuwen}, {Walton}, {Arenou}, {Bastian}, {Cropper},
  {Drimmel}, {Katz}, {Lattanzi}, {Bakker}, {Cacciari}, {Casta{\~n}eda},
  {Chaoul}, {Cheek}, {De Angeli}, {Fabricius}, {Guerra}, {Holl}, {Masana},
  {Messineo}, {Mowlavi}, {Nienartowicz}, {Panuzzo}, {Portell}, {Riello},
  {Seabroke}, {Tanga}, {Th{\'e}venin}, {Gracia-Abril}, {Comoretto},
  {Garcia-Reinaldos}, {Teyssier}, {Altmann}, {Andrae}, {Audard},
  {Bellas-Velidis}, {Benson}, {Berthier}, {Blomme}, {Burgess}, {Busso},
  {Carry}, {Cellino}, {Clementini}, {Clotet}, {Creevey}, {Davidson}, {De
  Ridder}, {Delchambre}, {Dell'Oro}, {Ducourant},
  {Fern{\'a}ndez-Hern{\'a}ndez}, {Fouesneau}, {Fr{\'e}mat}, {Galluccio},
  {Garc{\'\i}a-Torres}, {Gonz{\'a}lez-N{\'u}{\~n}ez}, {Gonz{\'a}lez-Vidal},
  {Gosset}, {Guy}, {Halbwachs}, {Hambly}, {Harrison}, {Hern{\'a}ndez},
  {Hestroffer}, {Hodgkin}, {Hutton}, {Jasniewicz}, {Jean-Antoine-Piccolo},
  {Jordan}, {Korn}, {Krone-Martins}, {Lanzafame}, {Lebzelter}, {L{\"o}ffler},
  {Manteiga}, {Marrese}, {Mart{\'\i}n-Fleitas}, {Moitinho}, {Mora}, {Muinonen},
  {Osinde}, {Pancino}, {Pauwels}, {Petit}, {Recio-Blanco}, {Richards},
  {Rimoldini}, {Robin}, {Sarro}, {Siopis}, {Smith}, {Sozzetti}, {S{\"u}veges},
  {Torra}, {van Reeven}, {Abbas}, {Abreu Aramburu}, {Accart}, {Aerts},
  {Altavilla}, {{\'A}lvarez}, {Alvarez}, {Alves}, {Anderson}, {Andrei},
  {Anglada Varela}, {Antiche}, {Antoja}, {Arcay}, {Astraatmadja}, {Bach},
  {Baker}, {Balaguer-N{\'u}{\~n}ez}, {Balm}, {Barache}, {Barata}, {Barbato},
  {Barblan}, {Barklem}, {Barrado}, {Barros}, {Barstow}, {Bartholom{\'e}
  Mu{\~n}oz}, {Bassilana}, {Becciani}, {Bellazzini}, {Berihuete}, {Bertone},
  {Bianchi}, {Bienaym{\'e}}, {Blanco-Cuaresma}, {Boch}, {Boeche}, {Bombrun},
  {Borrachero}, {Bossini}, {Bouquillon}, {Bourda}, {Bragaglia}, {Bramante},
  {Breddels}, {Bressan}, {Brouillet}, {Br{\"u}semeister}, {Brugaletta},
  {Bucciarelli}, {Burlacu}, {Busonero}, {Butkevich}, {Buzzi}, {Caffau},
  {Cancelliere}, {Cannizzaro}, {Cantat-Gaudin}, {Carballo}, {Carlucci},
  {Carrasco}, {Casamiquela}, {Castellani}, {Castro-Ginard}, {Charlot},
  {Chemin}, {Chiavassa}, {Cocozza}, {Costigan}, {Cowell}, {Crifo}, {Crosta},
  {Crowley}, {Cuypers}, {Dafonte}, {Damerdji}, {Dapergolas}, {David}, {David},
  {de Laverny}, {De Luise}, {De March}, {de Martino}, {de Souza}, {de Torres},
  {Debosscher}, {del Pozo}, {Delbo}, {Delgado}, {Delgado}, {Di Matteo},
  {Diakite}, {Diener}, {Distefano}, {Dolding}, {Drazinos}, {Dur{\'a}n},
  {Edvardsson}, {Enke}, {Eriksson}, {Esquej}, {Eynard Bontemps}, {Fabre},
  {Fabrizio}, {Faigler}, {Falc{\~a}o}, {Farr{\`a}s Casas}, {Federici},
  {Fedorets}, {Fernique}, {Figueras}, {Filippi}, {Findeisen}, {Fonti},
  {Fraile}, {Fraser}, {Fr{\'e}zouls}, {Gai}, {Galleti}, {Garabato},
  {Garc{\'\i}a-Sedano}, {Garofalo}, {Garralda}, {Gavel}, {Gavras}, {Gerssen},
  {Geyer}, {Giacobbe}, {Gilmore}, {Girona}, {Giuffrida}, {Glass}, {Gomes},
  {Granvik}, {Gueguen}, {Guerrier}, {Guiraud}, {Guti{\'e}rrez-S{\'a}nchez},
  {Haigron}, {Hatzidimitriou}, {Hauser}, {Haywood}, {Heiter}, {Helmi}, {Heu},
  {Hilger}, {Hobbs}, {Hofmann}, {Holland}, {Huckle}, {Hypki}, {Icardi},
  {Jan{\ss}en}, {Jevardat de Fombelle}, {Jonker}, {Juh{\'a}sz}, {Julbe},
  {Karampelas}, {Kewley}, {Klar}, {Kochoska}, {Kohley}, {Kolenberg},
  {Kontizas}, {Kontizas}, {Koposov}, {Kordopatis}, {Kostrzewa-Rutkowska},
  {Koubsky}, {Lambert}, {Lanza}, {Lasne}, {Lavigne}, {Le Fustec}, {Le
  Poncin-Lafitte}, {Lebreton}, {Leccia}, {Leclerc}, {Lecoeur-Taibi},
  {Lenhardt}, {Leroux}, {Liao}, {Licata}, {Lindstr{\o}m}, {Lister}, {Livanou},
  {Lobel}, {L{\'o}pez}, {Managau}, {Mann}, {Mantelet}, {Marchal}, {Marchant},
  {Marconi}, {Marinoni}, {Marschalk{\'o}}, {Marshall}, {Martino}, {Marton},
  {Mary}, {Massari}, {Matijevi{\v{c}}}, {Mazeh}, {McMillan}, {Messina},
  {Michalik}, {Millar}, {Molina}, {Molinaro}, {Moln{\'a}r}, {Montegriffo},
  {Mor}, {Morbidelli}, {Morel}, {Morris}, {Mulone}, {Muraveva}, {Musella},
  {Nelemans}, {Nicastro}, {Noval}, {O'Mullane}, {Ord{\'e}novic},
  {Ord{\'o}{\~n}ez-Blanco}, {Osborne}, {Pagani}, {Pagano}, {Pailler},
  {Palacin}, {Palaversa}, {Panahi}, {Pawlak}, {Piersimoni}, {Pineau}, {Plachy},
  {Plum}, {Poggio}, {Poujoulet}, {Pr{\v{s}}a}, {Pulone}, {Racero}, {Ragaini},
  {Rambaux}, {Ramos-Lerate}, {Regibo}, {Reyl{\'e}}, {Riclet}, {Ripepi}, {Riva},
  {Rivard}, {Rixon}, {Roegiers}, {Roelens}, {Romero-G{\'o}mez}, {Rowell},
  {Royer}, {Ruiz-Dern}, {Sadowski}, {Sagrist{\`a} Sell{\'e}s}, {Sahlmann},
  {Salgado}, {Salguero}, {Sanna}, {Santana-Ros}, {Sarasso}, {Savietto},
  {Schultheis}, {Sciacca}, {Segol}, {Segovia}, {S{\'e}gransan}, {Shih},
  {Siltala}, {Silva}, {Smart}, {Smith}, {Solano}, {Solitro}, {Sordo}, {Soria
  Nieto}, {Souchay}, {Spagna}, {Spoto}, {Stampa}, {Steele},
  {Steidelm{\"u}ller}, {Stephenson}, {Stoev}, {Suess}, {Surdej}, {Szabados},
  {Szegedi-Elek}, {Tapiador}, {Taris}, {Tauran}, {Taylor}, {Teixeira},
  {Terrett}, {Teyssandier}, {Thuillot}, {Titarenko}, {Torra Clotet}, {Turon},
  {Ulla}, {Utrilla}, {Uzzi}, {Vaillant}, {Valentini}, {Valette}, {van Elteren},
  {Van Hemelryck}, {van Leeuwen}, {Vaschetto}, {Vecchiato}, {Veljanoski},
  {Viala}, {Vicente}, {Vogt}, {von Essen}, {Voss}, {Votruba}, {Voutsinas},
  {Walmsley}, {Weiler}, {Wertz}, {Wevers}, {Wyrzykowski}, {Yoldas},
  {{\v{Z}}erjal}, {Ziaeepour}, {Zorec}, {Zschocke}, {Zucker}, {Zurbach}, \&
  {Zwitter}}]{GaiaDR2}
{Gaia Collaboration}, {Brown}, A.~G.~A., {Vallenari}, A., {et~al.} 2018, \aap,
  616, A1, \dodoi{10.1051/0004-6361/201833051}

\bibitem[{{Gaia Collaboration} {et~al.}(2022){Gaia Collaboration}, {Vallenari},
  {Brown}, {Prusti}, {de Bruijne}, {Arenou}, {Babusiaux}, {Biermann},
  {Creevey}, {Ducourant}, {Evans}, {Eyer}, {Guerra}, {Hutton}, {Jordi},
  {Klioner}, {Lammers}, {Lindegren}, {Luri}, {Mignard}, {Panem}, {Pourbaix},
  {Randich}, {Sartoretti}, {Soubiran}, {Tanga}, {Walton}, {Bailer-Jones},
  {Bastian}, {Drimmel}, {Jansen}, {Katz}, {Lattanzi}, {van Leeuwen}, {Bakker},
  {Cacciari}, {Casta{\~n}eda}, {De Angeli}, {Fabricius}, {Fouesneau},
  {Fr{\'e}mat}, {Galluccio}, {Guerrier}, {Heiter}, {Masana}, {Messineo},
  {Mowlavi}, {Nicolas}, {Nienartowicz}, {Pailler}, {Panuzzo}, {Riclet}, {Roux},
  {Seabroke}, {Sordo{\o}rcit}, {Th{\'e}venin}, {Gracia-Abril}, {Portell},
  {Teyssier}, {Altmann}, {Andrae}, {Audard}, {Bellas-Velidis}, {Benson},
  {Berthier}, {Blomme}, {Burgess}, {Busonero}, {Busso}, {C{\'a}novas}, {Carry},
  {Cellino}, {Cheek}, {Clementini}, {Damerdji}, {Davidson}, {de Teodoro},
  {Nu{\~n}ez Campos}, {Delchambre}, {Dell'Oro}, {Esquej},
  {Fern{\'a}ndez-Hern{\'a}ndez}, {Fraile}, {Garabato}, {Garc{\'\i}a-Lario},
  {Gosset}, {Haigron}, {Halbwachs}, {Hambly}, {Harrison}, {Hern{\'a}ndez},
  {Hestroffer}, {Hodgkin}, {Holl}, {Jan{\ss}en}, {Jevardat de Fombelle},
  {Jordan}, {Krone-Martins}, {Lanzafame}, {L{\"o}ffler}, {Marchal}, {Marrese},
  {Moitinho}, {Muinonen}, {Osborne}, {Pancino}, {Pauwels}, {Recio-Blanco},
  {Reyl{\'e}}, {Riello}, {Rimoldini}, {Roegiers}, {Rybizki}, {Sarro}, {Siopis},
  {Smith}, {Sozzetti}, {Utrilla}, {van Leeuwen}, {Abbas}, {{\'A}brah{\'a}m},
  {Abreu Aramburu}, {Aerts}, {Aguado}, {Ajaj}, {Aldea-Montero}, {Altavilla},
  {{\'A}lvarez}, {Alves}, {Anders}, {Anderson}, {Anglada Varela}, {Antoja},
  {Baines}, {Baker}, {Balaguer-N{\'u}{\~n}ez}, {Balbinot}, {Balog}, {Barache},
  {Barbato}, {Barros}, {Barstow}, {Bartolom{\'e}}, {Bassilana}, {Bauchet},
  {Becciani}, {Bellazzini}, {Berihuete}, {Bernet}, {Bertone}, {Bianchi},
  {Binnenfeld}, {Blanco-Cuaresma}, {Blazere}, {Boch}, {Bombrun}, {Bossini},
  {Bouquillon}, {Bragaglia}, {Bramante}, {Breedt}, {Bressan}, {Brouillet},
  {Brugaletta}, {Bucciarelli}, {Burlacu}, {Butkevich}, {Buzzi}, {Caffau},
  {Cancelliere}, {Cantat-Gaudin}, {Carballo}, {Carlucci}, {Carnerero},
  {Carrasco}, {Casamiquela}, {Castellani}, {Castro-Ginard}, {Chaoul},
  {Charlot}, {Chemin}, {Chiaramida}, {Chiavassa}, {Chornay}, {Comoretto},
  {Contursi}, {Cooper}, {Cornez}, {Cowell}, {Crifo}, {Cropper}, {Crosta},
  {Crowley}, {Dafonte}, {Dapergolas}, {David}, {David}, {de Laverny}, {De
  Luise}, {De March}, {De Ridder}, {de Souza}, {de Torres}, {del Peloso}, {del
  Pozo}, {Delbo}, {Delgado}, {Delisle}, {Demouchy}, {Dharmawardena}, {Di
  Matteo}, {Diakite}, {Diener}, {Distefano}, {Dolding}, {Edvardsson}, {Enke},
  {Fabre}, {Fabrizio}, {Faigler}, {Fedorets}, {Fernique}, {Fienga}, {Figueras},
  {Fournier}, {Fouron}, {Fragkoudi}, {Gai}, {Garcia-Gutierrez},
  {Garcia-Reinaldos}, {Garc{\'\i}a-Torres}, {Garofalo}, {Gavel}, {Gavras},
  {Gerlach}, {Geyer}, {Giacobbe}, {Gilmore}, {Girona}, {Giuffrida}, {Gomel},
  {Gomez}, {Gonz{\'a}lez-N{\'u}{\~n}ez}, {Gonz{\'a}lez-Santamar{\'\i}a},
  {Gonz{\'a}lez-Vidal}, {Granvik}, {Guillout}, {Guiraud},
  {Guti{\'e}rrez-S{\'a}nchez}, {Guy}, {Hatzidimitriou}, {Hauser}, {Haywood},
  {Helmer}, {Helmi}, {Sarmiento}, {Hidalgo}, {Hilger}, {H{\l}adczuk}, {Hobbs},
  {Holland}, {Huckle}, {Jardine}, {Jasniewicz}, {Jean-Antoine Piccolo},
  {Jim{\'e}nez-Arranz}, {Jorissen}, {Juaristi Campillo}, {Julbe}, {Karbevska},
  {Kervella}, {Khanna}, {Kontizas}, {Kordopatis}, {Korn}, {K{\'o}sp{\'a}l},
  {Kostrzewa-Rutkowska}, {Kruszy{\'n}ska}, {Kun}, {Laizeau}, {Lambert},
  {Lanza}, {Lasne}, {Le Campion}, {Lebreton}, {Lebzelter}, {Leccia}, {Leclerc},
  {Lecoeur-Taibi}, {Liao}, {Licata}, {Lindstr{\o}m}, {Lister}, {Livanou},
  {Lobel}, {Lorca}, {Loup}, {Madrero Pardo}, {Magdaleno Romeo}, {Managau},
  {Mann}, {Manteiga}, {Marchant}, {Marconi}, {Marcos}, {Marcos Santos},
  {Mar{\'\i}n Pina}, {Marinoni}, {Marocco}, {Marshall}, {Polo},
  {Mart{\'\i}n-Fleitas}, {Marton}, {Mary}, {Masip}, {Massari},
  {Mastrobuono-Battisti}, {Mazeh}, {McMillan}, {Messina}, {Michalik}, {Millar},
  {Mints}, {Molina}, {Molinaro}, {Moln{\'a}r}, {Monari}, {Mongui{\'o}},
  {Montegriffo}, {Montero}, {Mor}, {Mora}, {Morbidelli}, {Morel}, {Morris},
  {Muraveva}, {Murphy}, {Musella}, {Nagy}, {Noval}, {Oca{\~n}a}, {Ogden},
  {Ordenovic}, {Osinde}, {Pagani}, {Pagano}, {Palaversa}, {Palicio},
  {Pallas-Quintela}, {Panahi}, {Payne-Wardenaar}, {Pe{\~n}alosa Esteller},
  {Penttil{\"a}}, {Pichon}, {Piersimoni}, {Pineau}, {Plachy}, {Plum}, {Poggio},
  {Pr{\v{s}}a}, {Pulone}, {Racero}, {Ragaini}, {Rainer}, {Raiteri}, {Rambaux},
  {Ramos}, {Ramos-Lerate}, {Re Fiorentin}, {Regibo}, {Richards}, {Rios Diaz},
  {Ripepi}, {Riva}, {Rix}, {Rixon}, {Robichon}, {Robin}, {Robin}, {Roelens},
  {Rogues}, {Rohrbasser}, {Romero-G{\'o}mez}, {Rowell}, {Royer}, {Ruz Mieres},
  {Rybicki}, {Sadowski}, {S{\'a}ez N{\'u}{\~n}ez}, {Sagrist{\`a} Sell{\'e}s},
  {Sahlmann}, {Salguero}, {Samaras}, {Sanchez Gimenez}, {Sanna},
  {Santove{\~n}a}, {Sarasso}, {Schultheis}, {Sciacca}, {Segol}, {Segovia},
  {S{\'e}gransan}, {Semeux}, {Shahaf}, {Siddiqui}, {Siebert}, {Siltala},
  {Silvelo}, {Slezak}, {Slezak}, {Smart}, {Snaith}, {Solano}, {Solitro},
  {Souami}, {Souchay}, {Spagna}, {Spina}, {Spoto}, {Steele},
  {Steidelm{\"u}ller}, {Stephenson}, {S{\"u}veges}, {Surdej}, {Szabados},
  {Szegedi-Elek}, {Taris}, {Taylo}, {Teixeira}, {Tolomei}, {Tonello}, {Torra},
  {Torra}, {Torralba Elipe}, {Trabucchi}, {Tsounis}, {Turon}, {Ulla}, {Unger},
  {Vaillant}, {van Dillen}, {van Reeven}, {Vanel}, {Vecchiato}, {Viala},
  {Vicente}, {Voutsinas}, {Weiler}, {Wevers}, {Wyrzykowski}, {Yoldas}, {Yvard},
  {Zhao}, {Zorec}, {Zucker}, \& {Zwitter}}]{gaiadr3}
{Gaia Collaboration}, {Vallenari}, A., {Brown}, A.~G.~A., {et~al.} 2022, arXiv
  e-prints, arXiv:2208.00211, \dodoi{10.48550/arXiv.2208.00211}

\bibitem[{{Garling} {et~al.}(2018){Garling}, {Willman}, {Sand}, {Hargis},
  {Crnojevi{\'c}}, {Bechtol}, {Carlin}, {Strader}, {Zou}, {Zhou}, {Nie},
  {Zhang}, {Zhou}, \& {Peng}}]{Garling.etal.2018}
{Garling}, C., {Willman}, B., {Sand}, D.~J., {et~al.} 2018, \apj, 852, 44,
  \dodoi{10.3847/1538-4357/aa9bf1}

\bibitem[{{Geha} {et~al.}(2009){Geha}, {Willman}, {Simon}, {Strigari}, {Kirby},
  {Law}, \& {Strader}}]{geha.etal.2009}
{Geha}, M., {Willman}, B., {Simon}, J.~D., {et~al.} 2009, \apj, 692, 1464,
  \dodoi{10.1088/0004-637X/692/2/1464}

\bibitem[{{Hartley} {et~al.}(2022){Hartley}, {Choi}, {Amon}, {Gruendl},
  {Sheldon}, {Harrison}, {Bernstein}, {Sevilla-Noarbe}, {Yanny}, {Eckert},
  {Diehl}, {Alarcon}, {Banerji}, {Bechtol}, {Buchs}, {Cantu}, {Conselice},
  {Cordero}, {Davis}, {Davis}, {Dodelson}, {Drlica-Wagner}, {Everett},
  {Fert{\'e}}, {Gruen}, {Honscheid}, {Jarvis}, {Johnson}, {Kokron}, {MacCrann},
  {Myles}, {Pace}, {Palmese}, {Paz-Chinch{\'o}n}, {Pereira}, {Plazas}, {Prat},
  {Rodriguez-Monroy}, {Rykoff}, {Samuroff}, {S{\'a}nchez}, {Secco},
  {Tarsitano}, {Tong}, {Troxel}, {Vasquez}, {Wang}, {Zhou}, {Abbott}, {Aguena},
  {Allam}, {Annis}, {Bacon}, {Bertin}, {Bhargava}, {Brooks}, {Burke}, {Carnero
  Rosell}, {Carrasco Kind}, {Carretero}, {Castander}, {Costanzi}, {Crocce}, {da
  Costa}, {De Vicente}, {DeRose}, {Desai}, {Dietrich}, {Eifler}, {Elvin-Poole},
  {Ferrero}, {Flaugher}, {Fosalba}, {Garc{\'\i}a-Bellido}, {Gaztanaga},
  {Gerdes}, {Gschwend}, {Gutierrez}, {Hinton}, {Hollowood}, {Huterer}, {James},
  {Kent}, {Krause}, {Kuehn}, {Kuropatkin}, {Lahav}, {Lin}, {Maia}, {March},
  {Marshall}, {Martini}, {Melchior}, {Menanteau}, {Miquel}, {Mohr}, {Morgan},
  {Neilsen}, {Ogando}, {Pandey}, {Romer}, {Roodman}, {Sako}, {Sanchez},
  {Scarpine}, {Serrano}, {Smith}, {Soares-Santos}, {Suchyta}, {Swanson},
  {Tarle}, {Thomas}, {To}, {Varga}, {Walker}, {Wester}, {Wilkinson}, {Zuntz},
  {Zuntz}, \& {DES Collaboration}}]{Hartley.etal.2022}
{Hartley}, W.~G., {Choi}, A., {Amon}, A., {et~al.} 2022, \mnras, 509, 3547,
  \dodoi{10.1093/mnras/stab3055}

\bibitem[{{Hunter}(2007)}]{Hunter.2007}
{Hunter}, J.~D. 2007, Computing in Science and Engineering, 9, 90,
  \dodoi{10.1109/MCSE.2007.55}

\bibitem[{{Jenkins} {et~al.}(2021){Jenkins}, {Li}, {Pace}, {Ji}, {Koposov}, \&
  {Mutlu-Pakdil}}]{Jenkins.etal.2021}
{Jenkins}, S.~A., {Li}, T.~S., {Pace}, A.~B., {et~al.} 2021, \apj, 920, 92,
  \dodoi{10.3847/1538-4357/ac1353}

\bibitem[{{Jensen} {et~al.}(2024){Jensen}, {Hayes}, {Sestito}, {McConnachie},
  {Waller}, {Smith}, {Navarro}, \& {Venn}}]{Jensen.etal.2024}
{Jensen}, J., {Hayes}, C.~R., {Sestito}, F., {et~al.} 2024, \mnras, 527, 4209,
  \dodoi{10.1093/mnras/stad3322}

\bibitem[{{Ji} {et~al.}(2016){Ji}, {Frebel}, {Simon}, \& {Geha}}]{Ji.etal.2016}
{Ji}, A.~P., {Frebel}, A., {Simon}, J.~D., \& {Geha}, M. 2016, \apj, 817, 41,
  \dodoi{10.3847/0004-637X/817/1/41}

\bibitem[{Jones {et~al.}(2001)Jones, Oliphant, \& Peterson}]{Jones.etal.2001}
Jones, E., Oliphant, T., \& Peterson, P. 2001, {SciPy:} Open Source Scientific
  Tools for {Python}.
\newblock \url{http://www.scipy.org}

\bibitem[{{Keller} {et~al.}(2007){Keller}, {Schmidt}, {Bessell}, {Conroy},
  {Francis}, {Granlund}, {Kowald}, {Oates}, {Martin-Jones}, {Preston},
  {Tisserand}, {Vaccarella}, \& {Waterson}}]{Keller.etal.2007}
{Keller}, S.~C., {Schmidt}, B.~P., {Bessell}, M.~S., {et~al.} 2007, \pasa, 24,
  1, \dodoi{10.1071/AS07001}

\bibitem[{{Kim} {et~al.}(2018){Kim}, {Peter}, \& {Hargis}}]{Kim.etal.2018}
{Kim}, S.~Y., {Peter}, A. H.~G., \& {Hargis}, J.~R. 2018, \prl, 121, 211302,
  \dodoi{10.1103/PhysRevLett.121.211302}

\bibitem[{{Koch} {et~al.}(2009){Koch}, {Wilkinson}, {Kleyna}, {Irwin},
  {Zucker}, {Belokurov}, {Gilmore}, {Fellhauer}, \& {Evans}}]{Koch.etal.2009}
{Koch}, A., {Wilkinson}, M.~I., {Kleyna}, J.~T., {et~al.} 2009, \apj, 690, 453,
  \dodoi{10.1088/0004-637X/690/1/453}

\bibitem[{{Koposov} {et~al.}(2015){Koposov}, {Belokurov}, {Torrealba}, \&
  {Evans}}]{Koposov.etal.2015}
{Koposov}, S.~E., {Belokurov}, V., {Torrealba}, G., \& {Evans}, N.~W. 2015,
  \apj, 805, 130, \dodoi{10.1088/0004-637X/805/2/130}

\bibitem[{{Li} {et~al.}(2018){Li}, {Simon}, {Kuehn}, {Pace}, {Erkal},
  {Bechtol}, {Yanny}, {Drlica-Wagner}, {Marshall}, {Lidman}, {Balbinot},
  {Carollo}, {Jenkins}, {Mart{\'\i}nez-V{\'a}zquez}, {Shipp}, {Stringer},
  {Vivas}, {Walker}, {Wechsler}, {Abdalla}, {Allam}, {Annis}, {Avila},
  {Bertin}, {Brooks}, {Buckley-Geer}, {Burke}, {Carnero Rosell}, {Carrasco
  Kind}, {Carretero}, {Cunha}, {D'Andrea}, {da Costa}, {Davis}, {De Vicente},
  {Doel}, {Eifler}, {Evrard}, {Flaugher}, {Frieman}, {Garc{\'\i}a-Bellido},
  {Gaztanaga}, {Gerdes}, {Gruen}, {Gruendl}, {Gschwend}, {Gutierrez},
  {Hartley}, {Hollowood}, {Honscheid}, {James}, {Krause}, {Maia}, {March},
  {Menanteau}, {Miquel}, {Plazas}, {Sanchez}, {Santiago}, {Scarpine},
  {Schindler}, {Schubnell}, {Sevilla-Noarbe}, {Smith}, {Smith},
  {Soares-Santos}, {Sobreira}, {Suchyta}, {Swanson}, {Tarle}, {Tucker}, \& {DES
  Collaboration}}]{Li.etal.2018}
{Li}, T.~S., {Simon}, J.~D., {Kuehn}, K., {et~al.} 2018, \apj, 866, 22,
  \dodoi{10.3847/1538-4357/aadf91}

\bibitem[{{Li} {et~al.}(2022){Li}, {Ji}, {Pace}, {Erkal}, {Koposov}, {Shipp},
  {Da Costa}, {Cullinane}, {Kuehn}, {Lewis}, {Mackey}, {Simpson}, {Zucker},
  {Ferguson}, {Martell}, {Bland-Hawthorn}, {Balbinot}, {Tavangar},
  {Drlica-Wagner}, {De Silva}, \& {Simon}}]{Li.etal.2022}
{Li}, T.~S., {Ji}, A.~P., {Pace}, A.~B., {et~al.} 2022, \apj, 928, 30,
  \dodoi{10.3847/1538-4357/ac46d3}

\bibitem[{{Limberg} {et~al.}(2023){Limberg}, {Queiroz}, {Perottoni}, {Rossi},
  {Amarante}, {Santucci}, {Chiappini}, {P{\'e}rez-Villegas}, \&
  {Lee}}]{Limberg.etal.2023}
{Limberg}, G., {Queiroz}, A. B.~A., {Perottoni}, H.~D., {et~al.} 2023, \apj,
  946, 66, \dodoi{10.3847/1538-4357/acb694}

\bibitem[{{Longeard} {et~al.}(2021){Longeard}, {Martin}, {Ibata},
  {Starkenburg}, {Jablonka}, {Aguado}, {Carlberg}, {C{\^o}t{\'e}},
  {Gonz{\'a}lez Hern{\'a}ndez}, {Lucchesi}, {Malhan}, {Navarro},
  {S{\'a}nchez-Janssen}, {Thomas}, {Venn}, \&
  {McConnachie}}]{Longeard.etal.2021}
{Longeard}, N., {Martin}, N., {Ibata}, R.~A., {et~al.} 2021, \mnras, 503, 2754,
  \dodoi{10.1093/mnras/stab604}

\bibitem[{{Longeard} {et~al.}(2022){Longeard}, {Jablonka}, {Arentsen},
  {Thomas}, {Aguado}, {Carlberg}, {Lucchesi}, {Malhan}, {Martin},
  {McConnachie}, {Navarro}, {S{\'a}nchez-Janssen}, {Sestito}, {Starkenburg}, \&
  {Yuan}}]{Longeard.etal.2022}
{Longeard}, N., {Jablonka}, P., {Arentsen}, A., {et~al.} 2022, \mnras, 516,
  2348, \dodoi{10.1093/mnras/stac1827}

\bibitem[{{Longeard} {et~al.}(2023){Longeard}, {Jablonka}, {Battaglia},
  {Malhan}, {Martin}, {S{\'a}nchez-Janssen}, {Sestito}, {Starkenburg}, \&
  {Venn}}]{Longeard.etal.2023}
{Longeard}, N., {Jablonka}, P., {Battaglia}, G., {et~al.} 2023, arXiv e-prints,
  arXiv:2304.13046, \dodoi{10.48550/arXiv.2304.13046}

\bibitem[{{Malhan} {et~al.}(2022){Malhan}, {Valluri}, {Freese}, \&
  {Ibata}}]{Malhan.etal.2022}
{Malhan}, K., {Valluri}, M., {Freese}, K., \& {Ibata}, R.~A. 2022, \apjl, 941,
  L38, \dodoi{10.3847/2041-8213/aca6e5}

\bibitem[{{Martin} {et~al.}(2007){Martin}, {Ibata}, {Chapman}, {Irwin}, \&
  {Lewis}}]{Martin.etal.2007}
{Martin}, N.~F., {Ibata}, R.~A., {Chapman}, S.~C., {Irwin}, M., \& {Lewis},
  G.~F. 2007, \mnras, 380, 281, \dodoi{10.1111/j.1365-2966.2007.12055.x}

\bibitem[{{Mart{\'\i}nez-V{\'a}zquez}
  {et~al.}(2021){Mart{\'\i}nez-V{\'a}zquez}, {Cerny}, {Vivas}, {Drlica-Wagner},
  {Pace}, {Simon}, {Munoz}, {Walker}, {Allam}, {Tucker}, {Adam{\'o}w},
  {Carlin}, {Choi}, {Ferguson}, {Ji}, {Kuropatkin}, {Li},
  {Mart{\'\i}nez-Delgado}, {Mau}, {Mutlu-Pakdil}, {Nidever}, {Riley},
  {Sakowska}, {Sand}, {Stringfellow}, \&
  {Stringfellow}}]{Martinex-vazquez.etal.2021}
{Mart{\'\i}nez-V{\'a}zquez}, C.~E., {Cerny}, W., {Vivas}, A.~K., {et~al.} 2021,
  \aj, 162, 253, \dodoi{10.3847/1538-3881/ac2368}

\bibitem[{{Mau} {et~al.}(2022){Mau}, {Nadler}, {Wechsler}, {Drlica-Wagner},
  {Bechtol}, {Green}, {Huterer}, {Li}, {Mao}, {Mart{\'\i}nez-V{\'a}zquez},
  {McNanna}, {Mutlu-Pakdil}, {Pace}, {Peter}, {Riley}, {Strigari}, {Wang},
  {Aguena}, {Allam}, {Annis}, {Bacon}, {Bertin}, {Bocquet}, {Brooks}, {Burke},
  {Carnero Rosell}, {Carrasco Kind}, {Carretero}, {Costanzi}, {Crocce},
  {Pereira}, {Davis}, {De Vicente}, {Desai}, {Doel}, {Ferrero}, {Flaugher},
  {Frieman}, {Garc{\'\i}a-Bellido}, {Gatti}, {Giannini}, {Gruen}, {Gruendl},
  {Gschwend}, {Gutierrez}, {Hinton}, {Hollowood}, {Honscheid}, {James},
  {Kuehn}, {Lahav}, {Maia}, {Marshall}, {Miquel}, {Mohr}, {Morgan}, {Ogando},
  {Paz-Chinch{\'o}n}, {Pieres}, {Rodriguez-Monroy}, {Sanchez}, {Scarpine},
  {Serrano}, {Sevilla-Noarbe}, {Suchyta}, {Tarle}, {To}, {Tucker}, {Weller}, \&
  {DES Collaboration}}]{Mau.etal.2022}
{Mau}, S., {Nadler}, E.~O., {Wechsler}, R.~H., {et~al.} 2022, \apj, 932, 128,
  \dodoi{10.3847/1538-4357/ac6e65}

\bibitem[{{McConnachie}(2012)}]{McConnachie.etal.2012}
{McConnachie}, A.~W. 2012, \aj, 144, 4, \dodoi{10.1088/0004-6256/144/1/4}

\bibitem[{{McConnachie} \& {Venn}(2020)}]{McConnachie.Venn.2020}
{McConnachie}, A.~W., \& {Venn}, K.~A. 2020, Research Notes of the American
  Astronomical Society, 4, 229, \dodoi{10.3847/2515-5172/abd18b}

\bibitem[{{Morganson} {et~al.}(2018){Morganson}, {Gruendl}, {Menanteau},
  {Carrasco Kind}, {Chen}, {Daues}, {Drlica-Wagner}, {Friedel}, {Gower},
  {Johnson}, {Johnson}, {Kessler}, {Paz-Chinch{\'o}n}, {Petravick}, {Pond},
  {Yanny}, {Allam}, {Armstrong}, {Barkhouse}, {Bechtol}, {Benoit-L{\'e}vy},
  {Bernstein}, {Bertin}, {Buckley-Geer}, {Covarrubias}, {Desai}, {Diehl},
  {Goldstein}, {Gruen}, {Li}, {Lin}, {Marriner}, {Mohr}, {Neilsen}, {Ngeow},
  {Paech}, {Rykoff}, {Sako}, {Sevilla-Noarbe}, {Sheldon}, {Sobreira}, {Tucker},
  {Wester}, \& {DES Collaboration}}]{Morganson:2018}
{Morganson}, E., {Gruendl}, R.~A., {Menanteau}, F., {et~al.} 2018, \pasp, 130,
  074501, \dodoi{10.1088/1538-3873/aab4ef}

\bibitem[{{Mu{\~n}oz} {et~al.}(2018){Mu{\~n}oz}, {C{\^o}t{\'e}}, {Santana},
  {Geha}, {Simon}, {Oyarz{\'u}n}, {Stetson}, \& {Djorgovski}}]{Munoz.etal.2018}
{Mu{\~n}oz}, R.~R., {C{\^o}t{\'e}}, P., {Santana}, F.~A., {et~al.} 2018, \apj,
  860, 66, \dodoi{10.3847/1538-4357/aac16b}

\bibitem[{{Munshi} {et~al.}(2019){Munshi}, {Brooks}, {Christensen},
  {Applebaum}, {Holley-Bockelmann}, {Quinn}, \& {Wadsley}}]{Munshi.etal.2019}
{Munshi}, F., {Brooks}, A.~M., {Christensen}, C., {et~al.} 2019, \apj, 874, 40,
  \dodoi{10.3847/1538-4357/ab0085}

\bibitem[{{Mutlu-Pakdil} {et~al.}(2019){Mutlu-Pakdil}, {Sand}, {Walker},
  {Caldwell}, {Carlin}, {Collins}, {Crnojevi{\'c}}, {Mateo}, {Olszewski},
  {Seth}, {Strader}, {Willman}, \& {Zaritsky}}]{Mutlu-Pakdil.etal.2019}
{Mutlu-Pakdil}, B., {Sand}, D.~J., {Walker}, M.~G., {et~al.} 2019, \apj, 885,
  53, \dodoi{10.3847/1538-4357/ab45ec}

\bibitem[{{Nadler} {et~al.}(2021){Nadler}, {Drlica-Wagner}, {Bechtol}, {Mau},
  {Wechsler}, {Gluscevic}, {Boddy}, {Pace}, {Li}, {McNanna}, {Riley},
  {Garc{\'\i}a-Bellido}, {Mao}, {Green}, {Burke}, {Peter}, {Jain}, {Abbott},
  {Aguena}, {Allam}, {Annis}, {Avila}, {Brooks}, {Carrasco Kind}, {Carretero},
  {Costanzi}, {da Costa}, {De Vicente}, {Desai}, {Diehl}, {Doel}, {Everett},
  {Evrard}, {Flaugher}, {Frieman}, {Gerdes}, {Gruen}, {Gruendl}, {Gschwend},
  {Gutierrez}, {Hinton}, {Honscheid}, {Huterer}, {James}, {Krause}, {Kuehn},
  {Kuropatkin}, {Lahav}, {Maia}, {Marshall}, {Menanteau}, {Miquel}, {Palmese},
  {Paz-Chinch{\'o}n}, {Plazas}, {Romer}, {Sanchez}, {Scarpine}, {Serrano},
  {Sevilla-Noarbe}, {Smith}, {Soares-Santos}, {Suchyta}, {Swanson}, {Tarle},
  {Tucker}, {Walker}, {Wester}, \& {DES Collaboration}}]{Nadler.etal.2021}
{Nadler}, E.~O., {Drlica-Wagner}, A., {Bechtol}, K., {et~al.} 2021, \prl, 126,
  091101, \dodoi{10.1103/PhysRevLett.126.091101}

\bibitem[{{Norris} {et~al.}(2008){Norris}, {Gilmore}, {Wyse}, {Wilkinson},
  {Belokurov}, {Evans}, \& {Zucker}}]{Norris.etal.2008}
{Norris}, J.~E., {Gilmore}, G., {Wyse}, R. F.~G., {et~al.} 2008, \apjl, 689,
  L113, \dodoi{10.1086/595962}

\bibitem[{{Norris} {et~al.}(2010){Norris}, {Gilmore}, {Wyse}, {Yong}, \&
  {Frebel}}]{Norris.etal.2010}
{Norris}, J.~E., {Gilmore}, G., {Wyse}, R. F.~G., {Yong}, D., \& {Frebel}, A.
  2010, \apjl, 722, L104, \dodoi{10.1088/2041-8205/722/1/L104}

\bibitem[{{Orkney} {et~al.}(2021){Orkney}, {Read}, {Rey}, {Nasim}, {Pontzen},
  {Agertz}, {Kim}, {Delorme}, \& {Dehnen}}]{Orkney.etal.2021}
{Orkney}, M. D.~A., {Read}, J.~I., {Rey}, M.~P., {et~al.} 2021, \mnras, 504,
  3509, \dodoi{10.1093/mnras/stab1066}

\bibitem[{{Ou} {et~al.}(2024){Ou}, {Chiti}, {Shipp}, {Simon}, {Geha}, {Frebel},
  {Mardini}, {Erkal}, \& {Necib}}]{Ou.etal.2024}
{Ou}, X., {Chiti}, A., {Shipp}, N., {et~al.} 2024, arXiv e-prints,
  arXiv:2403.00921, \dodoi{10.48550/arXiv.2403.00921}

\bibitem[{{Pace} {et~al.}(2022){Pace}, {Erkal}, \& {Li}}]{Pace.Erkal.Li.2022}
{Pace}, A.~B., {Erkal}, D., \& {Li}, T.~S. 2022, \apj, 940, 136,
  \dodoi{10.3847/1538-4357/ac997b}

\bibitem[{{Pace} \& {Li}(2019)}]{Pace.Li.2019}
{Pace}, A.~B., \& {Li}, T.~S. 2019, \apj, 875, 77,
  \dodoi{10.3847/1538-4357/ab0aee}

\bibitem[{{Pan} \& {Kravtsov}(2023)}]{Pan.Kravtsov.2023}
{Pan}, Y., \& {Kravtsov}, A. 2023, arXiv e-prints, arXiv:2310.08636,
  \dodoi{10.48550/arXiv.2310.08636}

\bibitem[{{Pe{\~n}arrubia} {et~al.}(2008){Pe{\~n}arrubia}, {Navarro}, \&
  {McConnachie}}]{Penarrubia.etal.2008}
{Pe{\~n}arrubia}, J., {Navarro}, J.~F., \& {McConnachie}, A.~W. 2008, \apj,
  673, 226, \dodoi{10.1086/523686}

\bibitem[{{Plez}(2012)}]{Plez.2012}
{Plez}, B. 2012, {Turbospectrum: Code for spectral synthesis}, Astrophysics
  Source Code Library, record ascl:1205.004.
\newblock \doeprint{1205.004}

\bibitem[{{Pontzen} \& {Governato}(2012)}]{Pontzen.etal.2012}
{Pontzen}, A., \& {Governato}, F. 2012, \mnras, 421, 3464,
  \dodoi{10.1111/j.1365-2966.2012.20571.x}

\bibitem[{{Read} {et~al.}(2016){Read}, {Agertz}, \& {Collins}}]{Read.etal.2016}
{Read}, J.~I., {Agertz}, O., \& {Collins}, M.~L.~M. 2016, \mnras, 459, 2573,
  \dodoi{10.1093/mnras/stw713}

\bibitem[{{Rimoldini} {et~al.}(2023){Rimoldini}, {Holl}, {Gavras}, {Audard},
  {De Ridder}, {Mowlavi}, {Nienartowicz}, {Jevardat de Fombelle},
  {Lecoeur-Ta{\"\i}bi}, {Karbevska}, {Evans}, {{\'A}brah{\'a}m}, {Carnerero},
  {Clementini}, {Distefano}, {Garofalo}, {Garc{\'\i}a-Lario}, {Gomel},
  {Klioner}, {Kruszy{\'n}ska}, {Lanzafame}, {Lebzelter}, {Marton}, {Mazeh},
  {Molinaro}, {Panahi}, {Raiteri}, {Ripepi}, {Szabados}, {Teyssier},
  {Trabucchi}, {Wyrzykowski}, {Zucker}, \& {Eyer}}]{gaia.variable.2023}
{Rimoldini}, L., {Holl}, B., {Gavras}, P., {et~al.} 2023, \aap, 674, A14,
  \dodoi{10.1051/0004-6361/202245591}

\bibitem[{{Robin} {et~al.}(2003){Robin}, {Reyl{\'e}}, {Derri{\`e}re}, \&
  {Picaud}}]{Robin.etal.2003}
{Robin}, A.~C., {Reyl{\'e}}, C., {Derri{\`e}re}, S., \& {Picaud}, S. 2003,
  \aap, 409, 523, \dodoi{10.1051/0004-6361:20031117}

\bibitem[{{Roderick} {et~al.}(2016){Roderick}, {Mackey}, {Jerjen}, \& {Da
  Costa}}]{Roderick.etal.2016}
{Roderick}, T.~A., {Mackey}, A.~D., {Jerjen}, H., \& {Da Costa}, G.~S. 2016,
  \mnras, 461, 3702, \dodoi{10.1093/mnras/stw1541}

\bibitem[{{Schlegel} {et~al.}(1998){Schlegel}, {Finkbeiner}, \&
  {Davis}}]{Schlegel.Finkbeiner.Marc.1998}
{Schlegel}, D.~J., {Finkbeiner}, D.~P., \& {Davis}, M. 1998, \apj, 500, 525,
  \dodoi{10.1086/305772}

\bibitem[{{Simon}(2019)}]{Simon.2019}
{Simon}, J.~D. 2019, \araa, 57, 375,
  \dodoi{10.1146/annurev-astro-091918-104453}

\bibitem[{{Simon} \& {Geha}(2007)}]{Simon.Geha.2007}
{Simon}, J.~D., \& {Geha}, M. 2007, \apj, 670, 313, \dodoi{10.1086/521816}

\bibitem[{{Simon} {et~al.}(2011){Simon}, {Geha}, {Minor}, {Martinez}, {Kirby},
  {Bullock}, {Kaplinghat}, {Strigari}, {Willman}, {Choi}, {Tollerud}, \&
  {Wolf}}]{Simon.etal.2011}
{Simon}, J.~D., {Geha}, M., {Minor}, Q.~E., {et~al.} 2011, \apj, 733, 46,
  \dodoi{10.1088/0004-637X/733/1/46}

\bibitem[{{Simon} {et~al.}(2017){Simon}, {Li}, {Drlica-Wagner}, {Bechtol},
  {Marshall}, {James}, {Wang}, {Strigari}, {Balbinot}, {Kuehn}, {Walker},
  {Abbott}, {Allam}, {Annis}, {Benoit-L{\'e}vy}, {Brooks}, {Buckley-Geer},
  {Burke}, {Carnero Rosell}, {Carrasco Kind}, {Carretero}, {Cunha}, {D'Andrea},
  {da Costa}, {DePoy}, {Desai}, {Doel}, {Fernandez}, {Flaugher}, {Frieman},
  {Garc{\'\i}a-Bellido}, {Gaztanaga}, {Goldstein}, {Gruen}, {Gutierrez},
  {Kuropatkin}, {Maia}, {Martini}, {Menanteau}, {Miller}, {Miquel}, {Neilsen},
  {Nord}, {Ogando}, {Plazas}, {Romer}, {Rykoff}, {Sanchez}, {Santiago},
  {Scarpine}, {Schubnell}, {Sevilla-Noarbe}, {Smith}, {Sobreira}, {Suchyta},
  {Swanson}, {Tarle}, {Whiteway}, {Yanny}, \& {DES
  Collaboration}}]{Simon.etal.2017}
{Simon}, J.~D., {Li}, T.~S., {Drlica-Wagner}, A., {et~al.} 2017, \apj, 838, 11,
  \dodoi{10.3847/1538-4357/aa5be7}

\bibitem[{{Slater} {et~al.}(2020){Slater}, {Ivezi{\'c}}, \&
  {Lupton}}]{Slater.etal.2020}
{Slater}, C.~T., {Ivezi{\'c}}, {\v{Z}}., \& {Lupton}, R.~H. 2020, \aj, 159, 65,
  \dodoi{10.3847/1538-3881/ab6166}

\bibitem[{{Starkenburg} {et~al.}(2017){Starkenburg}, {Martin}, {Youakim},
  {Aguado}, {Allende Prieto}, {Arentsen}, {Bernard}, {Bonifacio}, {Caffau},
  {Carlberg}, {C{\^o}t{\'e}}, {Fouesneau}, {Fran{\c{c}}ois}, {Franke},
  {Gonz{\'a}lez Hern{\'a}ndez}, {Gwyn}, {Hill}, {Ibata}, {Jablonka},
  {Longeard}, {McConnachie}, {Navarro}, {S{\'a}nchez-Janssen}, {Tolstoy}, \&
  {Venn}}]{Starkenburg.etal.2017}
{Starkenburg}, E., {Martin}, N., {Youakim}, K., {et~al.} 2017, \mnras, 471,
  2587, \dodoi{10.1093/mnras/stx1068}

\bibitem[{Strutz(2010)}]{Strutz.2010}
Strutz, T. 2010, Data Fitting and Uncertainty (A practical introduction to
  weighted least squares and beyond)

\bibitem[{{Tarumi} {et~al.}(2021){Tarumi}, {Yoshida}, \&
  {Frebel}}]{tarumi.etal.2021}
{Tarumi}, Y., {Yoshida}, N., \& {Frebel}, A. 2021, \apjl, 914, L10,
  \dodoi{10.3847/2041-8213/ac024e}

\bibitem[{{Tau} {et~al.}(2024){Tau}, {Vivas}, \&
  {Mart{\'\i}nez-V{\'a}zquez}}]{Tau.etal.2024}
{Tau}, E.~A., {Vivas}, A.~K., \& {Mart{\'\i}nez-V{\'a}zquez}, C.~E. 2024, \aj,
  167, 57, \dodoi{10.3847/1538-3881/ad1509}

\bibitem[{{Taylor}(2006)}]{Taylor.2006}
{Taylor}, M.~B. 2006, in Astronomical Society of the Pacific Conference Series,
  Vol. 351, Astronomical Data Analysis Software and Systems XV, ed.
  C.~{Gabriel}, C.~{Arviset}, D.~{Ponz}, \& S.~{Enrique}, 666

\bibitem[{{Usman} {et~al.}(2024){Usman}, {Ji}, {Li}, {Pace}, {Cullinane}, {Da
  Costa}, {Koposov}, {Lewis}, {Zucker}, {Belokurov}, {Bland-Hawthorn},
  {Ferguson}, {Hansen}, {Limberg}, {Martell}, {McKenzie}, \& {S5
  Collaboration}}]{Usman.etal.2024}
{Usman}, S.~A., {Ji}, A.~P., {Li}, T.~S., {et~al.} 2024, \mnras,
  \dodoi{10.1093/mnras/stae185}

\bibitem[{{van der Walt} {et~al.}(2011){van der Walt}, {Colbert}, \&
  {Varoquaux}}]{van.der.Walt.etal.2011}
{van der Walt}, S., {Colbert}, S.~C., \& {Varoquaux}, G. 2011, Computing in
  Science and Engineering, 13, 22, \dodoi{10.1109/MCSE.2011.37}

\bibitem[{{Vivas} {et~al.}(2020){Vivas}, {Mart{\'\i}nez-V{\'a}zquez}, \&
  {Walker}}]{Vivas.etal.2020}
{Vivas}, A.~K., {Mart{\'\i}nez-V{\'a}zquez}, C., \& {Walker}, A.~R. 2020,
  \apjs, 247, 35, \dodoi{10.3847/1538-4365/ab67c0}

\bibitem[{{Walker} {et~al.}(2009){Walker}, {Mateo}, {Olszewski},
  {Pe{\~n}arrubia}, {Evans}, \& {Gilmore}}]{Walker.etal.2009b}
{Walker}, M.~G., {Mateo}, M., {Olszewski}, E.~W., {et~al.} 2009, \apj, 704,
  1274, \dodoi{10.1088/0004-637X/704/2/1274}

\bibitem[{{Waller} {et~al.}(2023){Waller}, {Venn}, {Sestito}, {Jensen},
  {Kielty}, {Borukhovetskaya}, {Hayes}, {McConnachie}, \&
  {Navarro}}]{Waller.etal.2023}
{Waller}, F., {Venn}, K.~A., {Sestito}, F., {et~al.} 2023, \mnras, 519, 1349,
  \dodoi{10.1093/mnras/stac3563}

\bibitem[{{Walsh} {et~al.}(2008){Walsh}, {Willman}, {Sand}, {Harris}, {Seth},
  {Zaritsky}, \& {Jerjen}}]{Walsh.etal.2008}
{Walsh}, S.~M., {Willman}, B., {Sand}, D., {et~al.} 2008, \apj, 688, 245,
  \dodoi{10.1086/592076}

\bibitem[{{Wheeler} {et~al.}(2019){Wheeler}, {Hopkins}, {Pace},
  {Garrison-Kimmel}, {Boylan-Kolchin}, {Wetzel}, {Bullock}, {Kere{\v{s}}},
  {Faucher-Gigu{\`e}re}, \& {Quataert}}]{Wheeler.etal.2019}
{Wheeler}, C., {Hopkins}, P.~F., {Pace}, A.~B., {et~al.} 2019, \mnras, 490,
  4447, \dodoi{10.1093/mnras/stz2887}

\bibitem[{{White} \& {Frenk}(1991)}]{White.Frenk.1991}
{White}, S. D.~M., \& {Frenk}, C.~S. 1991, \apj, 379, 52,
  \dodoi{10.1086/170483}

\bibitem[{{Willman} \& {Strader}(2012)}]{Willman.Strader.2012}
{Willman}, B., \& {Strader}, J. 2012, \aj, 144, 76,
  \dodoi{10.1088/0004-6256/144/3/76}

\bibitem[{{Willman} {et~al.}(2005{\natexlab{a}}){Willman}, {Blanton}, {West},
  {Dalcanton}, {Hogg}, {Schneider}, {Wherry}, {Yanny}, \&
  {Brinkmann}}]{Willman.etal.2005a}
{Willman}, B., {Blanton}, M.~R., {West}, A.~A., {et~al.} 2005{\natexlab{a}},
  \aj, 129, 2692, \dodoi{10.1086/430214}

\bibitem[{{Willman} {et~al.}(2005{\natexlab{b}}){Willman}, {Dalcanton},
  {Martinez-Delgado}, {West}, {Blanton}, {Hogg}, {Barentine}, {Brewington},
  {Harvanek}, {Kleinman}, {Krzesinski}, {Long}, {Neilsen}, {Nitta}, \&
  {Snedden}}]{Willman.etal.2005b}
{Willman}, B., {Dalcanton}, J.~J., {Martinez-Delgado}, D., {et~al.}
  2005{\natexlab{b}}, \apjl, 626, L85, \dodoi{10.1086/431760}

\bibitem[{{Wolf} {et~al.}(2010){Wolf}, {Martinez}, {Bullock}, {Kaplinghat},
  {Geha}, {Mu{\~n}oz}, {Simon}, \& {Avedo}}]{Wolf.etal.2010}
{Wolf}, J., {Martinez}, G.~D., {Bullock}, J.~S., {et~al.} 2010, \mnras, 406,
  1220, \dodoi{10.1111/j.1365-2966.2010.16753.x}

\end{thebibliography}
\bibliographystyle{aasjournal}
\end{document}